\documentclass[12pt]{article}
\usepackage[round,colon,authoryear]{natbib}
\usepackage{bibentry}
\RequirePackage[OT1]{fontenc}
\RequirePackage{amsthm,amsmath,amsfonts,amssymb,{mathtools}}
\RequirePackage[colorlinks]{hyperref}
\usepackage{bbm}
\hypersetup{
 colorlinks=false,
 citecolor=Violet,
 linkcolor=Red,
 urlcolor=Blue}
\RequirePackage{epsfig, graphicx, color}
\RequirePackage{latexsym}
\RequirePackage{psfrag}

\allowdisplaybreaks

\usepackage{fullpage}
\usepackage{graphics,graphicx}
\usepackage{subcaption}
\usepackage{booktabs}
\usepackage{verbatim}
\usepackage{algorithm,algorithmicx,algpseudocode}

\newtheorem{theorem}{Theorem}
\newtheorem{lemma}{Lemma}

\newtheorem{corollary}{Corollary}
\newtheorem{remark}{Remark}

\newcommand\undermat[2]{%
  \makebox[0pt][l]{$\smash{\underbrace{\phantom{%
    \begin{matrix}#2\end{matrix}}}_{\text{$#1$}}}$}#2}

\newcommand{\bfm}[1]{\ensuremath{\mathbf{#1}}}

   \def\bA{\bfm A}  
   \def\bB{\bfm B}  
   \def\bC{\bfm C}  
\def\bd{\bfm d}     
   \def\bE{\bfm E}  \def\EE{\mathbb{E}}

     \def\LL{\mathbb{L}}

     \def\OO{\mathbb{O}}
     \def\PP{\mathbb{P}}
     
     \def\RR{\mathbb{R}}
     
   \def\bT{\bfm T}  
   \def\bU{\bfm U}  
   \def\bV{\bfm V}  \def\VV{\mathbb{V}}

   \def\bY{\bfm Y}

\def\calC{{\cal  C}} 
 
\def\calE{{\cal  E}} 
 
\def\calG{{\cal  G}}

\def\calL{{\cal  L}} 
\def\calM{{\cal  M}}

\def\calP{{\cal  P}}

\def\calU{{\cal  U}} 
\def\calV{{\cal  V}}

\newcommand{\bfsym}[1]{\ensuremath{\boldsymbol{#1}}}

            \def\bDelta {\bfsym {\Delta}}

           \def\bOmega {\bfsym {\Omega}}

\DeclareMathOperator{\argmin}{argmin}

\newdimen\biblioindent    \biblioindent=30pt

 at 8truept

\newcommand{\beq}{\begin{equation}}
  \newcommand{\eeq}{\end{equation}}
\newcommand{\beqn}{\begin{eqnarray}}
  \newcommand{\eeqn}{\end{eqnarray}}
\newcommand{\beqnn}{\begin{eqnarray*}}
  \newcommand{\eeqnn}{\end{eqnarray*}}

\numberwithin{equation}{section}

\usepackage {tikz}
\usetikzlibrary {positioning}

\usepackage{mathtools}

\newtheorem{claim}{Claim}
\newtheorem{condition}{Condition}
\newtheorem{definition}{Definition}
\newcounter{CondCounter}

\begin{document}
\title{Community Detection on Mixture Multi-layer Networks via Regularized Tensor Decomposition}
\author{Bing-Yi Jing, Ting Li, Zhongyuan Lyu and Dong Xia\\
Hong Kong University of Science and Technology
}
\date{(\today)}

\maketitle

\footnotetext[1]{
Jing and Li's research is partially supported by the Hong Kong RGC Grants GRF 16304419 and GRF 16305616. 
Lyu and Xia's research is partially supported by the Hong Kong RGC Grant ECS 26302019 and WeBank-HKUST project WEB19EG01-g.}

\begin{abstract}
We study the problem of community detection in multi-layer networks, where pairs of nodes can be related in multiple modalities. We introduce a general framework, i.e., mixture multi-layer stochastic block model (MMSBM), which includes many earlier models as special cases. We propose a tensor-based algorithm (TWIST) to reveal both global/local memberships of nodes, and memberships of layers. We show that the TWIST procedure can accurately detect the communities with small misclassification error as the number of nodes and/or number of layers increases. Numerical studies confirm our theoretical findings. To our best knowledge,  this is the first systematic study on the mixture multi-layer networks using tensor decomposition. The method is applied to two real datasets: worldwide trading networks and malaria parasite genes networks, yielding new and interesting findings.
\end{abstract}

\section{Introduction}

Networks arise in many areas of research and applications, which come in all shapes and sizes.  The most studied and best understood are static network models. Many other network models are also in existence, but have been less studied. One such example is the multi-layer networks, which are a powerful representation of relational data, and commonly encountered in contemporary data analysis (\cite{kivela2014multilayer}). The nodes in a multi-layer network represent the entities of interest and the edges in different layers indicate the multiple relations among those entities. Examples include brain connectivity networks, world trading networks, gene-gene interactive networks and so on. 
In this paper, we focus on the multi-layer networks with the same nodes set of each layer and there are no edges between two different layers. 

The study on multi-layer networks has received an increasing interest. Considering the dependency among the different layers, \cite{paul2016consistent} derives consistency results for the community assignments from the maximum likelihood estimators in two models. Consistency properties of various methods for community detection under the multi-layer stochastic block model are investigated in \cite{paul2017spectral}. Three different matrix factorization-based algorithms are employed in \cite{tang2009clustering}, \cite{nickel2011three} and \cite{dong2012clustering} separately. Common community structures for multiple networks are identified via two spectral clustering algorithms with theoretical guarantee in \cite{bhattacharyya2018spectral}.  In \cite{arroyo2019inference}, authors introduce the common subspace independent-edge multiple random graph model to describe a heterogeneous collection of networks with a shared latent structure and propose a joint spectral embedding of adjacency matrices to simultaneously and consistently estimate underlying parameters for each graph. Consistency results for a least squares estimation of memberships under the multi-layer stochastic block model framework are derived in \cite{lei2019consistent}. Several literature focus on recovering the network from a collection of networks with edge contamination. The original network is estimated from multiple noisy realizations utilizing community structure in \cite{le2018estimating} and low-rank expectation in \cite{levin2019recovering}. A weighted latent position graph model contaminated via an edge weight gross error model is proposed in \cite{tang2017robust} with an estimation methodology based on robust $L_q$ estimation followed by low-rank adjacency spectral decomposition. 

In applications, a random effects stochastic block model is proposed by \cite{paul2018random} for the neuroimaging data and a statistical framework with a significance and a robustness test for detecting common modules in the Drosophila melanogaster dynamic gene regulation network is proposed in \cite{zhang2017finding}. 

Most of the literature about community detection in multi-layer networks is limited to consistent membership setting, which means all the layers carry information about the same community assignment. However, in reality, different layers may have different community structures. For instance, in a social network, layers related with sports (people connected with the same sport hobbies) may have different community structure with layers about movie taste (people connected with similar movie taste). Understanding the large-scale structure of multi-layer networks is made difficult by the fact that the patterns of one type of link may be similar to, uncorrelated with, or different from the patterns of another type of link. These differences from layer to layer may exist at the level of individual links, connectivity patterns among groups of nodes, or even the hidden groups themselves to which each node belongs. In \cite{de2017community}, authors pointed out that, in order to do community detection on multi-layer networks, it is crucial to know which layers have related structure and which layer are unrelated, since redundant information across layers may provide stronger evidence for clear communities than each layer would on its own. Such situation is not clearly discussed in the works mentioned above. Although, in \cite{matias2017statistical}, authors introduced community structure variety as time varying, it is hard to be applied in general multi-layer networks without time ordering.

\begin{figure}
\centering
	\includegraphics[scale=.4]{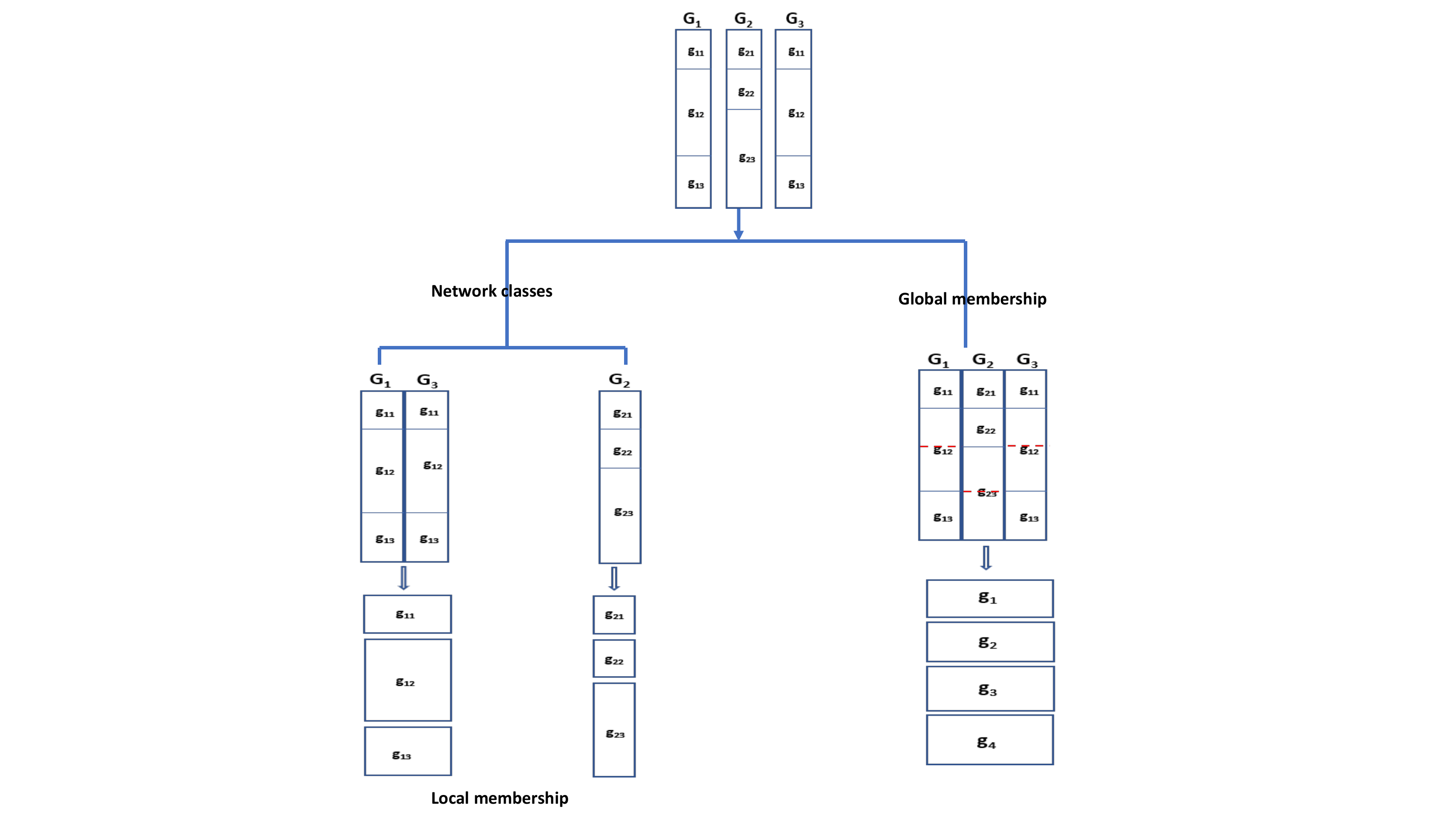}
	\caption{A toy example.} 
	\label{fig:toyexample}
\end{figure}

In this paper, we introduce a general framework, i.e., mixture multi-layer stochastic block model (MMSBM), and propose a tensor-based algorithm (TWIST) to reveal both global/local memberships of nodes, and memberships of layers.  
To fix ideas, we start with a simple motivating example, illustrated in Figure \ref{fig:toyexample}.  We have $L=3$ layers of networks $\{{\cal G}_1,{\cal G}_2,{\cal G}_3\}$, each containing 3 local communities. The 3 networks are of $m=2$ types: $\{{\cal G}_2\}$ and $\{{\cal G}_1,{\cal G}_3\}$. The community structure differs between  $\{{\cal G}_2\}$ and the other two networks as some members in the third community $g_{23}$ are in the second one $g_{12}$ in $\{{\cal G}_1\}$ and $\{{\cal G}_3\}$.  
Viewing the 3 layers of networks together, we notice that there are 4 global communities, in which members stay in all layers throughout. Clearly, the global communities are related to, but different from the local ones in each network.  Our interest lies in detecting both local as well as global community structures, which are of great value in theory and practice.  There is an increasing literature on the global community structure as mention earlier. However, to the best of our knowledge, there is no systematic investigation into detecting local and global community structure together.

\begin{figure}
\centering
	\includegraphics[scale=.4]{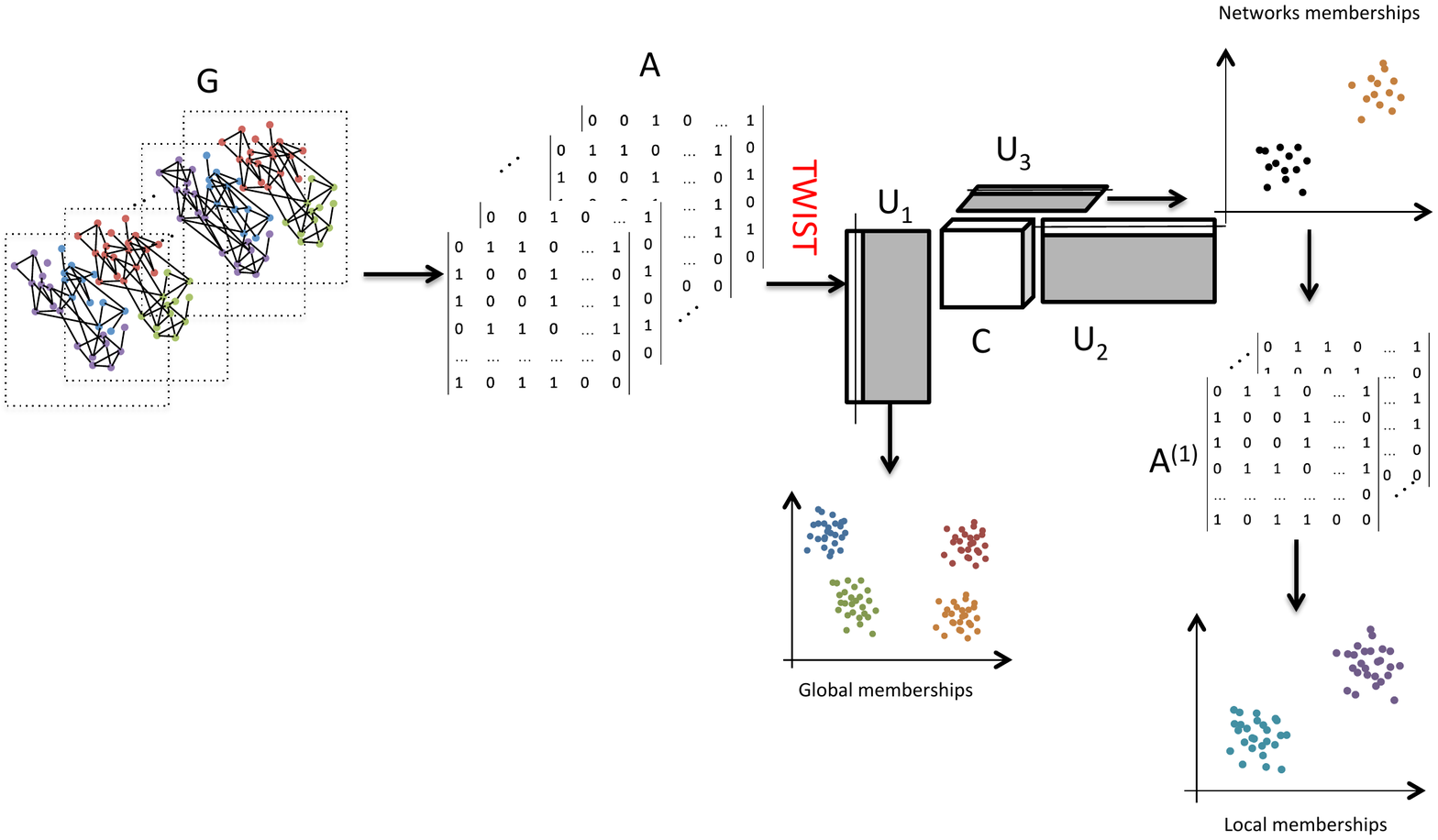}
	\caption{The general procedure of TWIST. } 
	\label{fig:process}
\end{figure}

Our line of attack can be illustrated via the following diagram in Figure \ref{fig:process}. First, we pool adjacency matrices from all layers of networks to form a tensor (multi-way array), and then apply the TWIST (to be introduced later) to obtain the global community structure as well as labels of each layer. We then group the layers of networks with the same labels, which will be used to detect local community structures. Details will be unfolded next. 

The main contributions of this paper are summarized as follows. 

First, we propose a very general model to handle the type of problems discussed above. To be more specific, we will introduce the so-called mixture multi-layer stochastic block model (MMSBM), which can characterize the different community structures among different layers of the multi-layer network. In some way, the MMSBM resembles the relatively well studied multi-layer stochastic block model (MLSBM) \cite{paul2017spectral, han2015consistent, paul2016consistent}. However, the MMSBM is more general in that it allows the multi-layer network to contain different block structures. Thus, the MMSBM not only allows each layer to have different community structures, but also can maintain the consistent structure in the network.  

Secondly, we propose a tensor-based method to study the MMSBM. The approach is referred to as the Tucker decomposition with integrated SVD transformation (TWIST). Unlike earlier approaches for multi-layer network analysis, TWIST can uncover the clusters of layers, the local and global membership of nodes simultaneously. On the theoretical front, we prove for MMSBM that TWIST can consistently recover the layer labels and global memberships of nodes under near optimal network sparsity conditions. In addition, network labels can be exactly recovered under a slightly stronger network sparsity condition. 
To the best of our knowledge, this is the first systematic study on statistical guarantees about community detection in a mixture multi-layer networks using tensor decomposition. Our primary technical tool is a sharp concentration inequality of sparse tensors which might be of independent interest.

Finally, two real-world applications of the proposed methodology demonstrated to be a powerful tool in analysing multi-layer networks. The algorithm is easy to use and can help practitioners  quickly uncover interesting findings, which would otherwise be difficult by using other tools. 

The rest of the paper is organized as follows. Section \ref{sec:model} introduces the mixture multi-layer stochastic block model (MMSBM) for describing the mixture structure. A new algorithm, the TWIST, is proposed in Section \ref{sec:method}. We explore the theoretical properties of the TWIST under the MMSBM in Section \ref{sec:theory}. Moreover, we make comparisons between our main results and the cutting-edge theoretical results. The advantages of the proposed method is numerically evaluated with several simulations in Section \ref{sec:simulation} and two real data examples in Section \ref{sec:realdata}. Section \ref{sec:discussion} gives concluding remarks and discussions. All the proofs are shown in the supplement.

\section{Model framework} 
\label{sec:model}
\subsection{Mixture multi-layer stochastic block model (MMSBM)}

The observed data contains $L$-layers of networks on the same set of vertices: $\calV=[n]:=\{1,2,\cdots,n\}$:
$$
 \calG = \{{\cal G}_l: \ l =1,\cdots, L\}. 
 $$
Assume that these networks are generated from a mixture of $m$ latent networks with probability $\pi = (\pi_1,...,\pi_m)$.  Denoting $\ell_l \in \{1,\cdots,m\}$ as a latent label of ${\cal G}_l$ with $1\le l \le L$, then $$ \PP(\ell_l=j)=\pi_j,  \quad \mbox{with}  \ \ \sum_{j=1}^m\pi_j=1$$

Assume that each of the $m$ classes of networks  satisfies the {\it stochastic block model} (SBM). More specifically, for $j\in[m]$, the $j$-th class SBM is described by the {\it membership matrix} $Z_j\in\{0,1\}^{n\times K_j}$ and the probability matrix $B_j\in[0,1]^{K_j\times K_j}$ where $K_j$ is the number of communities.  Each row of $Z_j$ has exactly one entry which is non-zero. For simplicity, we denote

\begin{itemize}
\item SBM$(Z_j, B_j) = $ the $j$-th SBM with parameter $Z_j$ and $B_j$, $j=1,...,m$.

\item $\calV^j_k = $ the $k$-th community in the $j$-th SBM.  So $\calV^j_k\subset \calV$ and $\cup_{k=1}^{K_j} \calV_k^j = \calV$. 

\item $L_j = \#\{l: \ell_l=j, 1\leq l\leq L\} =$ the number of layers generated by SBM$(Z_j, B_j)$. 
Clearly, $L=\sum_{j=1}^m L_j$. 

\item $\mathring{K}=K_1+\cdots+K_m$ and $\LL=\{\ell_l\}_{l=1}^L$ and $\VV^j:=\{\calV^j_k\}_{k=1}^{K_j}$.
\end{itemize}

The observed adjacency matrix $A_l \in\{0,1\}^{n\times n}$ of $\calG_l$ obeys Bernoulli distribution:
\begin{equation}\label{eq:Al_bern}
A_{l}(i_1,i_2)\sim_{iid} {\rm Bern}\big(Z_{\ell_l}(i_1,:)B_{\ell_l}Z_{\ell_l}(i_2,:)^{\top}\big)
 \  \Longleftrightarrow  \  
A_l \sim_{iid} {\rm Bern}\big(Z_{\ell_l}B_{\ell_l}Z_{\ell_l}^{\top}\big) 
\end{equation}
for all $i_1\leq  i_2 \in [n]$, where $Z(i,:)$ denotes the $i$-th row of $Z$. 

The resulting model is referred to as ``mixture multi-layer stochastic block model" (MMSBM).

\subsection{Adjacency tensor and its decomposition}
Observing the $L$ layers of networks, we define the adjacency tensor $\bA\in\RR^{n\times n\times L}$ so that $\bA$'s $l$-th slice
$$
A(:,:,l)=A_l,\quad \forall 1\leq l\leq L.
$$  
See \cite{kolda2009tensor} for an introduction to tensor and applications. 
It follows from (\ref{eq:Al_bern}) that 
$$
\EE (A_l\vert \ell_l)=Z_{\ell_l}B_{\ell_l}Z_{\ell_l}^{\top},\quad \forall 1\leq l\leq L,
$$
from which we can derive the following tensor representation, whose proof is in the Appendix. 

\begin{lemma}[Tensor representation]  \label{lem:EA_decomp}
We have
\begin{equation} \label{eq:barZ}
\EE(\bA |\LL)=\bB  \times_1 \bar{Z} \times_2 \bar{Z} \times_3 W,
\end{equation}
where $\LL=\{\ell_l\}_{l=1}^L$ and 
\begin{itemize}
\item $\bar{Z}=(Z_1,Z_2,\cdots,Z_m)\in\{0,1\}^{n\times \mathring{K}}$ is the {\bf global} membership matrix, whereas each $Z_j$ is the {\bf local} membership matrix,

\item $W=(e_{\ell_1},e_{\ell_2},\cdots,e_{\ell_L})^{\top} \in\{0,1\}^{L\times m}$ is the network label matrix 
with each row of $W$ having exactly one non-zero entry, and $e_j\in\RR^m$ being the $j$-th canonical basis vector,

\item  $\bB \in\RR^{\mathring{K}\times \mathring{K}\times m}$ is a 3-way probability tensor whose $j$-th frontal slide is
$$
B(:,:,j)={\rm} diag(0_{K_1},\cdots,0_{K_{j-1}},B_j,0_{K_{j+1}},\cdots,0_{K_m}),\quad 1\leq j\leq m
$$
with $0_K$ being a $K\times K$ zero matrix. 
\end{itemize}
\end{lemma}

 \subsection{Local versus global memberships via Tucker decomposition}
 
 The matrix $\bar{Z}$ defined in Lemma~\ref{lem:EA_decomp} suggests the existence of {\it global community structures}. 
We say that two nodes $i_1$ and $i_2$ belong to the same global community if and only if they belong to the same local community for all the $m$ classes of SBM, i.e., 
$$ \bar{Z}(i_1,:)=\bar{Z}(i_2,:). $$
Let $\bar{K}$ denote the number of global communities. Clearly, $\max_j K_j\leq \bar{K}\leq \prod_j K_j$. Denote $\bar{\VV}=\{\bar{\calV}_k\}_{k=1}^{\bar{K}}$ the global community clusters such that $\cup_{k=1}^{\bar{K}}\bar{\calV}_k=\calV$. Therefore, for two nodes $i_1 \ne i_2$, 
 \begin{equation}\label{eq:barVk}
 \{i_1, i_2\} \in\bar{\calV}_k \qquad \Longleftrightarrow \qquad \{i_1, i_2\} \in \calV^j_{k_j},   \quad k_j\in[K_j], \ \ \forall j\in[m], 
 \end{equation}

 Let $r={\rm rank}(\bar{Z})$ denote the rank of $\bar{Z}$.  We hereby write the thin SVD of $\bar{Z}$ as
 \begin{equation}\label{eq:barZ_svd}
 \bar{Z}=\bar{U}\bar{D}\bar{R}^{\top}
 \end{equation}
where $\bar{U}\in\RR^{n\times r}$, $\bar{R} \in \RR^{(\mathring{K})\times r}$ have orthonormal columns, and $\bar D$ is the singular value diagonal matrix 
$$
\bar{D} = diag(\sigma_1(\bar D), \cdots, \sigma_r(\bar D))  \in \RR^{r\times r},  
\qquad \sigma_1(\bar D)\geq \cdots\geq \sigma_r(\bar D)>0. 
$$
The global community structure can be checked by $\bar{U}$ as in Lemma~\ref{lem:barU_sep}.
 \begin{lemma}\label{lem:barU_sep}
 For $i_1\in\bar{\calV}_{k_1}$ and $i_2\in\bar{\calV}_{k_2}$ with $k_1\neq k_2$, then,
 $$
 \|\bar{U}(i_1,:)-\bar{U}(i_2,:)\|_{\ell_2}\geq \frac{1}{\sigma_1(\bar{D})}.
 $$
 \end{lemma}
 
 By (\ref{eq:barZ_svd}), the population adjacency tensor $\EE(\bA|\LL)$ admits the {\it Tucker decomposition} as 
 \begin{equation}\label{eq:A_tucker}
 \EE(\bA|\LL)=\bar{\bC}\times_1\bar{U}\times_2\bar{U}\times_3\bar{W}
 \end{equation}
 where the core tensor $\bar \bC\in\RR^{r\times r\times m}$ is defined by
 \begin{equation}\label{eq:barC}
 \bar{\bC}=\bB\times_1(\bar{D}\bar{R}^{\top})\times_2 (\bar{D}\bar{R}^{\top})\times_3 D_L^{1/2}
 \end{equation}
 and $\bar{W}=WD_L^{-1/2} \in \RR^{L\times m}$ so that $\bar{W}^{\top}\bar{W}=I_m$, and the diagonal matrix 
 $$
 D_L={\rm diag}(L_1,L_2,\cdots,L_m).
 $$  
We assume that $\bar{\bC}$ has Tucker ranks $(r,r,m)$. Further assume that $m\leq r$, which is reasonable as one new type of network will introduce at least one new global community. 
  
The decomposition (\ref{eq:A_tucker}) shows that the singular vectors of $\EE(\bA|\LL)$ contain the latent network information.
More exactly, the singular vectors in the $1$-st dimension of $\EE(\bA|\LL)$ could identify the global community structures and singular vectors in the $3$-rd dimension could identify the latent network labels.  After identifying the latent network labels, a post-processing procedure can identify the local community structures. 
 
\section{Methodology: TWIST} \label{sec:method}

By observing the multi-layer networks $\{\calG_l\}_{l=1}^L$ satisfying model (\ref{eq:Al_bern}), our goals are to: 
\begin{enumerate}
\item[(1)] recover the global community structures of vertices $\{\bar{\calV}_k\}_{k=1}^{\bar{K}}$; 
\item[(2)] identify network classes $\{\ell_l\}_{l=1}^L$, and grouping networks with the same class;
\item[(3)] recover the local community structures of the vertices $\VV^j:=\{\calV_{k}^j:  k\in [K_j]\}$ for all $j\in[m]$. 
\end{enumerate}
Note that in order to efficiently recover the local community structures, it is necessary to first identify the network classes. As a result, task (3) usually follows from task (2). 

By the decomposition of oracle tensor (\ref{eq:A_tucker}), the singular vectors $\bar{U}$ contains information of global memberships since its column space comes from $\bar{Z}$. Additionally, the singular vectors $\bar{W}$ contains information of network classes. Therefore, task (1) and task (2) are both related with the Tucker decomposition of oracle tensor $\EE(\bA|\LL)$. Since the oracle is unavailable, we seek a low-rank approximation of $\bA$.

\subsection{Tucker decomposition with integrated SVD transformation (TWIST)}  

In order to utilize the low rank structure of the tensor and the non-negative property of the elements, we propose a new algorithm called Tucker decomposition with integrated SVD transformation (TWIST). The general procedure is summarized below and illustrated in Figure~\ref{fig:process}. 

\begin{itemize}
\item {\bf Step 1: Decomposition of adjacency tensor} \\
Apply the regularized tensor power iterations to $\bA$ to obtain its low-rank approximation. The outputs are $\widehat U$ and $\widehat W$. Details are given in Algorithm~\ref{algo:Power Iterations}. 

\smallskip

\item {\bf Step 2: Global memberships} \\
Apply the standard K-means algorithm on the rows of $\widehat U$ to identify the global community memberships and output $\widehat{\bar \VV}=\{\widehat{\bar{\calV}}_k\}_{k=1}^{\bar{K}}$.
\smallskip

\item {\bf Step 3: Network classes}  \\
Use the rows of $\widehat W$ to identify the network classes and output the network classes: $\widehat\LL=\{\hat\ell_l\in[m]\}_{l=1}^L$. We can use either the standard K-means or the sup-norm related algorithm (Algorithm~\ref{algo:network}).

\smallskip

\item {\bf Step 4: Local memberships}  \\
We can find the local membership $\VV^j=\{\calV_{k}^j\}$ by focusing on networks with the same labels (\cite{lei2015consistency, rohe2011spectral}).  More precisely, for each $j\in\{1,...,m\}$, we can apply K-means either
\begin{itemize}
\item to the sum of those networks with the same label $\sum_{l: \hat\ell_l=j} A_l$, or
\item to the sub-tensor $A(:,:,\{l: \hat\ell_l=j\})$), those slides with the same labels. 
\end{itemize}
Outputs are $\widehat\VV^j=\{\widehat{\calV}^j_{k}\}_{k=1}^{K_j}$. 
\end{itemize}

\smallskip

\begin{algorithm}
	\caption{Regularized power iterations for sparse tensor decomposition}
	\label{algo:Power Iterations}
	\begin{algorithmic}[20]
		\Require
		$\bA_{(n\times n \times L)}$, warm initialization $\widehat U^{(0)}$ and $\widehat W^{(0)}$\\
		\hspace{1.5cm}  maximum iterations ${\rm iter}_{\max}$ and regularization parameters $\delta_1,\delta_2>0.$
		\Ensure
		$\widehat{U}$ and $\widehat W$
		\State Set counter ${\rm iter} = 0.$
        \While{${\rm iter} < {\rm iter}_{\max}$}
         \State Regularization: $\widetilde U^{({\rm iter})}\leftarrow \calP_{\delta_1}(\widehat{U}^{({\rm iter})})$ and $\widetilde W^{({\rm iter})}\leftarrow \calP_{\delta_2}(\widehat{W}^{({\rm iter})})$ by (\ref{eq:reg}).
        \State ${\rm iter} \leftarrow {\rm iter} + 1$ 
          \State Set $\widehat U^{({\rm iter})}$ to be the top $r$ left singular vectors of 
          $\calM_1\big(\bA \times_{2}\widetilde{U}^{({\rm iter}-1)\top} \times_3 \widetilde{W}^{({\rm iter}-1)\top}\big).$ 
          \State set $\widehat W^{({\rm iter})}$ to be the top $m$ left singular vectors of $\calM_3\big(\bA \times_1 \widetilde{U}^{({\rm iter}-1)\top} \times_2 \widetilde{U}^{({\rm iter}-1)\top} \big)$.
        \EndWhile
        \State Return $\widehat{U}\leftarrow \widehat{U}^{({\rm iter})}$ and $\widehat{W}\leftarrow\widehat{W}^{({\rm iter})}$.
    	\end{algorithmic}
\end{algorithm}

\begin{algorithm}
\caption{Network clustering by sup-norm K-means}
\label{algo:network}
\begin{algorithmic}[20]
\Require $\widehat W$, number of clusters $m$ and threshold $\varepsilon\in(0,1)$
\Ensure  Network labels $\widehat\LL=\{\hat\ell_l\}_{l=1}^L$
\State Initiate $\calC\leftarrow\{1\}$, $\hat\ell_1\leftarrow1$, $k\leftarrow 1$ and $l\leftarrow 2$.
\While{$l\leq L$}
\State Compute $j\leftarrow \argmin_{j\in\calC} \|\widehat W(l,:)-\widehat{W}(j,:)\|$
\If{$\|\widehat W(l,:)-\widehat W(j,:)\|>\varepsilon$}
\State $k\leftarrow k+1$
\State $\hat\ell_l\leftarrow k$
\State $\calC\leftarrow \calC\cup \{l\}$
\Else
\State $\hat\ell_l\leftarrow \hat\ell_j$
\EndIf
\State $l\leftarrow l+1$
\EndWhile
\If{$k>m$ (or $k<m$)}
\State  Set $\varepsilon\leftarrow 2\varepsilon$ (or set $\varepsilon\leftarrow \varepsilon/2$); Re-run the algorithm. 
\Else
\State Output $\widehat\LL=\{\hat\ell_l\}_{l=1}^L$
\EndIf
\end{algorithmic}
\end{algorithm}

\subsection{Features about TWIST}

There are several key features concerning the TWIST. 

\subsubsection*{Warm starts for $\widehat U^{(0)}$ and $\widehat W^{(0)}$ in Algorithm 1}

Computing the optimal low-rank approximation of a tensor $\bA$ is NP-hard  in general; see \cite{hillar2013most}. 
Algorithms with random initializations are almost always trapped in non-informative local minimals which can be nearly orthogonal to the truth, see \cite{arous2019landscape}.  To avoid these issues, tensor decomposition algorithms usually run from a warm starting point \cite{zhang2018tensor, xia2019polynomial, xia2017statistically, richard2014statistical, jain2014provable, ke2019community, zhang2019cross, sun2017provable, cai2019nonconvex, wang2018learning}.

In Section~\ref{sec:init}, we will introduce a warm initialization algorithm, obtained by applying a spectral method for initializing $\widehat{U}^{(0)}$ by summing up all the network layers. 
Initialization of $\widehat W^{(0)}$ is easy whenever $\widehat U^{(0)}$ is available. We show in Lemma \ref{lem:initialization} that these initializations can indeed improve estimation accuracy. 

\subsubsection*{Regularized power iterations for sparse tensor decomposition}

Adjacency matrices from some layers are often very sparse and the individual layers are even disconnected graphs. For example, in the Malaria parasite genes networks given in Section~\ref{sec:realdata},  three out of nine networks are very sparse and and disconnected.  Under these circumstances, the popular tensor power iteration algorithm, i.e., high-order orthogonal iterations (HOOI, see \cite{sheehan2007higher}) may not work. In fact, its statistical optimality was proved by \cite{zhang2018tensor} only for dense tensors, while its properties on sparse random tensors remain much more challenging.

To handle sparse random tensors, we employ a regularized tensor power iteration algorithm in Algorithm~\ref{algo:Power Iterations}, which was used in \cite{ke2019community} to deal with sparse hypergraph networks.  Regularizations to singular vectors $\widehat U^{(t)}, \widehat W^{(t)}$ are applied before each power iteration. We take $\widehat U^{(t)}$ for example as $\widehat W^{(t)}$ can be treated similarly. The regularization $\calP_{\delta}(U)$ is done by
\begin{equation}\label{eq:reg}
\calP_{\delta}(U)={\rm SVD}_r(U_{\star}) \quad {\rm where}\quad U_{\star}(i,:):=U(i,:)\cdot\frac{\min\{\delta,\|U(i,:)\|\}}{\|U(i,:)\|}\quad i\in[n],
\end{equation}
The effect of regularization is to dampen the influence of ``large" rows of $\widehat U^{(t)}$, which is due to the communities of small sizes, besides stochastic errors.  
Following Lemma~\ref{lem:incoh}, the true singular vectors $\bar{U}$ are incoherent with $\max_j\|e_j^{\top}\bar{U}\|=O(\sqrt{r/n})$ if community sizes are balanced. In practice, we suggest 
\begin{equation}\label{eq:delta_algo}
\hat\delta_1 =2\sqrt{r}\cdot \frac{\max_{1\leq i\leq n}{\rm deg}_i}{\sqrt{\sum_i {\rm deg}_i^2}}
\quad {\rm and}\quad\hat \delta_2=2\sqrt{m}\cdot \frac{\max_{1\leq l\leq L} {\rm neg}_l}{\sqrt{\sum_l {\rm neg}_l^2}}
\end{equation}
where the node degree ${\rm deg}_i=\sum_{j, l}A(i,j,l)$ and layer degree ${\rm neg}_l=\sum_{i,j}A(i,j,l)$.

\subsubsection*{K-means with sup-norm distance}

Clearly, the accuracy of local membership clustering (Step 4) hinges on the reliability of layer labelling. In Algorithm~\ref{algo:network}, 
a sup-norm version of K-means is applied to the singular vectors $\widehat W$ obtained from Algorithm~\ref{algo:Power Iterations}, and then outputs the network labels $\widehat\LL=\{\hat\ell_l\}_{l=1}^L$. The sup-norm K-means has recently been extensively investigated and shown to perform well in network community detection. See, e.g. \cite{chen2019spectral,abbe2017entrywise, kim2018stochastic} and references therein. 

The rationale of Algorithm~\ref{algo:network} is that when the rows of $\bar{W}$ are well-separated (similar to Lemma~\ref{lem:barU_sep}), a row-wise screening of $\widehat W$ can immediately recover the true network labels as long as the row-wise perturbation bound of $\widehat W-W$ is small enough. As shown in Section~\ref{sec:main_theory}, Algorithm~\ref{algo:network} guarantees exact clustering of networks under weak conditions. 
 
In Steps 2-4, one could use alternative methods other than K-means clustering, which might improve its performances. For example, we can use DBSCAN, Gaussian mixture model, the SCORE method (\cite{jin2015fast, jin2017estimating, ke2019community}).

\section{Preliminary Results} \label{sec:theory}

\subsection{Notations and definitions}

For ease of exposition, we introduce the following notations. 
\begin{itemize}
\item Denote $c, c_j, C, C_j,C_j',\ j\ge 1$ as generic constants, which may vary from line to line. 
\item Denote $e_k$ as the $k$-th canonical basis vector in Euclidean space (i.e., with only the $k$-th entry equal to 1 and others 0), 
whose dimension depends on each context. 
 
\item For a matrix $M=\{m_{ij}\}$, let 

$\sigma_i(M) = $ the $i$-th largest singular value of $M$,

$\|M\|_{\max} = \max_{i,j} |m_{ij}|$, the maximal absolute value of all entries of $M$,

$\|M\|=\max\{\sigma_i(M^TM)\}^{1/2}$, the spectral norm (Euclidean norm for vectors).

\item For a $d_1\times d_2\times d_3$ tensor $\bT$, let $\calM_j(\bT)$ be a $d_j\times (d_1d_2d_3/d_j)$ matrix by unfolding $\bT$ in the $j$-th dimension.
\end{itemize}


\subsection{Signal Strengths}

Recall the low-rank decomposition of $\EE(\bA|\LL)$ in (\ref{eq:A_tucker}) with 
$$
\EE(\bA|\LL)=\bar{\bC}\times_1 \bar{U}\times_2 \bar{U}\times_3 \bar{W}
$$
where $\LL=\{\ell_l\}_{l=1}^L$ and the core tensor 
\begin{eqnarray*}
\bar{\bC} 
&=& \bB\times_1(\bar{D}\bar{R}^{\top})\times_2 (\bar{D}\bar{R}^{\top})\times_3 D_L^{1/2} \\ 
&=& \bar{\bB}\times_1 \bar{D}\times_2 \bar{D}\times_3 D_L^{1/2}
\end{eqnarray*}
where 
\begin{equation}\label{eq:barB}
\bar{\bB}=\bB\times_1 \bar{R}^{\top}\times_2 \bar{R}^{\top} \in\RR^{r\times r\times m}
\end{equation}
Denote the signal strengths of $\bar{\bC}$ and $\bar{\bB}$ by $\sigma_{\min}(\bar{\bC})$ and $\sigma_{\min}(\bar{\bB})$, respectively, where 
\begin{equation} \label{eq:ss}
\sigma_{\min}({\bf T})=\min\big\{\sigma_{r_j}\big(\calM_j({\bf T})\big): j=1,2,3\big\}
\end{equation}
if $\bT$ has Tucker ranks $(r_1,r_2,r_3)$. 

\smallskip
 
We give a lower bound for the signal strength in the population adjacency tensor. The following conditions can sometimes greatly simplify our presentation.
\begin{condition}
Assume that
\begin{itemize} 
     \item $(A1)$: \ $\sigma_{\min}(\bar{\bB})\geq c_1p_{\max}$, where $p_{\text{max}}=\max_{i,j,l}[\mathbb{E}A_l]_{ij}$,
     \item $(A2)$: \ $\bar{D}$ is well-conditioned, i.e. \
                         $ \sigma_1(\bar{D})\leq \kappa_0  \sigma_r(\bar{D}).$
  \item (A3): \ Minimal network balance condition:  $L_{\min} \asymp L/m$, where $L_{\min}=\min_{1\le j \le m} L_j$. 
     \item (A4): \ Maximal network balance condition:  $L_{\max} \asymp L/m$, where $L_{\max}=\max_{1\le j \le m} L_j$.
\end{itemize} 
\end{condition}
\begin{lemma}[signal strength] \label{lem:ss}
If conditions (A1) and (A2) hold, then we have  
$$
\sigma_{\min}(\bar{\bC})\geq r^{-1}\kappa_0^{-2}\cdot m p_{\max} n \sqrt{L_{\min}}.
$$ 
Further if (A3) holds and $m, r, \kappa_0$ are fixed, then 
$$\sigma_{\min}(\bar{\bC}) \geq C \sqrt{L} np_{\max}. $$ 
\end{lemma}

Condition (A1) is mild since we assumed that $\bar{\bC}$ has Tucker ranks $(r,r,m)$ implying that $\bar{\bB}$ also has the same Tucker ranks. By Lemma~\ref{lem:ss}, the signal strength of $\bar{\bC}$ is characterized by the overall network sparsity. 

\smallskip

\subsection{Incoherence Property}
Theoretically, the ideal regularization parameters in Algorithm~\ref{algo:Power Iterations} are
$$
\delta_1=\max_j\|e_j^{\top}\bar{U}\|, \qquad \mbox{and} \qquad \delta_2=\max_j\|e_j^{\top}\bar{W}\|.
$$ 
Incoherence property ensures that singular vectors $\bar{U}$ and $\bar{W}$ are not too correlated with or incoherent to the standard basis ${e_j}$'s, as stated in the next lemma,
which the sharp convergence rates of regularized tensor power iteration algorithm rely crucially on.  
\begin{lemma}[Incoherence of $\bar{U}$ and $\bar{W}$]  \label{lem:incoh}
If conditions (A1) and (A2) hold, we have  
$$
\delta_1 \leq \kappa_0\sqrt{\frac{r}{n}} \quad {\rm and} \quad 
\delta_2 \leq \frac{\kappa_0 r}{m\sqrt{L_{\min}}}.
$$
\end{lemma}

\smallskip

Denote $A\asymp B$ iff  $A =O(B)$ and $B=O(A)$. \ 
Then under conditions (A1)-(A3), it follows from Lemma \ref{lem:incoh}:
$$
\delta_2 \leq C \kappa_0r/\sqrt{mL}.
$$ 
 
\smallskip

\subsection{Tensor incoherent norms and a concentration inequality}
Given a random tensor $\mathbf{A}$, we write 
$$
\bA=\EE\bA+(\bA-\EE\bA) = \mbox{the signal +  noise part}.
$$ 
A tensor norm is needed to measure the size of the noise. In order to deal with extremely sparse tensors, we will adopt the following definition, first introduced in \cite{yuan2017incoherent}. 
\begin{definition}[Tensor incoherent norm \cite{yuan2017incoherent}] For $\delta\in(0,1]$ and $k=1,2,3$, define
$$
\Vert\mathbf{A}-\mathbb{E}\mathbf{A}\Vert_{k,\delta}:=\sup\nolimits_{\bU\in\mathcal{U}_k(\delta)}\ \langle\mathbf{A}-\mathbb{E}\mathbf{A},\bU\rangle
$$
where $\mathcal{U}_k(\delta):=\{u_1\otimes u_2\otimes u_3:\Vert u_j\Vert_{\ell_2}\le1, \forall j; \Vert u_k \Vert_{\ell_\infty}\le\delta \}$, and $\Vert u \Vert_{\ell_p}$ is the $l_p$-norm of $u$. 
\end{definition}

We now present a concentration inequality for tensor incoherent norms for sparse random tensors, which is essential in proving  Theorem~\ref{thm:power_iteration} and Corollary~\ref{cor:PI} later.  

\begin{theorem}[A concentration inequality for tensor incoherent norm] \label{thm:tensor_CI} 
Suppose that $L\leq n$ and $Lnp_{\text{max}}\ge\log n$. Denote $n_1=n_2=n$ and $n_3=L$. Then for $k=1,2,3$, we have
\begin{equation*}
\begin{aligned}
\mathbb{P} \left\{ \Vert \mathbf{A} - \mathbb{E} \mathbf{A} \Vert_{k,\delta} \ge 3t \right\}
\le \frac{2}{n^2}+10(\log n)^2\lceil \log_2\delta^2n_k \rceil 
\left[\exp\left(-\frac{t^2}{C_3p_{\text{max}}}\right)+\exp\left(-\frac{3t}{C_4\delta}\right)\right]
\end{aligned}
\end{equation*}
provided 
\begin{equation*}
t \ge \max\left\{C_1 , C_2 \delta \sqrt{n_k} \log(n) \right\} \sqrt{np_{\text{max}}} \log(\delta^2n_k) \log(n). 
\end{equation*}
\end{theorem}

\smallskip

We make several remarks concerning the inequality. 

\begin{enumerate}
 \item  The bound in Theorem~\ref{thm:tensor_CI} is sharper than that in \cite{yuan2017incoherent}, in order to deal with extremely sparse networks. Analogous results were previously established for sparse hypergraph networks (\cite{ke2019community}), however, the dimension sizes ($n$ and $L$) in our model can be drastically different from (\cite{ke2019community}), which needs more careful treatments. 

 \item   Clearly, if $\delta=1$, $\|\bA-\EE\bA\|_{k,\delta}$ reduces to the standard tensor operator norm $\|\bA-\EE\bA\|$. 
From Lemma~\ref{lem:ss}, the signal strength is of the order $\sqrt{L}np_{\max}$. By using the standard operator norm to control $\bA-\EE\bA$, the success of power iterations then requires, with high probability, that
\begin{align}\label{eq:weak_cond}
\sqrt{L}np_{\max}\geq C_1'\|\bA-\EE\bA\|.
\end{align}
Under the condition $Lnp_{\max}\gg 1$, it is easy to check (by the maximum number of non-zero entries on the fibers of $\bA-\EE\bA$) that $\|\bA-\EE\bA\|\gg 1$ with high probability (see \cite[Theorem~2]{lei2019consistent}).  Consequently, by using the  standard tensor operator norm to control the size of noise part, condition (\ref{eq:weak_cond}) requires that $\sqrt{L}np_{\max}\gg 1$ rather than the optimal condition $Lnp_{\max}\gg 1$.    Indeed, if $\delta_1=O(1/\sqrt{n})$ and $\delta_2=O(1/\sqrt{L})$, Theorem~\ref{thm:tensor_CI} shows that $\|\bA-\EE\bA\|_{1,\delta_1}, \|\bA-\EE\bA\|_{3,\delta_2}=O_p(\sqrt{np_{\max}})$ up to some logarithmic factor. It can be much smaller than $\|\bA-\EE\bA\|\gg1$ (w.h.p.) when $L$ is large. 

\item To apply tensor incoherent norms to analyze the convergence property of power iterations, it is necessary to prove that $\{\widehat U^{(t)}\}$ and $\{\widehat W^{(t)}\}$ are incoherent. It is possible to generalize the methods in \cite{koltchinskii2016perturbation, xia2019sup, xia2019statistical, cai2019subspace} for this purpose whose actual proof can be very involved. For simplicity, we adopt an auxiliary regularization step (\ref{eq:reg}) to truncate those singular vectors. 

\end{enumerate}

\section{Main results}\label{sec:main_theory}

\subsection{Error Bound of Regularized Power Iteration}

Theorem~\ref{thm:power_iteration} states that regularized power iteration method (Algorithm~\ref{algo:Power Iterations}) works if we have a warm initialization and a strong enough signal-to-noise ratio. 
These conditions are typically required (see, e.g., \cite{zhang2018tensor, xia2017statistically, ke2019community, xia2019polynomial}) and generally unavoidable (\cite{zhang2018tensor}). 

For $\widehat{V}, V\in\OO_{p,r} = \{V\in\mathbb{R}^{p\times r}:V^TV=I_r \}$, the distance between their column spaces is 
$$
{\rm d}(\hat{V},V):=\inf_{O\in\mathbb{O}_{r,r}}\Vert\hat{V}-VO\Vert
$$
Define 
$$
{\rm Err}(t)=\max\{{\rm d}(\widehat U^{(t)}, \bar U),\ {\rm d}(\widehat W^{(t)},\bar{W}) \}. 
$$ 
We have the following result. 
\begin{theorem}[General convergence results of regularized power iterations]\label{thm:power_iteration}
Assume that 
\begin{itemize} 
\item the initializations $\widehat{U}^{(0)}$ and $\widehat{W}^{(0)}$ are warm, i.e., ${\rm Err}(0) \leq 1/4$,
\item the signal strength of $\bar{\bC}$ satisfies
$$
\frac{\sigma_{\min}(\bar{\bC})}{\sqrt{r\wedge 2m}}
\geq \Big(C_1 +C_2 \big((\delta_1\sqrt{n})\vee (\delta_2\sqrt{L})\big)\log n\Big)  \log\big(\delta_1^2n\vee \delta_2^2L\big) \sqrt{np_{\max}}\log(n).
$$
\end{itemize} 
Then with probability at least $1-2n^{-2}$, 
\begin{enumerate} 
\item for all $t \le t_{\max}$, we have
$$
{\rm Err}(t) \leq \frac{1}{2}\cdot{\rm Err}({t-1})+C_3\frac{\sqrt{np_{\max}\log n}+\delta_1\delta_2\log n}{\sigma_{\min}(\bar{\bC})}
$$

\item  for $t \ge t_{\max}= C \log\big(\sigma_{\min}(\bar{\bC})/(\sqrt{np_{\max}}+\delta_1\delta_2)\big) $ iterations, we have
$$
{\rm Err}({t_{\max}}) \leq C_3\frac{\sqrt{np_{\max}\log n}+\delta_1\delta_2\log n}{\sigma_{\min}(\bar{\bC})}.
$$
\end{enumerate} 
\end{theorem}

Theorem~\ref{thm:power_iteration} holds true on general tensor structures. Specializing Theorem~\ref{thm:power_iteration} to the mixture multi-layer network model (Section~\ref{sec:model}) yields the following corollary. 

\begin{corollary}\label{cor:PI}
Assume that (A1)-(A3) hold. Further assume 
\begin{itemize} 
\item the initializations $\widehat{U}^{(0)}$ and $\widehat{W}^{(0)}$ are warm, i.e., ${\rm Err}(0) \leq 1/4$,
\item the signal strength of $\bar{\bC}$ satisfies
\begin{equation} \label{eq:sparsity_cond}
\sqrt{Lnp_{\max}} \geq \big( C_1 +C_2\kappa_0^2 (r/\sqrt{m}) \log(n) \big) r \kappa_0^2 \log(\kappa_0r) \log(n).
\end{equation}
\end{itemize} 
Then with probability at least $1-2n^{-2}$, after at most $t_{\max}=O(\log(n))$ iterations,
$$
{\rm Err}({t_{\max}}) \leq C_3 \frac{\kappa_0^2 r }{\sqrt{m}} \cdot \sqrt{ \frac{\log n}{Lnp_{\max}} }. 
$$
\end{corollary}
By Corollary~\ref{cor:PI}, if $\kappa_0, r, m$ are fixed and the network sparsity satisfies $Lnp_{\max} \geq C\log^4 n$, then 
\begin{equation}\label{eq:rate_1}
{\rm Err}({t_{\max}}) = O_p(\sqrt{\log(n)/(Lnp_{\max}})). 
\end{equation}
The convergence rate (\ref{eq:rate_1}) is optimal up to the logarithmic factor.

\subsection{Consistency of recovering global memberships}

Recall from Section~\ref{sec:model} that the global community structure is denoted as $\calV=\cup_{k=1}^{\bar{K}}\bar{\calV}_k$ with disjoint communities $\{\bar{\calV}_j\}_{j=1}^{\bar{K}}$ where node $i_1$ and $i_2$ belong to the same global community if and only if $(e_{i_1}-e_{i_1})^{\top}\bar{Z}=0$. 
%
In the TWIST algorithm, after applying the K-means to the rows of $\widehat U$, we get the vertices' global membership 
$\widehat{\bar \VV} = \{ \widehat{\bar{\calV}}_k, k\in[\bar{K}] \}$. 

We measure the performance by the Hamming error of clustering:
$$
\calL(\widehat{\bar \VV},\bar \VV)
=\min_{\tau: \textrm{ a permutation on }[\bar{K}]} \sum_{i=1}^n {\bf 1}\big(i\in \bar{\calV}_k, i\notin \widehat{\bar{\calV}}_{\tau(k)}\big)
$$
where $\bar\VV=\{\bar{\calV}_k, k\in[\bar{K}]\}$.

\begin{theorem}[{\bf Consistency of global clustering}]\label{thm:global_consistency}
Assume that (A1)-(A4) hold, and that $\min_k|\bar{\calV}_k|\asymp n/\bar{K}$.  Then with probability at least $1-n^{-2}$, we have
$$
n^{-1} \cdot\calL(\widehat{\bar \VV}, \bar{\VV})
 \le C_3\kappa_0^6\frac{r^2\log n}{Lnp_{\max}}
$$
provided that the network sparsity satisfies
\begin{equation} \label{eq:global_cluster_cond1}
\sqrt{Lnp_{\max}} 
\geq \big( C_1 \sqrt{\bar{K}} + C_2\kappa_0 r \log(n) \big) (\kappa_0^3 r/\sqrt m)  \log(\kappa_0r) \log(n), 
\end{equation}
\end{theorem}

From Theorem~\ref{thm:global_consistency}, it follows that the relative clustering error is $O_p\big(\log^{-3}(n)\big)$ when $\bar{K}, m, \kappa_0$ are bounded. Therefore, vertices' global memberships can be consistently recovered under near optimal network sparsity conditions. 

We now compare our method with some other available ones in the literature. 
\begin{itemize}
\item 
A special case when $m=1$ was considered by \cite{lei2019consistent}, who showed that their algorithm is able to consistently recover the communities if 
$np_{\max}\sqrt{L}\gg \log^{3/2} n$. On the other hand, our result deals with more general mixture multi-layer model, is computationally more efficient, and requires weaker network sparsity: $np_{\max}L\gg \log^4(n)$. The dependence of $L$ in (\ref{eq:global_cluster_cond1}) is optimal, if we ignore the logarithmic term. This improvement is due to a sharper concentration inequality of $\bA-\EE\bA$ in terms of tensor incoherent norm.

\item 
A joint matrix factorization method (Co-reg) was proposed in \cite{paul2017spectral} for a special case with $m=1$ and different $B_j$s, in which they proved that their method can consistently recover the vertices memberships if $Lnp_{\max}\gg \log n$ and the signal strengths of $B_j$s are similar. Their network sparsity condition is similar to  (\ref{eq:global_cluster_cond1}) above up to the logarithmic factor. On the other hand, our approach differs from \cite{paul2017spectral} in several aspects. Our method can perform vertices clustering and network clustering simultaneously when $m>1$.  Computationally, \cite{paul2017spectral} employed a BFGS algorithm to solve the non-convex programming, which is more computationally intensive than TWIST. 
More will be discussed later in the paper after Lemma~\ref{lem:initialization}. 
\end{itemize}

\subsection{Network classification}

We now show that the standard K-means algorithm on $\widehat W$ can consistently uncover the network classes of $L$ layers under the near optimal network sparsity condition (\ref{eq:global_cluster_cond1}). Further under a slightly stronger network sparsity condition (\ref{eq:network_sparsity_cond2}), we can apply Algorithm~\ref{algo:network} to exactly recover the layer labels with high probability. This shows that more layers will provide more information about layer structure and be very helpful in exact clustering of networks. Similarly, we denote
$$
\calL(\hat\LL, \LL)=\min_{\tau: \textrm{ permutation of }[m]}\sum_{l=1}^L {\bf 1}(\ell_l\neq \tau(\hat\ell_l)).
$$
\begin{theorem}[{\bf Consistency and exact recovery of network classes}] \label{thm:exact_recovery}
Let $\widetilde\LL=\{\tilde \ell_l\}_{l=1}^L$ be the output of the standard K-means algorithm applied to $\widehat W$. 
\begin{enumerate}
\item Under the same conditions in Theorem~\ref{thm:global_consistency}, we have, with probability at least $1-n^{-2}$,
$$
L^{-1}\cdot \calL\big(\widetilde\LL, \LL\big)
\leq C_3\kappa_0^4\frac{(r^2/m)\log n}{Lnp_{\max}}
$$ 
where $\LL=\{\ell_l\}_{l=1}^L$.

\item 
We further assume
\begin{equation}\label{eq:network_sparsity_cond2}
\sqrt{L}np_{\max}\geq C_1m^{-1}\kappa_0^5r^{5/2} \log(r\kappa_0) \log^{5/2}(n).
\end{equation}
Then with probability at least $1-3n^{-2}$, we have
$$
\calL(\widehat\LL, \LL)=0
$$ 
where $\widehat\LL=\{\hat{\ell}_l\}_{l=1}^L$ is the output of Algorithm~\ref{algo:network} with parameters $m$ and $\varepsilon\in[c_1, c_2] \sqrt{m/L}$. 
\end{enumerate}
\end{theorem}
By Theorem~\ref{thm:exact_recovery}, in the case $r,m,\kappa_0=O(1)$, Algorithm~\ref{algo:network} is capable to exactly recover the network classes $\LL=\{\ell_l\}_{l=1}^L$ with appropriately chosen parameter $\varepsilon$ if the network sparsity (\ref{eq:network_sparsity_cond2}) satisfies $\sqrt{L}np_{\max}\gg\log^{5/2}n$.  On the other hand, consistent network clustering requires, by (\ref{eq:global_cluster_cond1}), network sparsity $Lnp_{\max}\gg \log^4n$. Therefore, condition (\ref{eq:network_sparsity_cond2}) is stronger with respect to the number of layers $L$. 
\begin{remark}
Clustering of networks is essentially a problem of clustering or classifying high-dimensional data. Without exploring network structures, a simple method for clustering high-dimensional data is by spectral clustering on the left singular vectors of $\calM_3(\bA)$ which is a highly rectangular matrix. A simple fact is that $\|\calM_3(\bA-\EE\bA)\|=O_p(n\sqrt{p_{\max}})$ implying that a naive spectral clustering on $\calM_3(\bA)$ requires very strong condition of network sparsity for consistent network clustering. The cause of such a sub-optimality is due to the ignorance of matrix structures of rows of $\calM_3(\bA)$. 
\end{remark}

\subsection{Consistency of local clustering}

After obtaining the network classes, it suffices to apply spectral clustering on $\sum_{l: \hat{\ell}_l=j}A_l$ for all $j\in[m]$ to recover the local memberships $\VV^j=\{\calV_k^j\}_{k=1}^{K_j}$. Its consistency can be directly proved by existing results in the literature (see, e.g., \cite{lei2015consistency}). We hereby omit the proof of the following theorem.
\begin{theorem}\label{thm:local_consistency}
Suppose that the conditions of Theorem~\ref{thm:exact_recovery} and (\ref{eq:network_sparsity_cond2}) hold. For all $j\in[m]$, let $\widehat\VV^j=\{\hat\calV_k^j\}_{k=1}^{K_j}$ denote the output of K-means algorithm on $\sum_{l: \hat{\ell}_l=j}A_l$. If $\sigma_{\min}(B_j)\geq c_1p_{\max}$ for some absolute constant $c_1>0$ and $|\calV_k^j|\asymp n/K_j$ for all $k\in[K_j]$, then with probability at least $1-3n^{-2}$,
$$
n^{-1}\cdot\calL(\widehat\VV^j, \VV^j)\leq C_1m^{-1}\kappa_0^4\cdot\frac{K_j^2\log n}{Lnp_{\max}}. 
$$
\end{theorem}

\subsection{Warm initialization for regularized power iteration}\label{sec:init}
An important condition for the success of Algorithm~\ref{algo:Power Iterations} is the existence of warm initialization 
$$
{\rm Err}(0) = \max\{{\rm d}(\widehat{U}^{(0)},\bar{U}), {\rm d}(\widehat{W}^{(0)},\bar{W})\}\leq 1/4.
$$ 
In this section, we introduce a spectral method for initializing $\widehat{U}^{(0)}$ by summing up all the layers of networks. After that, we initialize $\widehat{W}^{(0)}$ by taking the left singular vectors of $\calM_3(\bA)(\widetilde{U}^{(0)}\otimes\widetilde{U}^{(0)})$ where $\widetilde{U}^{(0)}=\calP_{\delta_1}(\widehat{U}^{(0)})$. The following lemma shows that these initializations are indeed close to the truth under reasonable conditions.
\begin{lemma}[Initialization]\label{lem:initialization}
Let $\widehat{U}^{(0)}$ denote the top-$r$ left singular vectors of $\sum_{l=1}^{L}A_l$ and let $\widehat{W}^{(0)}$ be the top-$r$ left singular vectors of $\calM_3(\bA)(\widetilde{U}^{(0)}\otimes\widetilde{U}^{(0)})$ where $\widetilde{U}^{(0)}=\calP_{\delta_1}(\widehat{U}^{(0)})$ with $\delta_1=\max_j\|e_j^{\top}\bar{U}\|$. Then with probability at least $1-3n^{-2}$,
\begin{equation}\label{eq:initU}
{\rm d}(\widehat{U}^{(0)},\bar{U})\leq \min\Big\{\frac{C_3\sqrt{np_{\max}}\log^2n}{\sigma_r\big(\bar{\bC}\times_3(\bd_L/L)^{1/2}\big)},\ 2\Big\}
\end{equation}
where $\bd_L=(L_1,\cdots,L_m)$. If $\delta_1=O(\sqrt{r/n})$ and
\begin{equation}\label{eq:init_cond} 
\sigma_r\big(\bar{\bC}\times_3(\bd_L/L)^{1/2}\big)\geq 4C_3\sqrt{np_{\max}}\log^2n,
\end{equation}
then with same probability,
$$
{\rm d}(\widehat{W}^{(0)},\bar{W})\leq \min\Big\{\frac{C_4\sqrt{mrnp_{\max}}\log^2(n)\log(r)}{\sigma_{\min}(\bar{\bC})},\ 2\Big\}.
$$
\end{lemma}
It is possible to improve $\log^2n$ to $\sqrt{\log n}$ in (\ref{eq:initU}) as in \cite{paul2016consistent}. 
Comparing the rate of initialization (\ref{eq:initU}) and the rate after regularized power iterations in Theorem~\ref{thm:power_iteration}, Algorithm~\ref{algo:Power Iterations} improve the estimation error by a ratio of $\sigma_{\min}(\bar{\bC})$ and $\sigma_r\big(\bar{\bC}\times_3(\bd_L/L)^{1/2}\big)$. In special cases, such an improvement can be significant. For instance, consider $m=r=2$ and $L_1=L_2$ and $\bar{\bC}\in\RR^{2\times 2\times 2}$ with 
$$
\bar{C}(:,:,1)=\left(\begin{array}{cc}1+\varepsilon&0\\0&1+\varepsilon\end{array}\right)\quad {\rm and}\quad \bar{C}(:,:,2)=\left(\begin{array}{cc}0&1-\varepsilon\\1-\varepsilon&0\end{array}\right)
$$
for some small number $\varepsilon\in(0,1)$. It is easy to check that $\sigma_{\min}(\bar{\bC})=\sigma_2\big(\calM_3(\bar{\bC})\big)=\sqrt{2}(1-\varepsilon)$. On the other hand, 
$$
\sigma_{2}\big(\bar{\bC}\times_3(\bd_L/L)^{1/2}\big)=\sqrt{2}\varepsilon.
$$
Moreover, if $\varepsilon=0$, then $\bar{\bC}\times_3(\bd_L/L)^{1/2}$ is rank deficient implying that simply projecting the multi-layer networks into a graph can potentially cause serious  information loss. See more details in \cite{ke2019community} and a similar discussion in \cite{lei2019consistent}.  

It is worthwhile pointing out that one could use other methods to initialize $\widehat{U}^{(0)}$ (the initialization of $\widehat{W}^{(0)}$ is easy once it is done for $\widehat{U}^{(0)}$). Examples include the HOSVD by extracting the top-$r$ left singular vectors of $\calM_1(\bA)$, the joint matrix factorization method in \cite{paul2017spectral}, and random projection \cite{ke2019community}.

\section{Simulation studies}\label{sec:simulation}

We conduct several simulations to test the performance of the TWIST on the MMSBM with different choices of network sparsity, "out-in" ratio, number of layers and the size of each layer. We use K-means as the clustering algorithm. 
The evaluation criterion is the mis-clustering rate. All the experiments are replicated 100 times and the average performance across the repetitions is reported.

We generate the data according to the MMSBM in Section \ref{sec:model} in the following fashion. The underlying class $\ell_l$ for each layer $l$ is generated from the multinomial distribution with $\PP(\ell_l=j)=1/m,\ j=1,\cdots,m.$  The membership $z_i^j$ for each node $i$ in layer type $j$ is generated from the multinomial distribution with $\PP(z_i^j=s)=1/K,\ s=1,\cdots,K.$ We choose the connection matrix as 
$B=pI+q(\mathbf{1}\mathbf{1}^T-I),$ where $\mathbf{1}$ is a $K$-dimensional vector with all elements being $1.$ Let $\alpha=q/p$ be the out-in ratio.

\subsection{Global memberships}
First, we consider the task to detect the global memberships defined in Section~\ref{sec:model}. We compare the performance of the TWIST with Tucker decomposition with HOSVD initialization (HOSVD-Tucker), and  we also adopt a baseline method by performing spectral clustering on the sum of the adjacency matrices from all layers (Sum-Adj). Sum-Adj has been considered in literature (\cite{paul2017spectral}, \cite{dong2012clustering}, \cite{tang2009clustering}) as a simple but effective procedure \cite{kumar2011co}. The function "\textit{tucker}" from the R package "\textit{rTensor}" \cite{li2018rtensor} is adopted to apply Tucker decomposition for HOSVD-Tucker. 

In Simulation 1, the networks are generated with the number of node $n=600$, the number of layers $L=20$, number of types of networks $m=3,$ number of communities of each network $K=2$ and out-in ratio of each layer $\alpha=0.4.$ The average degree $d$ of each layer varies from $2$ to $20.$ 

In Simulation 2, the setting is the same as in Simulation 1 except the average degree of each layer is fixed at $d=10$ and the out-in ration $\alpha$ of each layer varies from $0.1$ to $0.8.$

In Simulation 3,  the setting is the same as in Simulation 1, except that the out-in ratio $\alpha=0.6$ and the number of layers $L$ varies from $10$ to $60.$

In Simulation 4, the setting is the same as that in in Simulation 3, except that the out-in ratio is $L=20$ and the number of nodes $n$ varies from $100$ to $1200.$

\begin{figure}
	\centering
	\begin{subfigure}[b]{.4\linewidth}
		\includegraphics[width=\linewidth]{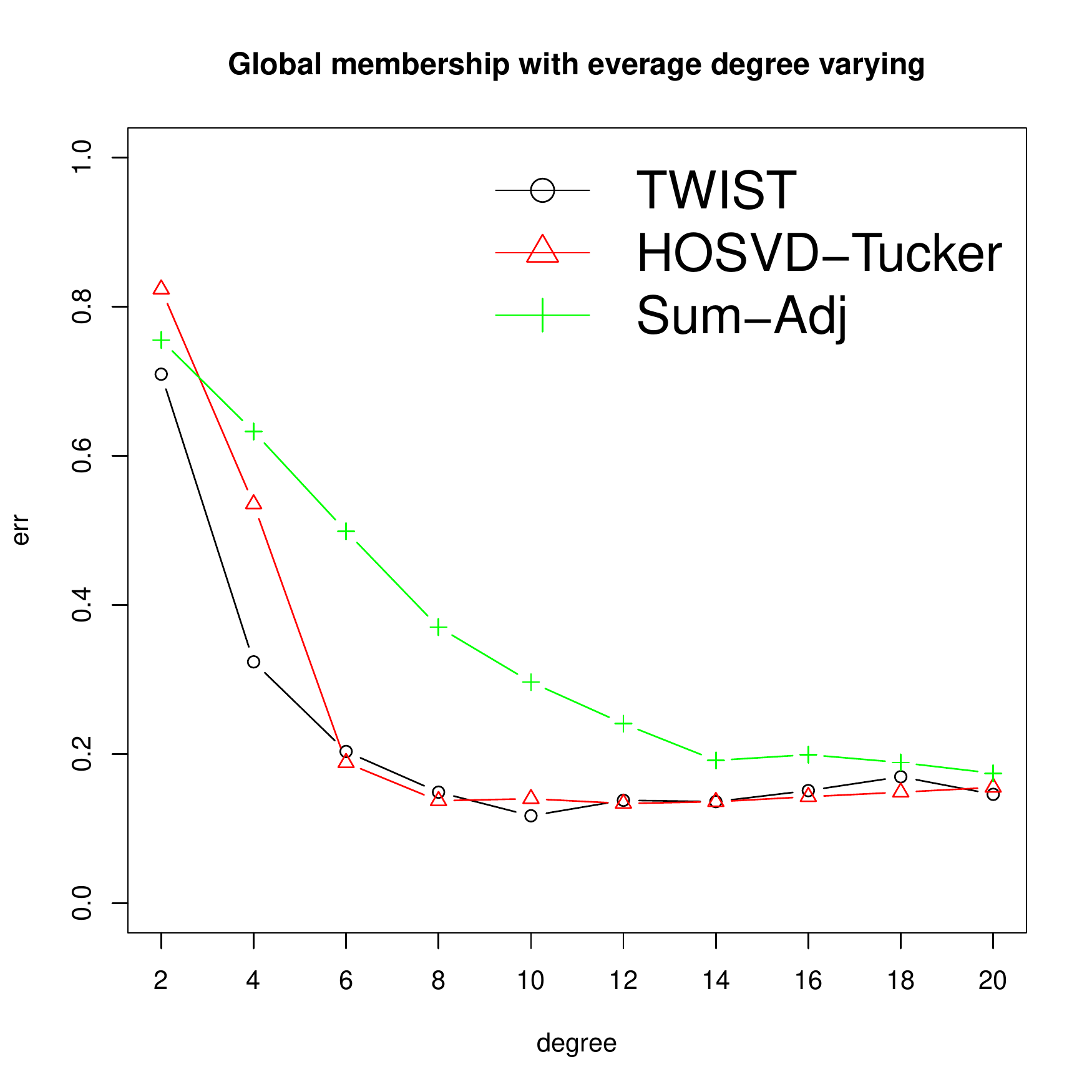}
		\caption{\tiny{The result of Simulation 1: $n=600$, $K=2$, $m=3$, $L=20$, $\alpha=0.4$, varying $d.$ }}
	\end{subfigure}
	\begin{subfigure}[b]{.4\linewidth}
		\includegraphics[width=\linewidth]{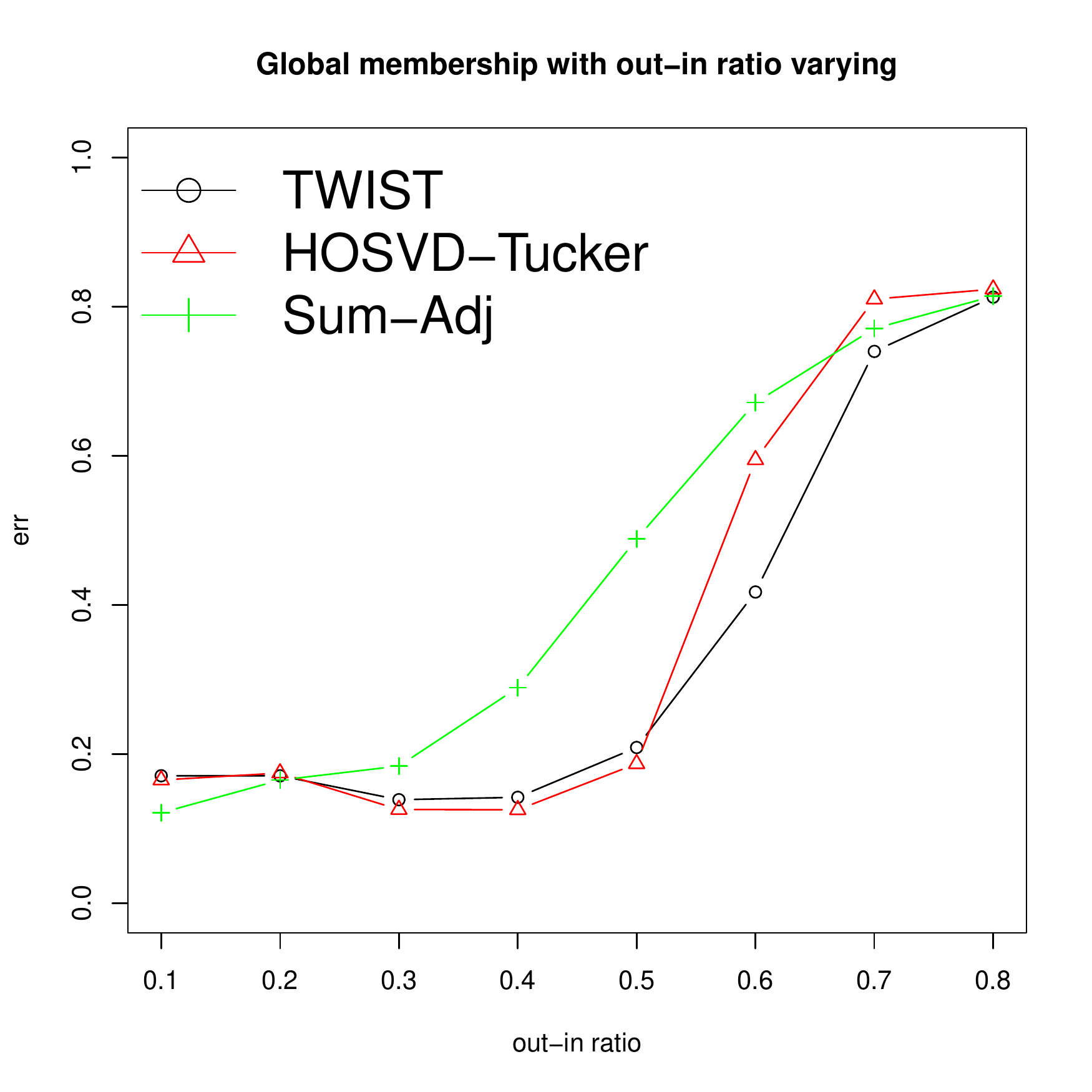}
		\caption{\tiny{The result of Simulation 2: $n=600$, $K=2$, $m=3$, $L=20$, $d=10$, varying $\alpha.$} }
	\end{subfigure}
	\begin{subfigure}[b]{.4\linewidth}
		\includegraphics[width=\linewidth]{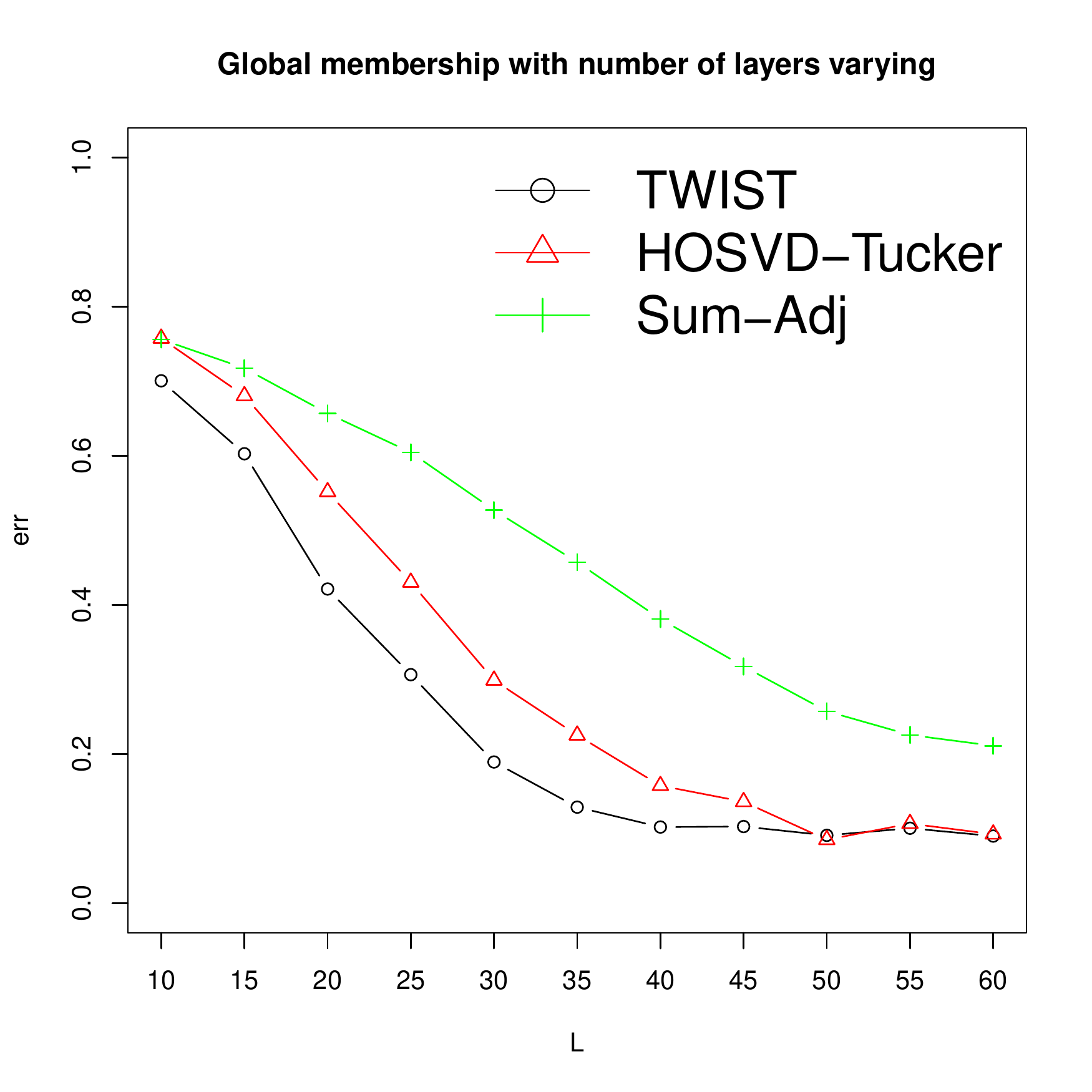}
		\caption{\tiny{The result of Simulation 3: $n=600$, $K=2$, $m=3$, $d=10$, $\alpha=0.6$, varying $L.$ }}
	\end{subfigure}
	\begin{subfigure}[b]{.4\linewidth}
	\includegraphics[width=\linewidth]{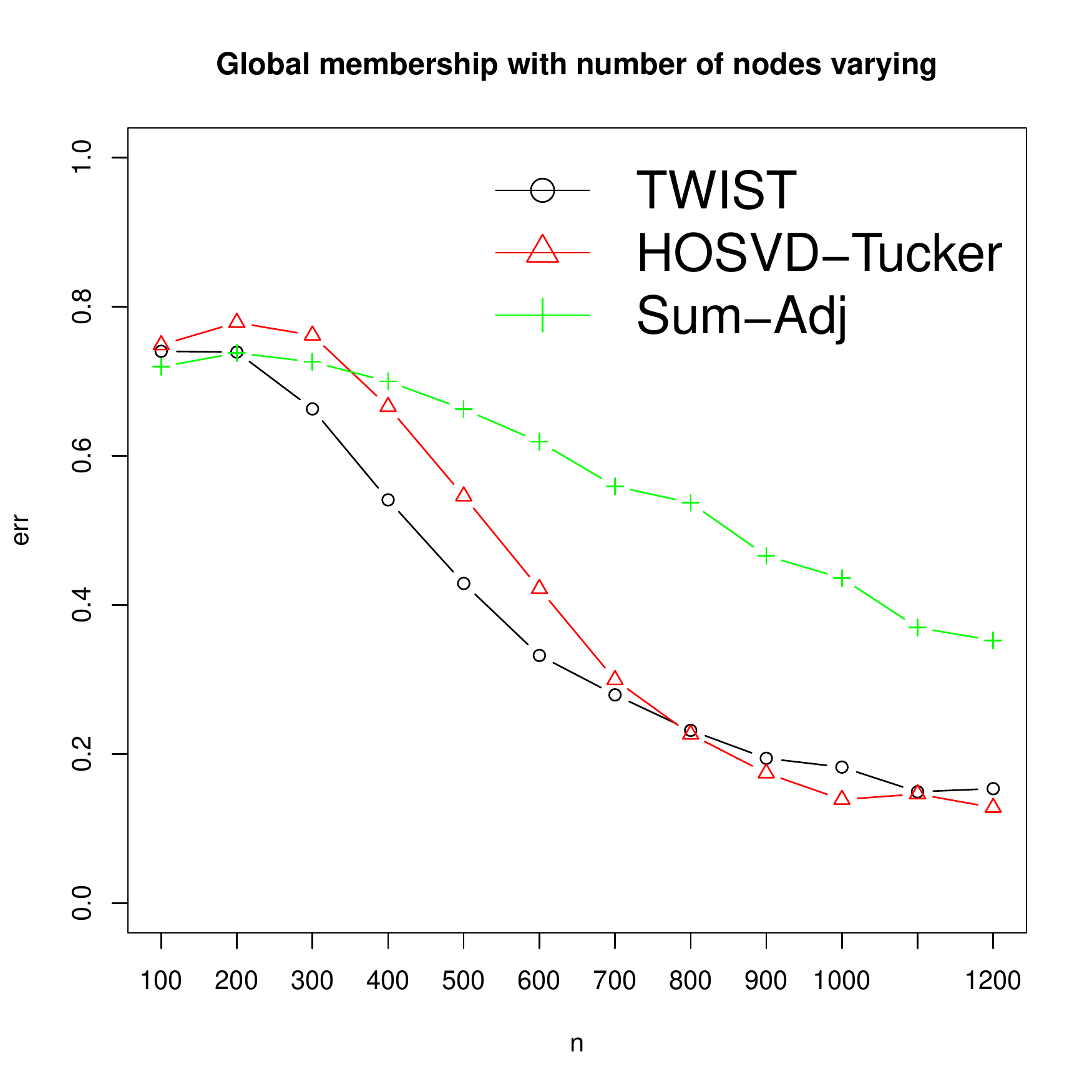}
	\caption{\tiny{The result of Simulation 3: $K=2$, $m=3$, $d=10$, $L=20$, $\alpha=0.6$, varying $n.$ }}
    \end{subfigure}
	\caption{Overall, the TWIST and the HOSVD-Tucker perform much better than Sum-Adj. The TWIST outperforms the HOSVD-Tucker
	 when the signal is not strong enough, for instance $d<6$ in (a), $\alpha>0.5$ in (b), $L<50$ in (c) and $n<800$ in (d).}
	\label{fig:global membership}
\end{figure}

The results of Simulations 1-4 are given in Figure \ref{fig:global membership}. 
\begin{enumerate}
\item Clearly, the mis-clustering rate of all the methods decreases as the average degree of each layer increases, the out-in ratio of each layer decreases and the number of layers increases. This is consistent with our theoretical findings. 

\item TWIST and HOSVD-Tucker, both utilizing tensor structure, perform much better than Sum-Adj, which only uses the matrix structure. 
The mis-clustering rate of the TWIST and the HOSVD-Tucker decreases more rapidly. 

\item TWIST outperforms HOSVD-Tucker when the signal is not strong enough, e.g., for $d<6$ in Simulation 1; for $\alpha>0.5$ in Simulation 2; for $L<50$ in Simulation 3; and for $n<800$ in Simulation 4. 
\end{enumerate}

\subsection{Layers' labels}
We now explore the task of clustering different types of layers. We compare the TWIST with the HOSVD-Tucker and spectral clustering applied to the mode-3 flatting of \bA (M3-SC). 

In Simulation 5, the networks are generated with the number of node $n=600$, the number of layers $L=20$, number of types of networks $m=3,$ number of communities of each network $K=3$ and out-in ratio of each layer $\alpha=0.6.$ The average degree $d$ of each layer varies from $3$ to $30.$ 

In Simulation 6, the networks are generated as in Simulation 5, except that the average degree of each layer $d=10,$ the number of layers $L=30$ and the out-in ration $r$ of each layer varies from $0.1$ to $0.9.$

In Simulation 7, the networks are the same as in Simulation 6, except that the out-in ratio $\alpha=0.6$ and the number of layers $L$ varies from $20$ to $80.$

In Simulation 8, the networks are the same as in Simulation 7, except that the average degree of each layer $d=0.02n$ and the the size of each layer $n$ varies from $100$ to $1200.$

\begin{figure}
	\centering
	\begin{subfigure}[b]{.4\linewidth}
		\includegraphics[width=\linewidth]{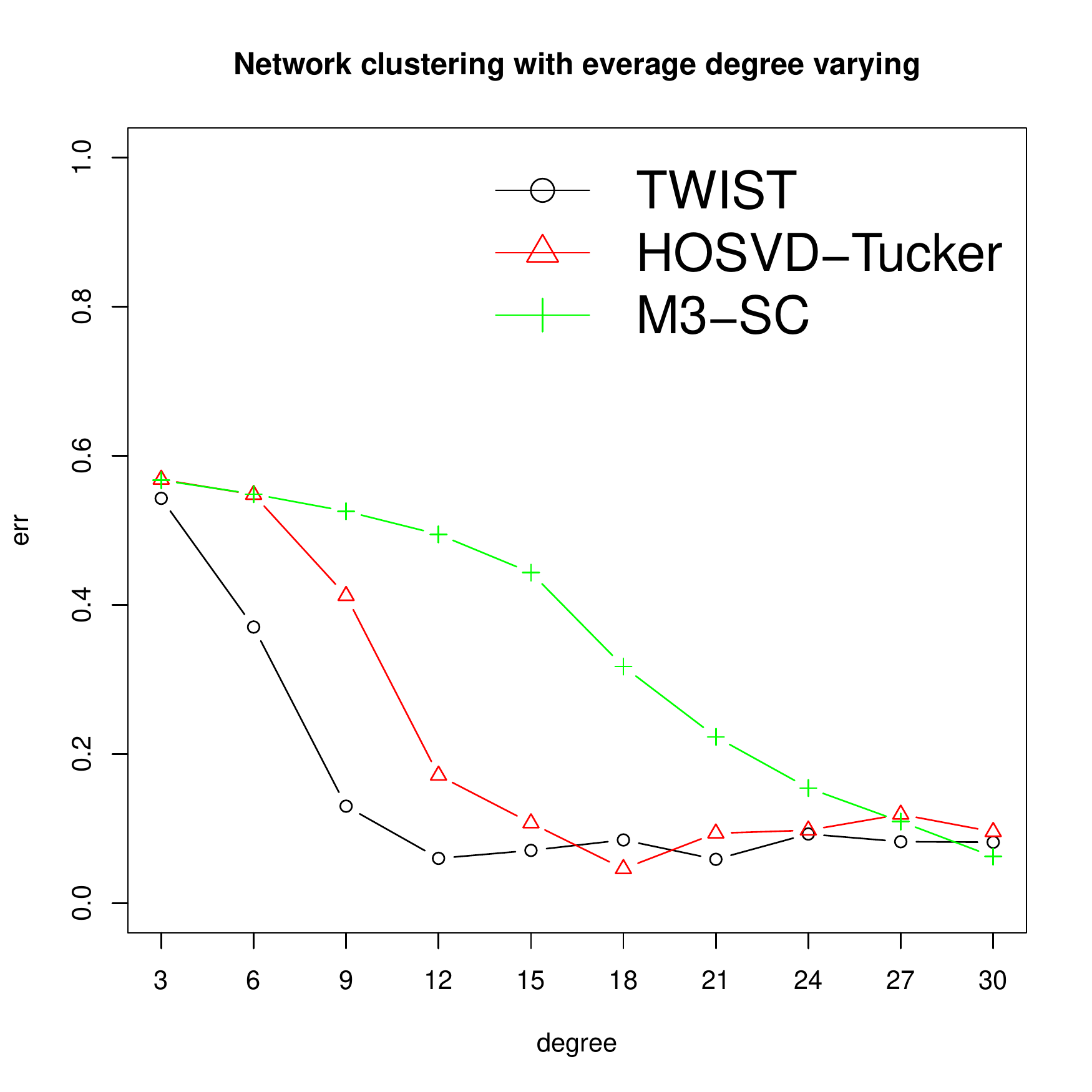}
		\caption{\tiny{The result of Simulation 4: $n=600$, $K=3$, $m=3$, $L=20$, $\alpha=0.6$, varying $d.$ }}
	\end{subfigure}
	\begin{subfigure}[b]{.4\linewidth}
		\includegraphics[width=\linewidth]{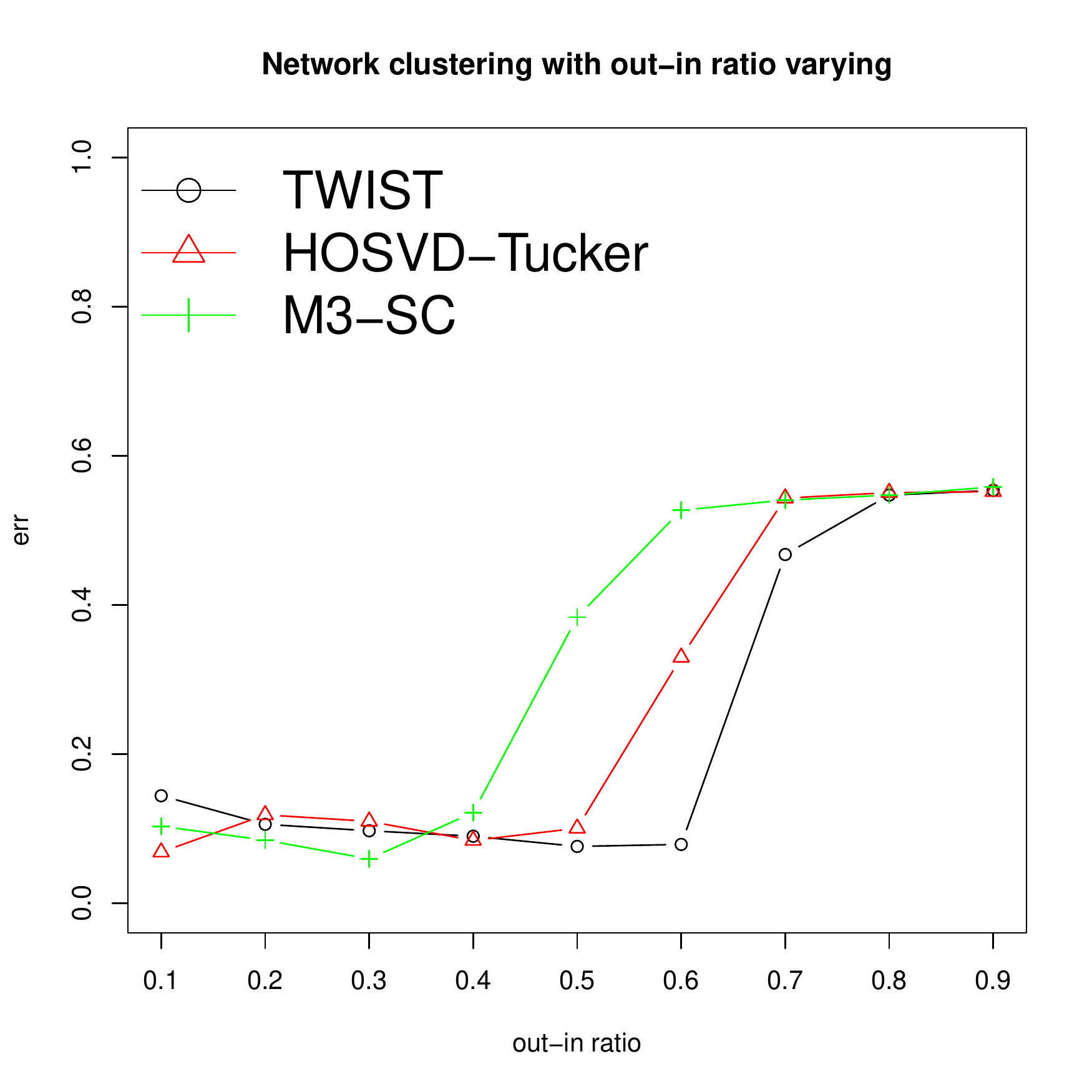}
		\caption{\tiny{The result of Simulation 5: $n=600$, $K=3$, $m=3$, $d=10$, $L=30$, varying $\alpha.$ }}
	\end{subfigure}
	\begin{subfigure}[b]{.4\linewidth}
		\includegraphics[width=\linewidth]{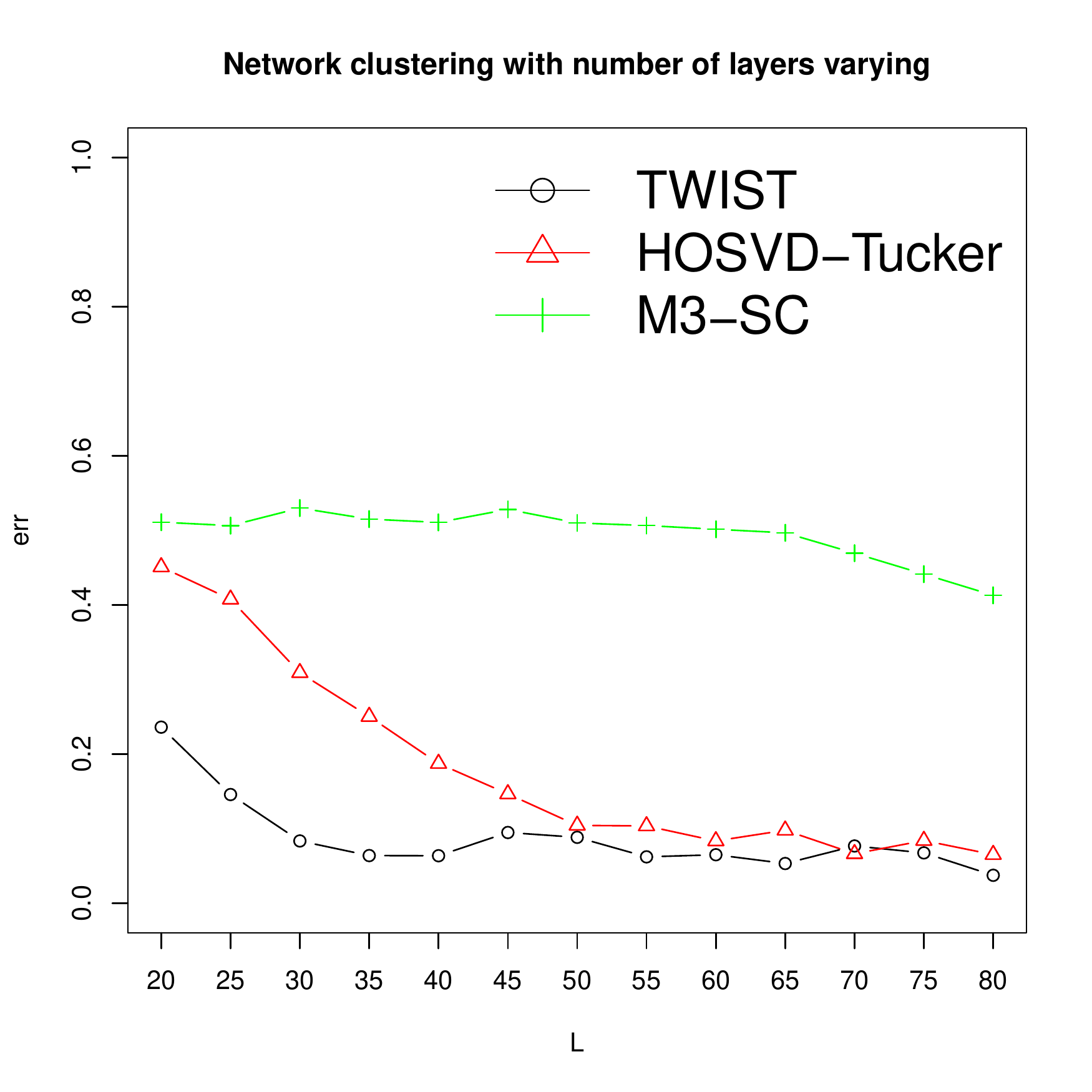}
		\caption{\tiny{The result of Simulation 6: $n=600$, $K=3$, $m=3$, $d=10$, $\alpha=0.6$, varying $L.$ }}
	\end{subfigure}
	\begin{subfigure}[b]{.4\linewidth}
	\includegraphics[width=\linewidth]{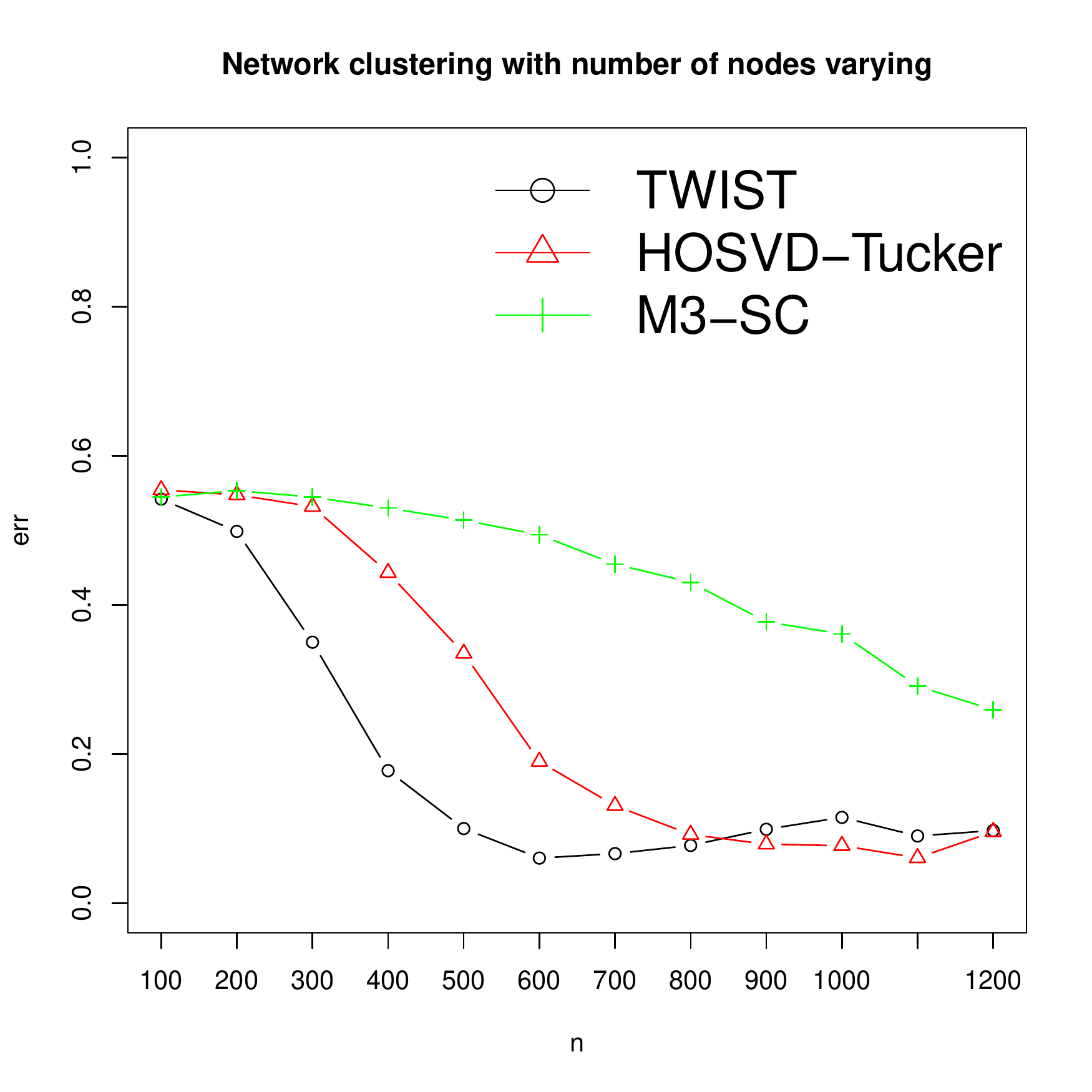}
		\caption{\tiny{The result of Simulation 7: $K=3$, $m=3$, $d=0.02n$, $\alpha=0.6$, $L=30,$ varying $n.$ }}
    \end{subfigure}
	\caption{The TWIST is the best overall, particularly when the signal is not strong enough, for instance, $d<15$ in (a), $r>0.4$ in (b), $L<50$ in (c) and $n<800$ in (d). From Simulation 7, the naive method M3-SC hardly changes as the number of layers increases.
	}
	\label{fig:network membership}
\end{figure}

The results are presented in Figure \ref{fig:network membership}. We make the following observations. 
\begin{enumerate}
\item The mis-clustering rates of all three methods decrease as the average degree of each layer increases, the out-in ratio of each layer decreases, the number of layers increases and the size of each layer increases. This agrees with our theoretical results.  

\item From Simulation 7, the naive method M3-SC shows no response to the increase of the number of layers, as might be expected. 

\item Overall, the TWIST performs the best among the three methods. This can be clearly seen when the signal is not strong enough, for instance, $d<15$ in Simulation 5, $\alpha>0.4$ in Simulation 6, $L<50$ in Simulation 7 and $n<800$ in Simulation 8. 
\end{enumerate}

\section{Real data analysis}\label{sec:realdata}

In this section, we apply the TWIST to two real data sets: worldwide food trading networks and Malaria parasite genes networks. The two datasets have been studied in the literature before. However, with the TWIST, we are able to make some new, interesting, and sometimes surprising findings, which the earlier methods have failed to do so. 

\subsection{Malaria parasite genes networks}

\begin{figure}
	\centering
	\begin{subfigure}[b]{.6\linewidth}
		\includegraphics[width=\linewidth]{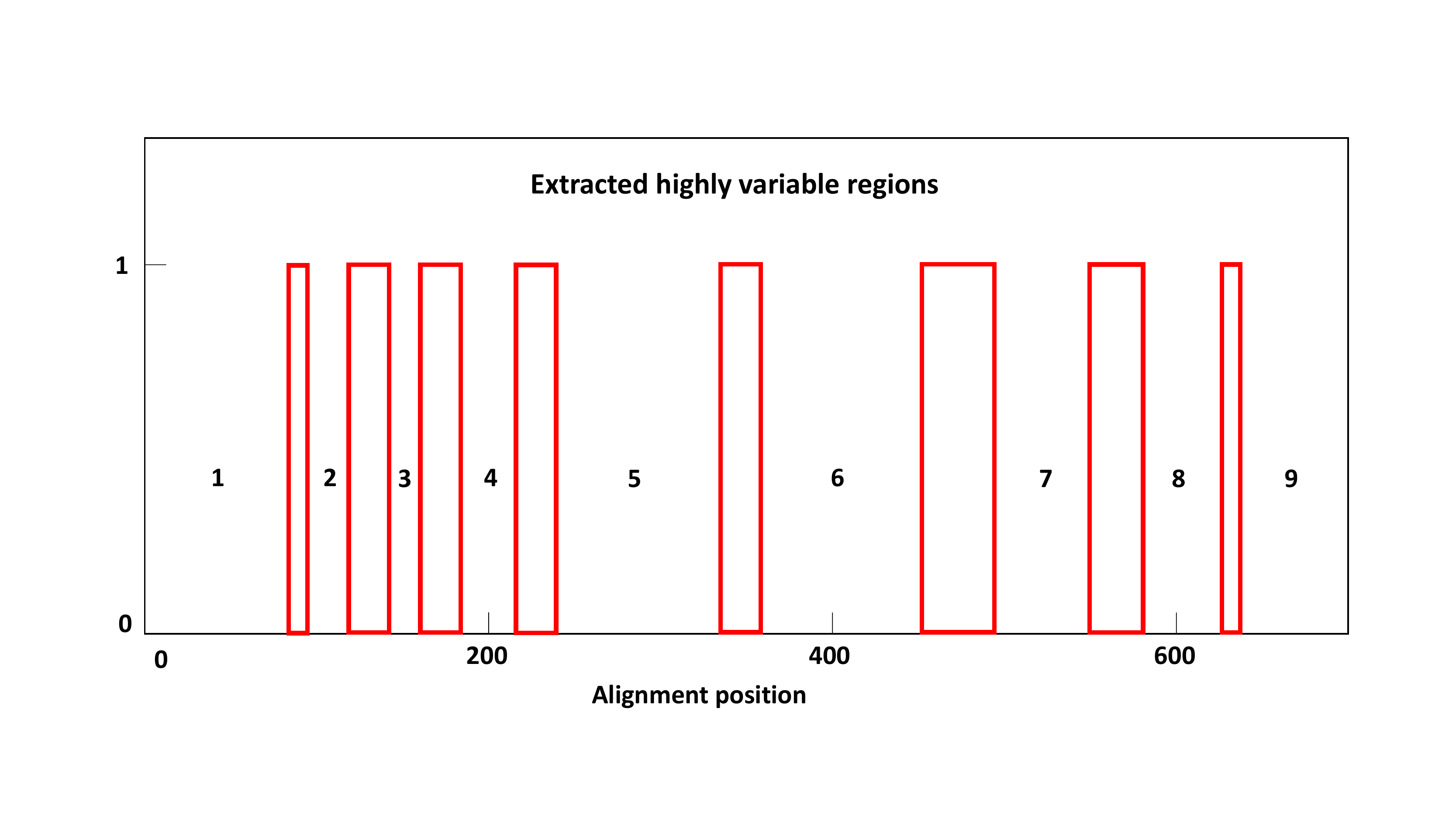}
		\caption{Extracted highly variable regions.}
	\end{subfigure}
\\
	\begin{subfigure}[b]{.18\linewidth}
		\includegraphics[width=\linewidth]{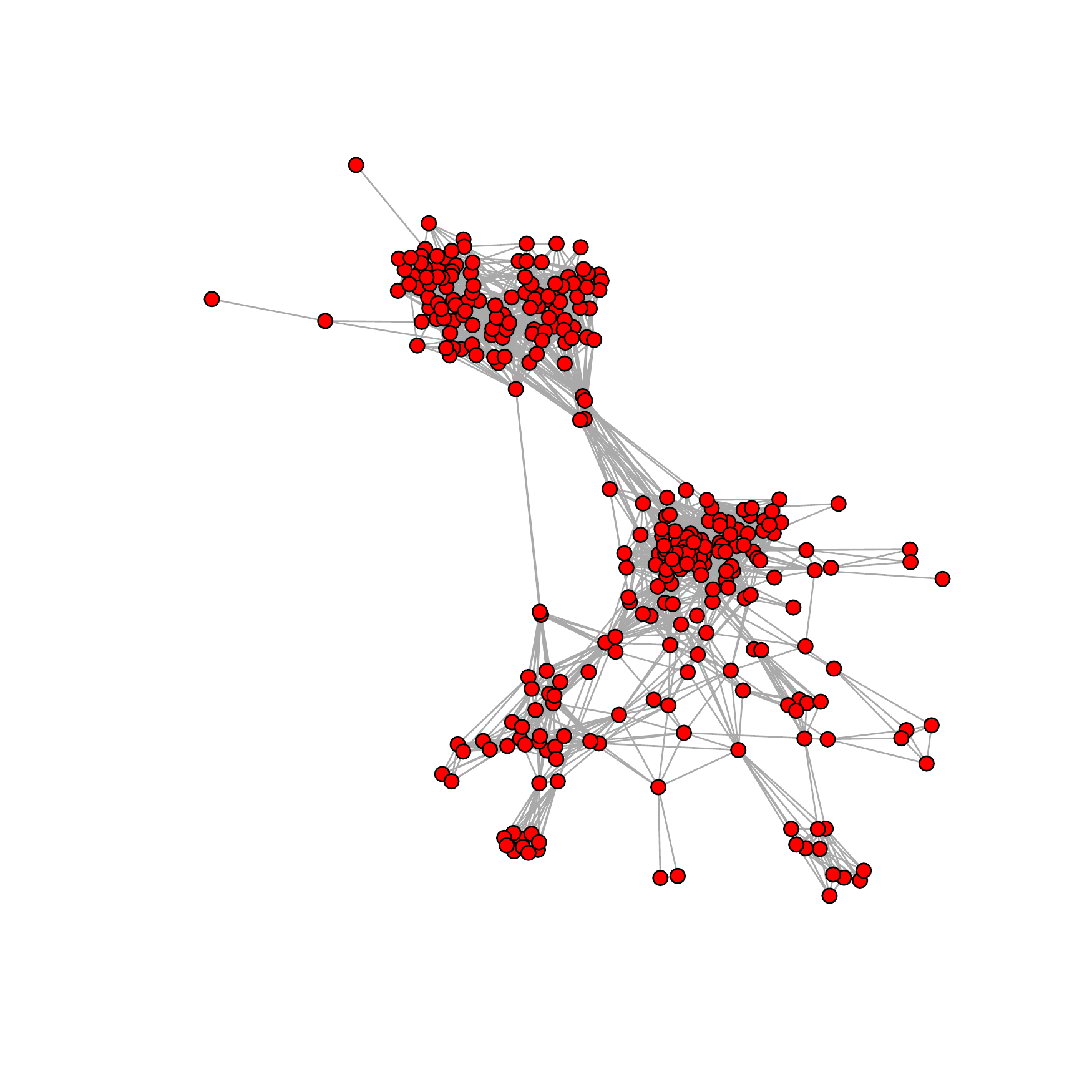}
		\caption{HVR1.}
	\end{subfigure}
	\begin{subfigure}[b]{.18\linewidth}
		\includegraphics[width=\linewidth]{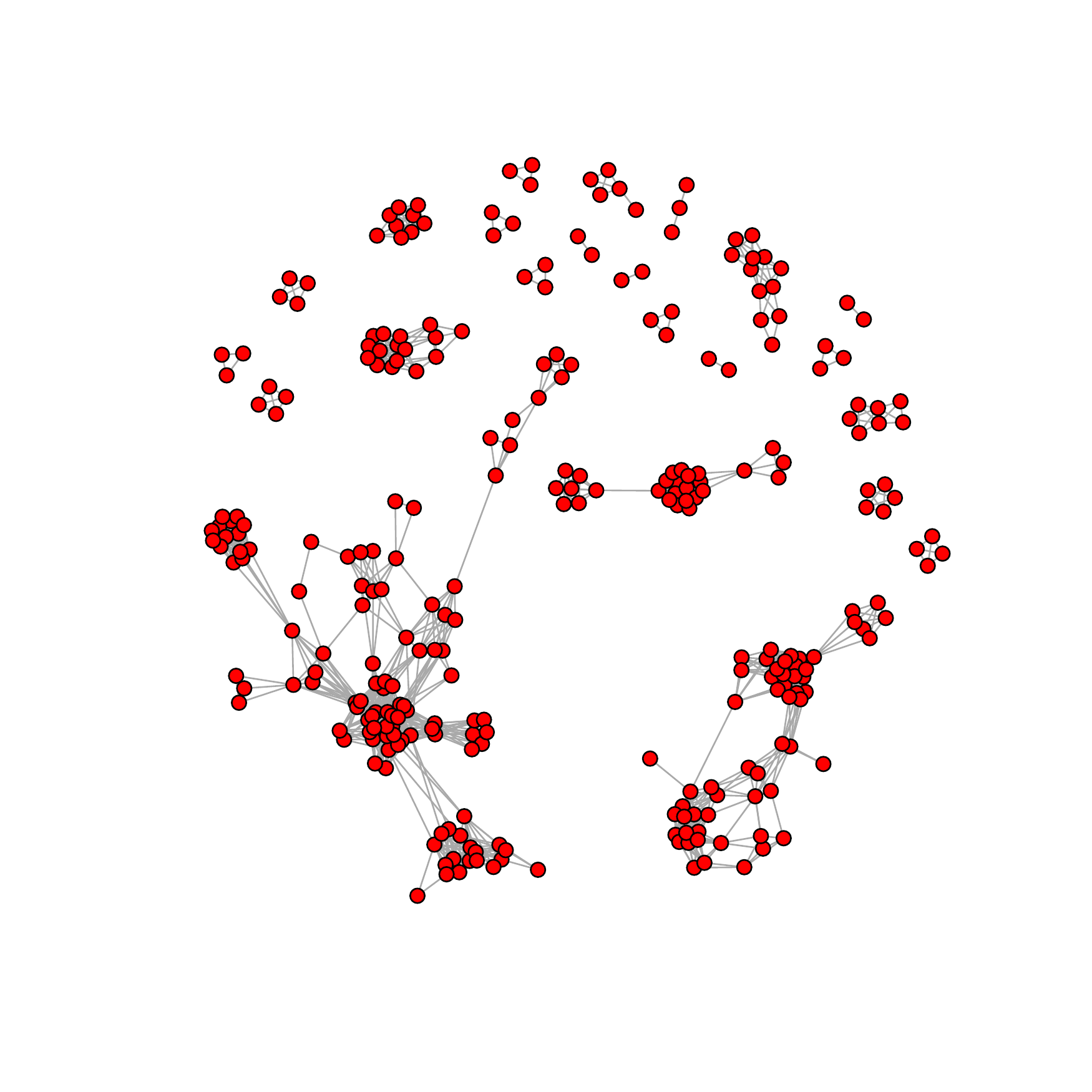}
		\caption{HVR2.}
	\end{subfigure}
	\begin{subfigure}[b]{.18\linewidth}
		\includegraphics[width=\linewidth]{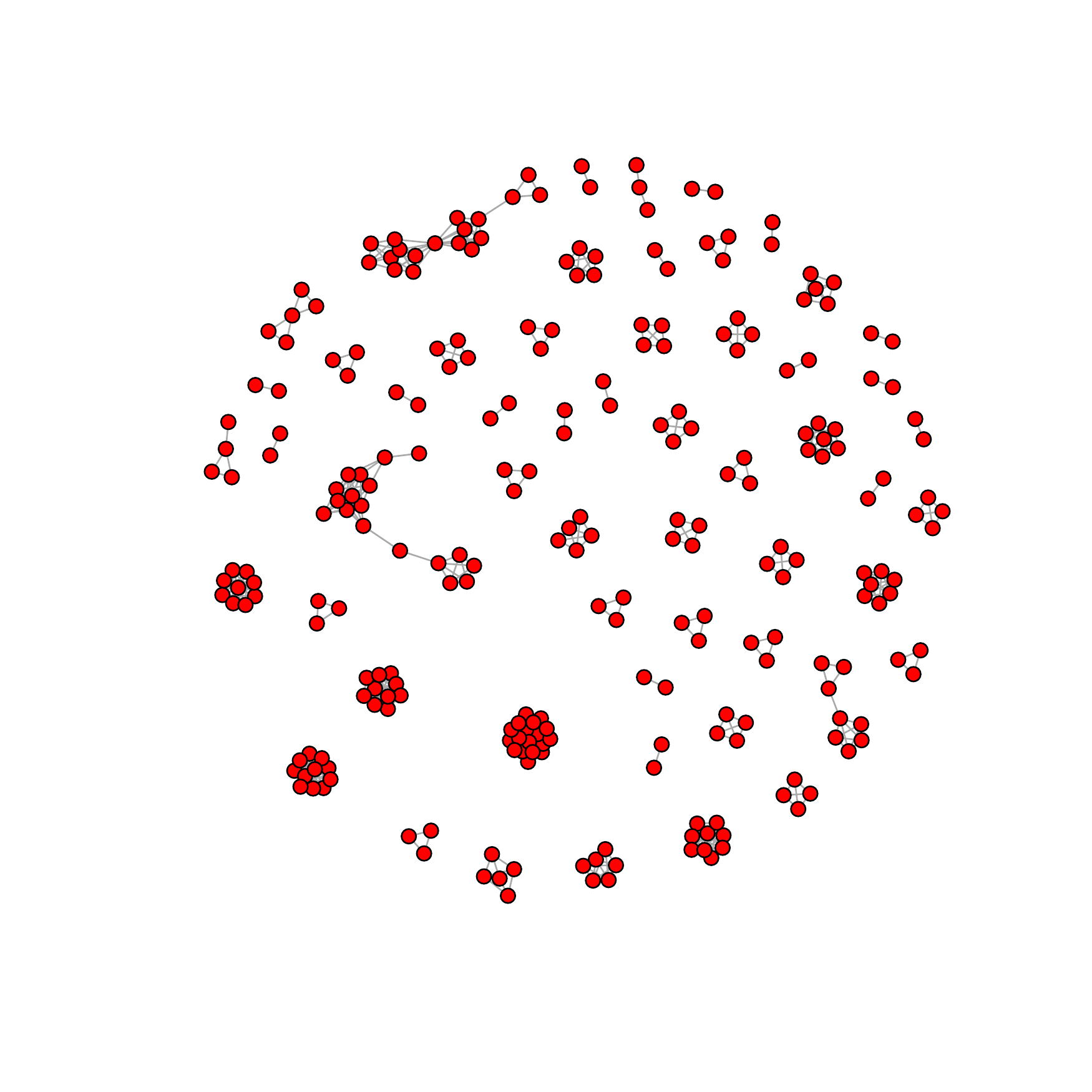}
		\caption{HVR3.}
	\end{subfigure}
	\begin{subfigure}[b]{.18\linewidth}
		\includegraphics[width=\linewidth]{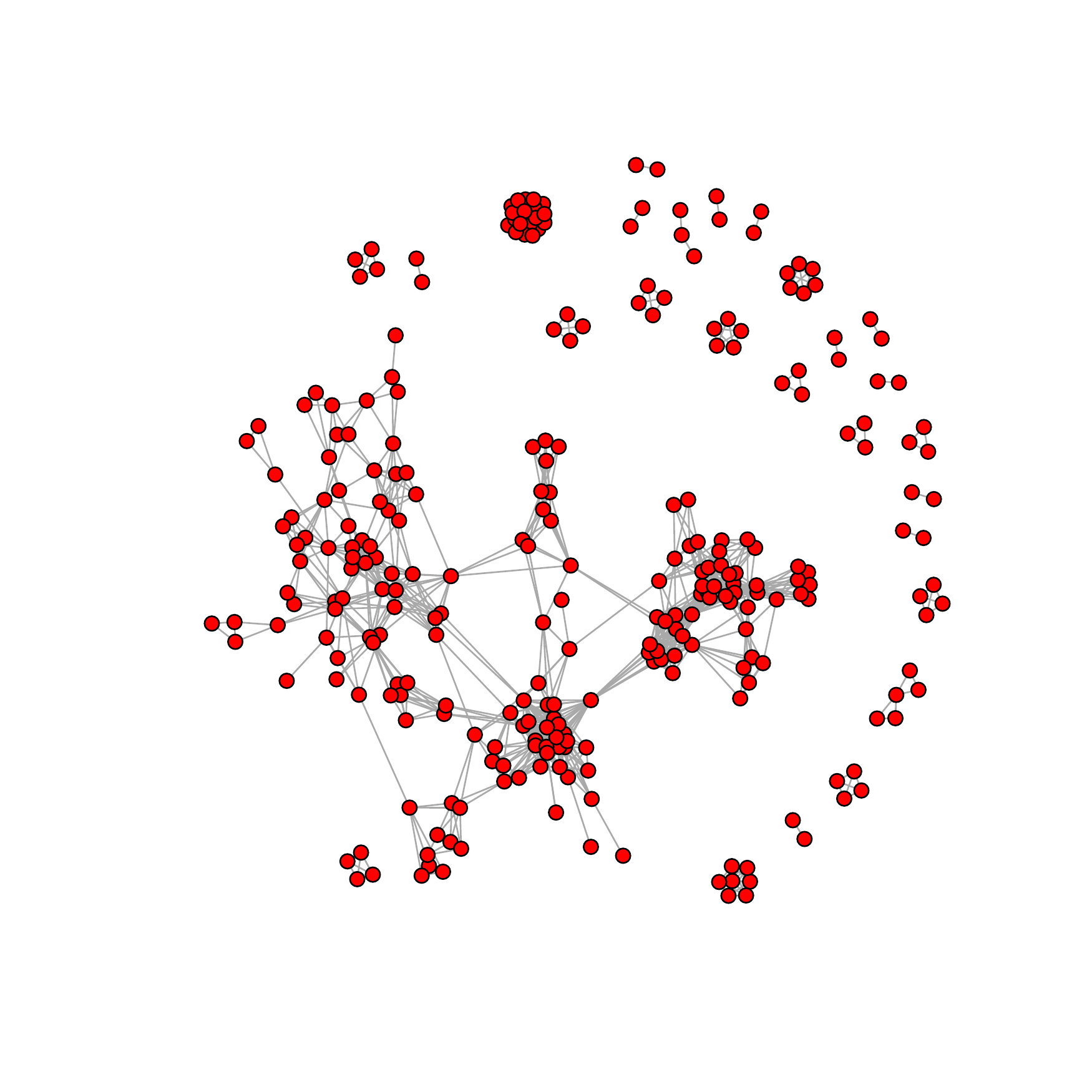}
		\caption{HVR4.}
	\end{subfigure}
	\begin{subfigure}[b]{.18\linewidth}
		\includegraphics[width=\linewidth]{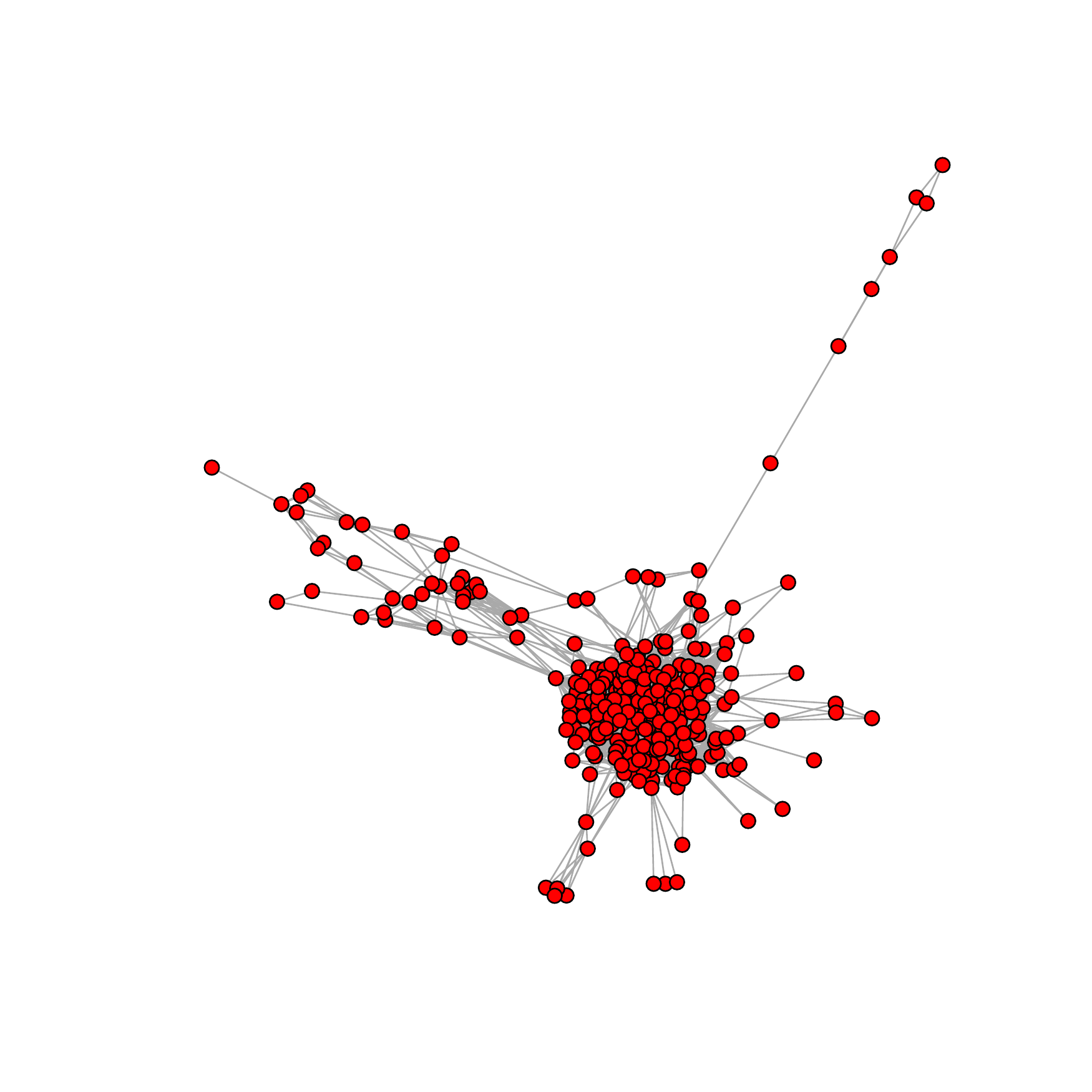}
		\caption{HVR5.}
	\end{subfigure}
	\begin{subfigure}[b]{.18\linewidth}
		\includegraphics[width=\linewidth]{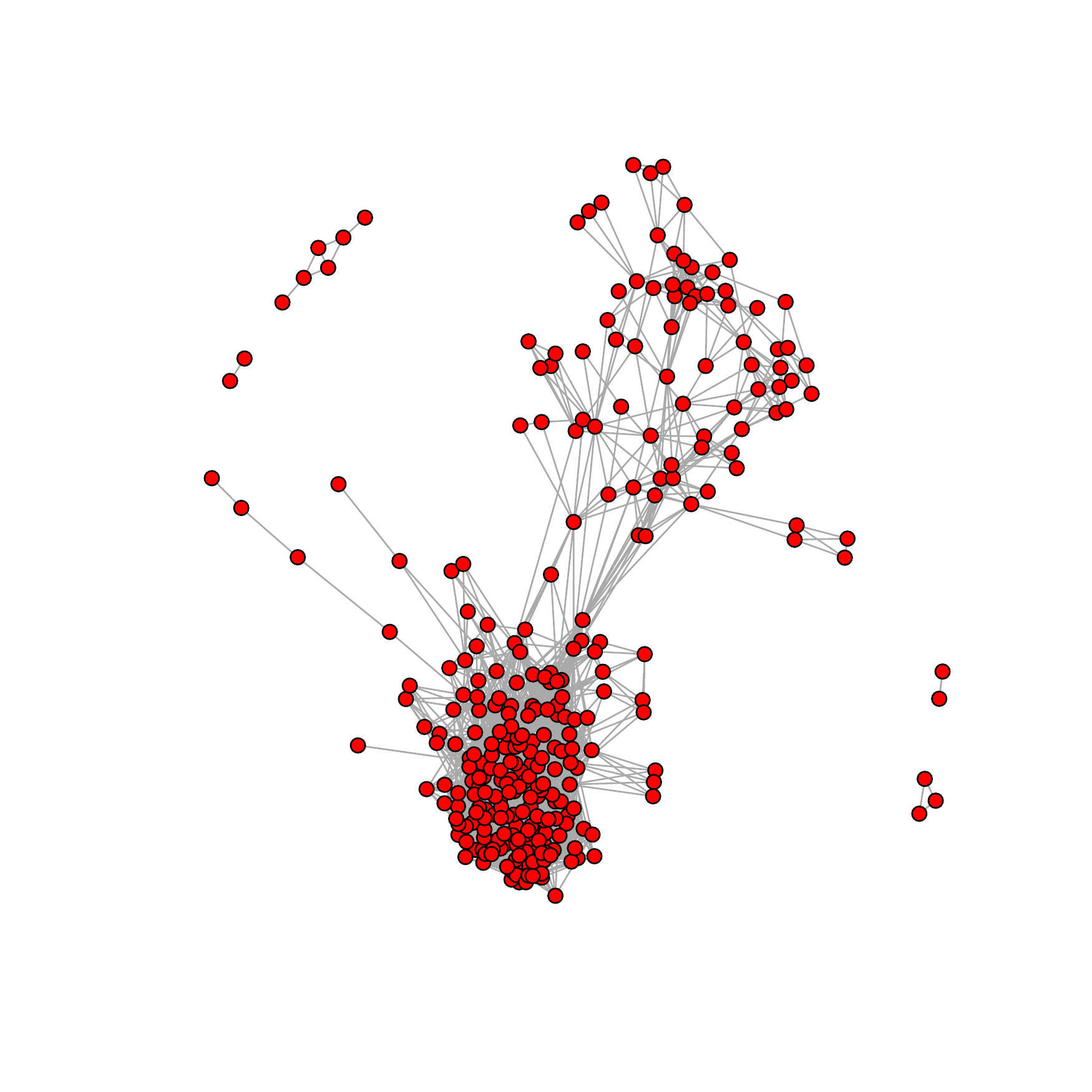}
		\caption{HVR6.}
	\end{subfigure}
	\begin{subfigure}[b]{.18\linewidth}
		\includegraphics[width=\linewidth]{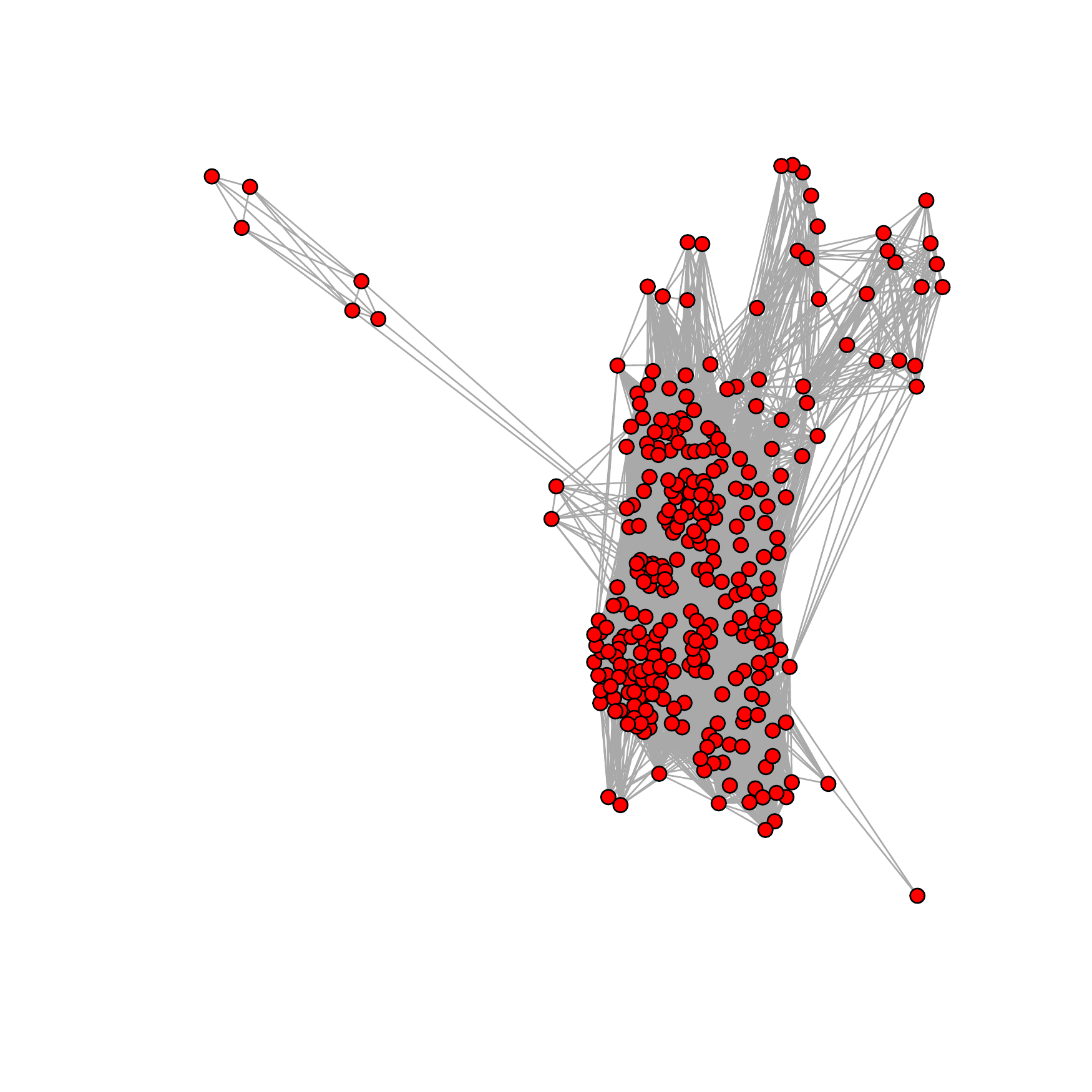}
		\caption{HVR7.}
	\end{subfigure}
	\begin{subfigure}[b]{.18\linewidth}
		\includegraphics[width=\linewidth]{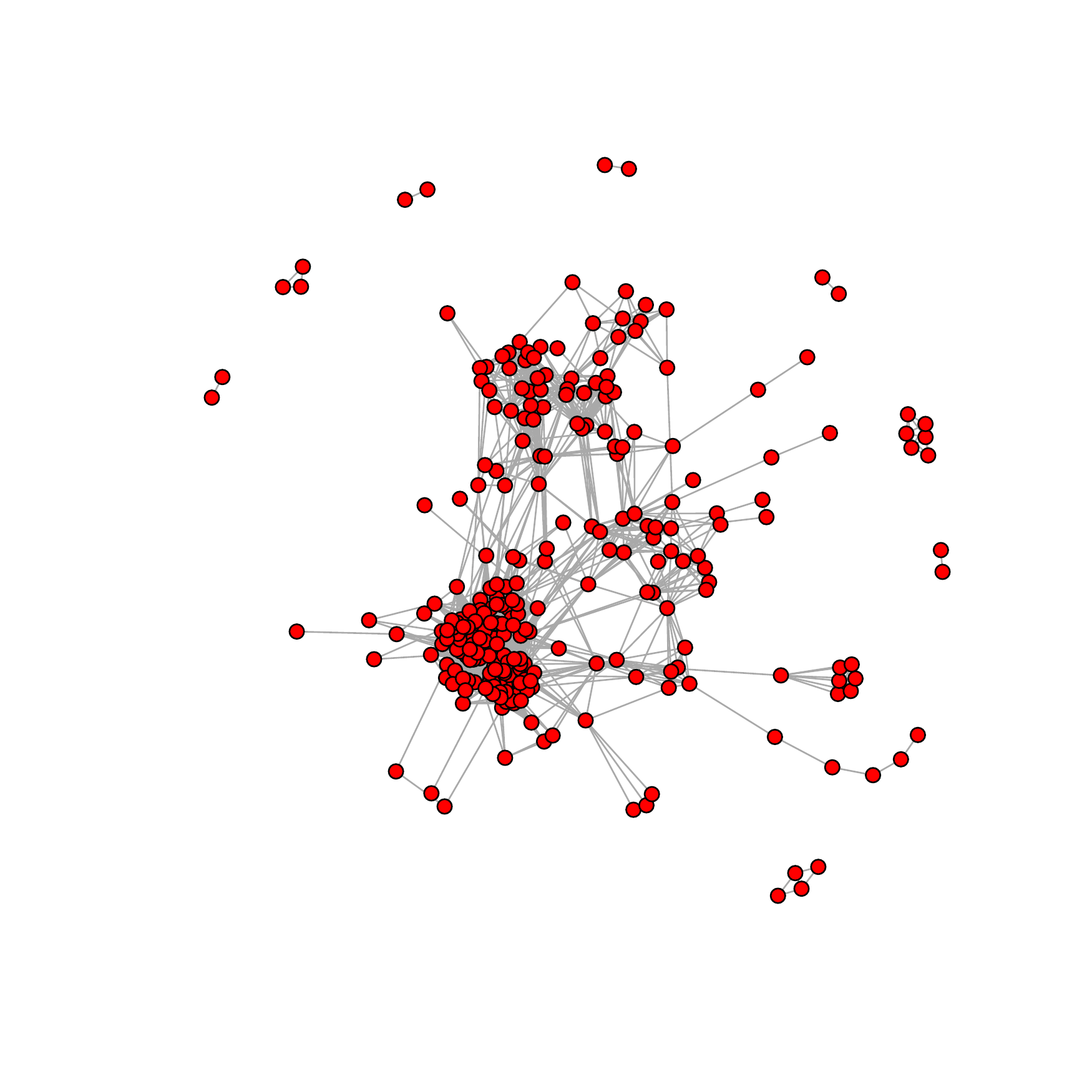}
		\caption{HVR8.}
	\end{subfigure}
	\begin{subfigure}[b]{.18\linewidth}
		\includegraphics[width=\linewidth]{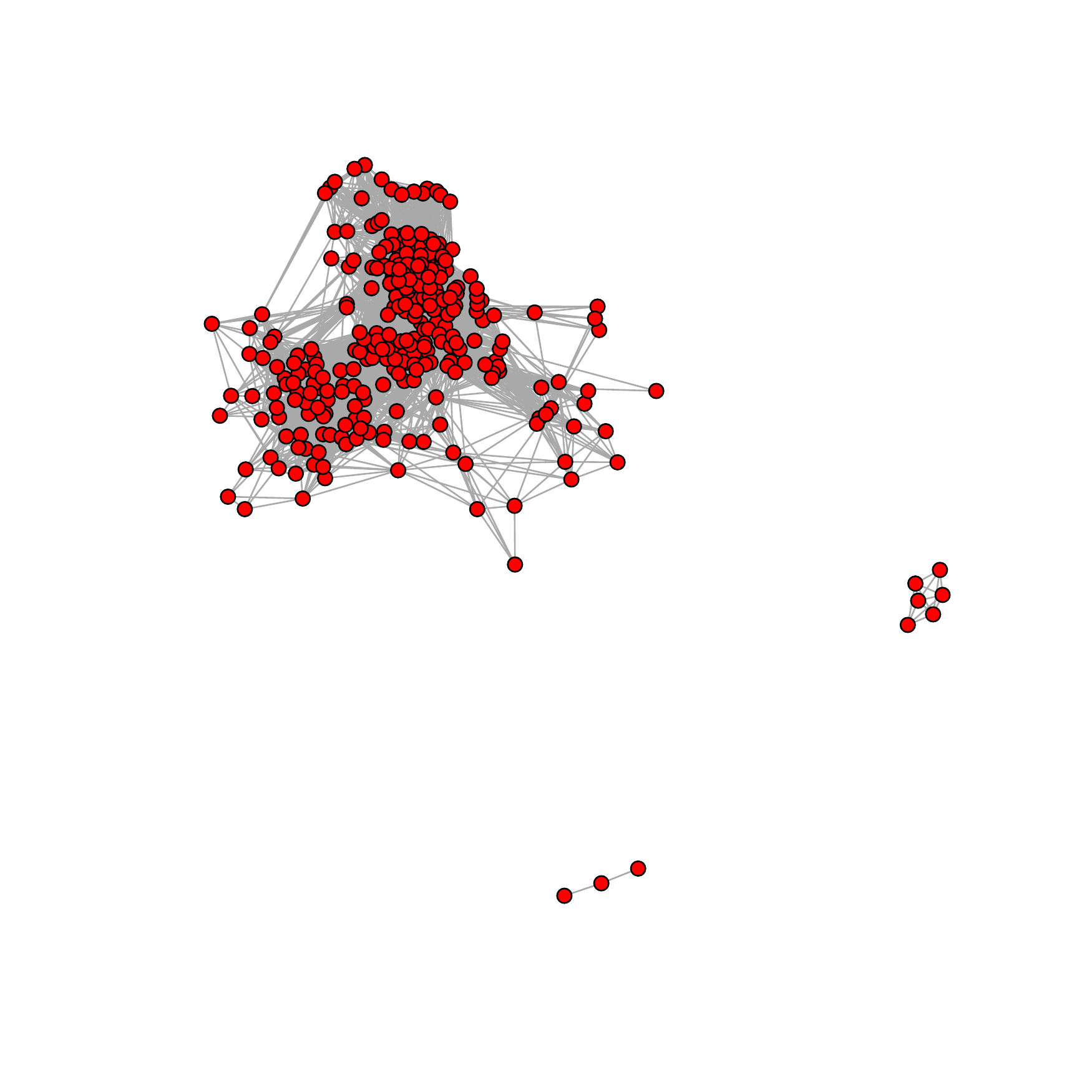}
		\caption{HVR9.}
	\end{subfigure}
\\
	\begin{subfigure}[b]{.6\linewidth}
		\includegraphics[width=\linewidth]{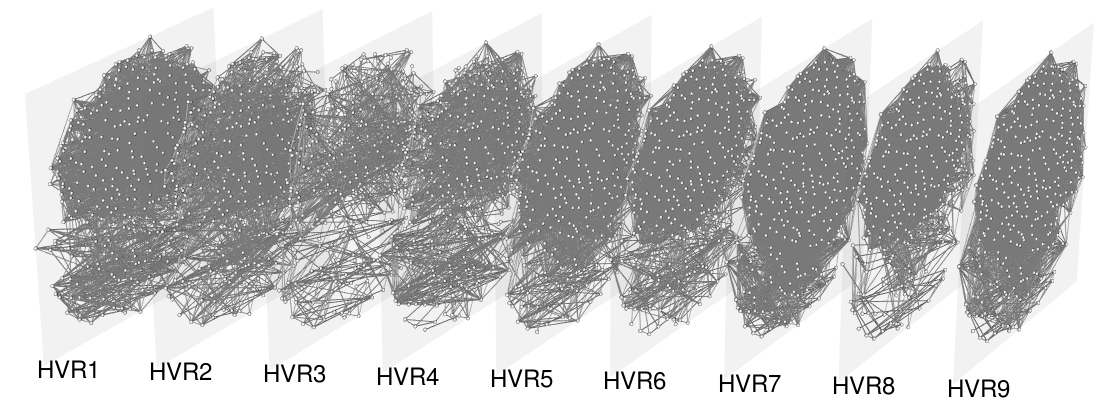}
		\caption{The $212\times 212\times 9$ mixture multi-layer tensor.}
	\end{subfigure}
	\caption{9 extracted highly variable regions and corresponding networks. HVR2, HVR3 and HVR4 are quite sparse and not mainly connected with only some small components.}
	\label{fig:HVR}
\end{figure}

The var genes of the human malaria parasite Plasmodium falciparum present a challenge to population geneticists due to their extreme diversity, which is generated by high rates of recombination. Var gene sequences are characterized by pronounced mosaicism, precluding the use of traditional phylogenetic tools. \cite{larremore2013network} identifies 9 highly variable regions (HVRs), and then maps each HVR to a complex network, see Figure \ref{fig:HVR}. They showed that the recombinational constraints of some HVRs are correlated, while others are independent, suggesting that this micromodular structuring facilitates independent evolutionary trajectories of neighboring mosaic regions, allowing the parasite to retain protein function while generating enormous sequence diversity. 

Despite the innovative network approach, there are still some drawbacks in \cite{larremore2013network}. 
\begin{enumerate}
	\item Even though 9 HVRs have been identified, only 6 HVRs have been used in the analysis, while the other three HRVs are discarded due to their sparse structures, as seen in Figure \ref{fig:HVR} (c,d,e). However, these sparse networks still contain valuable information, which would be of great interest to researchers and practitioners. 
	\item Community structures are identified individually for each network, and then compared with each other to identify similar structures. This is not only very demanding and tedious computationally, but also involves much human intervention. This becomes increasing undesirable as the number of networks grows bigger. 
\end{enumerate}

Here we propose to employ the TWIST to the problem, in order to overcome the above difficulties.  
The data under investigation are the 9 highly variable regions (HVRs) used in \cite{larremore2013network}. Each network is  derived from the same set of 307 genetic sequences from var genes of malaria parasites. A node represents a specific gene and an edge is generated by comparing sequences pair-wisely within each HVR. More information about the data and data pre-processing could be found in \cite{larremore2013network}. 
In our study, we consider 212 nodes which appear on all 9 layers. This results in a $212\times 212\times 9$ mixture multi-layer tensor, as shown in Figure \ref{fig:HVR} (k).  

\begin{figure}
	\centering
	\begin{subfigure}[b]{.4\linewidth}
		\includegraphics[width=\linewidth]{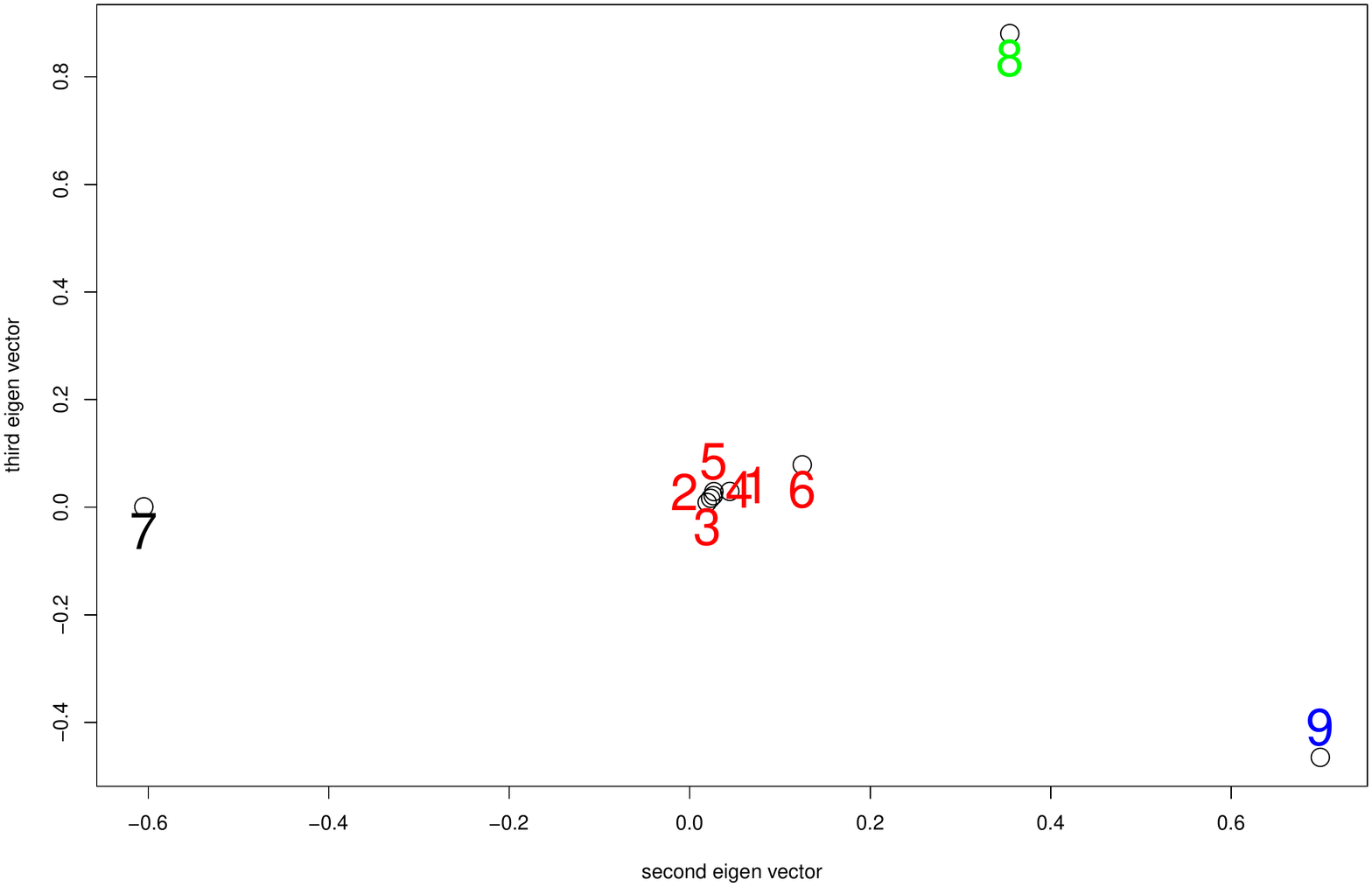}
		\caption{Embedding of each network.}
	\end{subfigure}
	\begin{subfigure}[b]{.4\linewidth}
		\includegraphics[width=\linewidth]{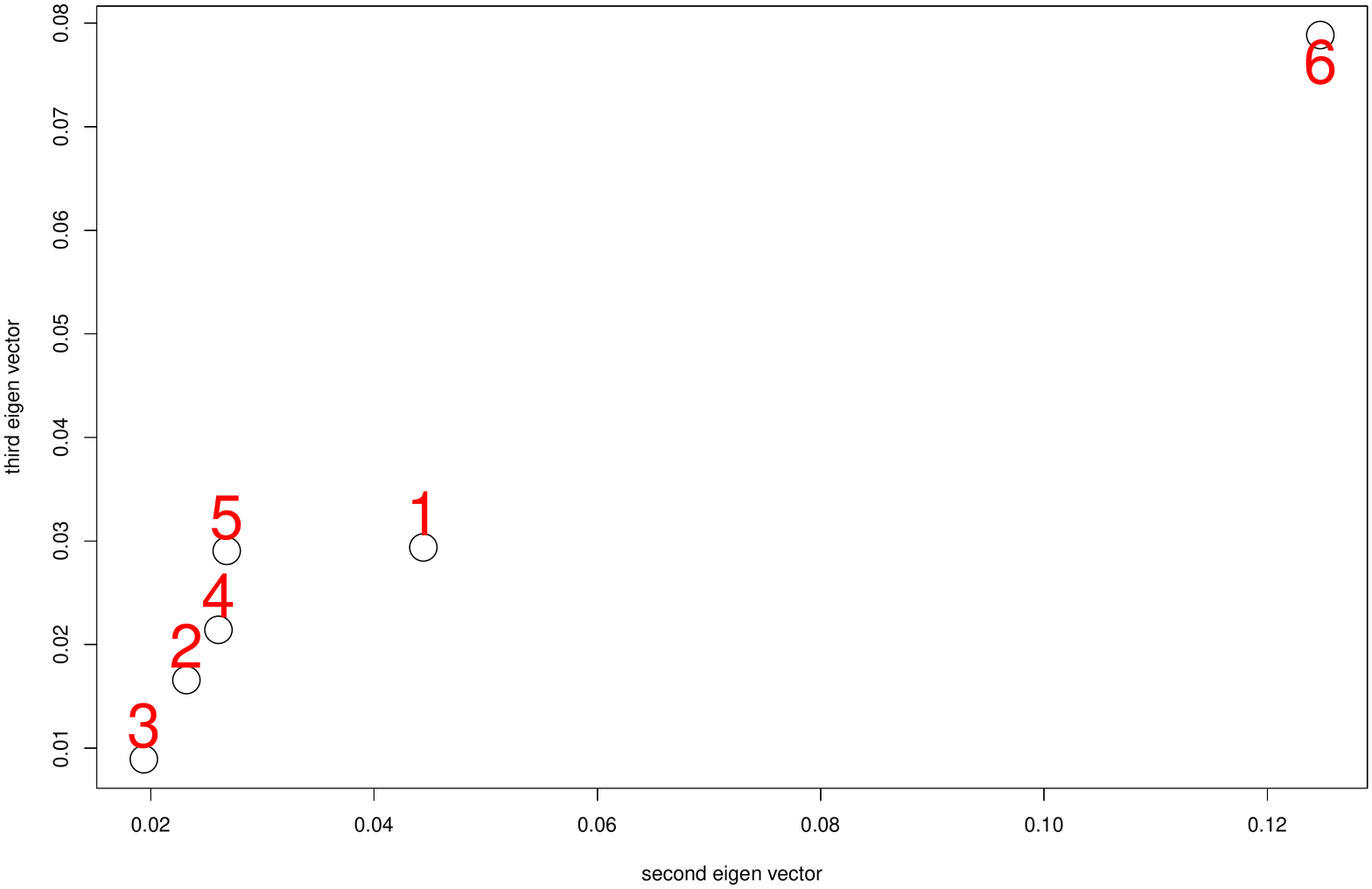}
		\caption{Zoom center area in (a).}
	\end{subfigure}
	\begin{subfigure}[b]{.4\linewidth}
	\includegraphics[width=\linewidth]{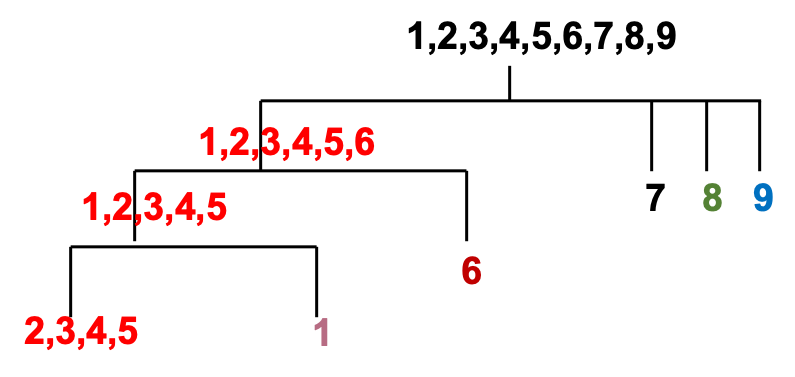}
	\caption{Hierarchical structure of network classes.}
\end{subfigure}
	\caption{Embedding of each layer in malaria parasite genes networks and hierarchical structure of network classes.} \label{fig:genes_layer_embedding}
\end{figure}

\begin{figure}
	\centering
	\begin{subfigure}[b]{.50\linewidth}
		\includegraphics[width=\linewidth]{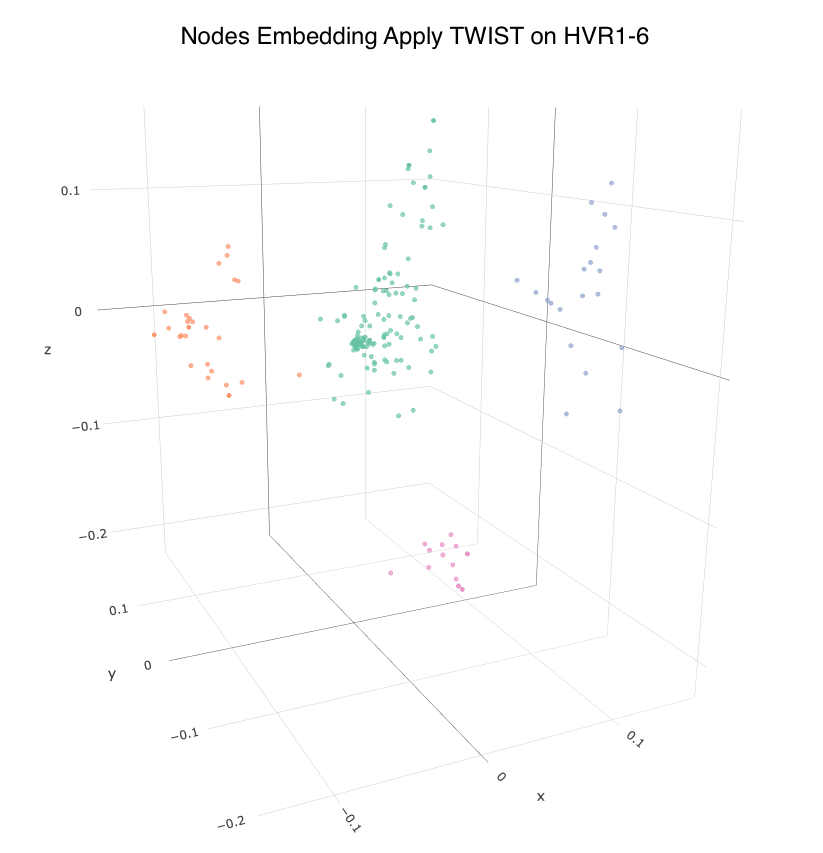}
		\caption{TWIST applied to HVRs 1-6}
	\end{subfigure}
	\begin{subfigure}[b]{.44\linewidth}
		\includegraphics[width=\linewidth]{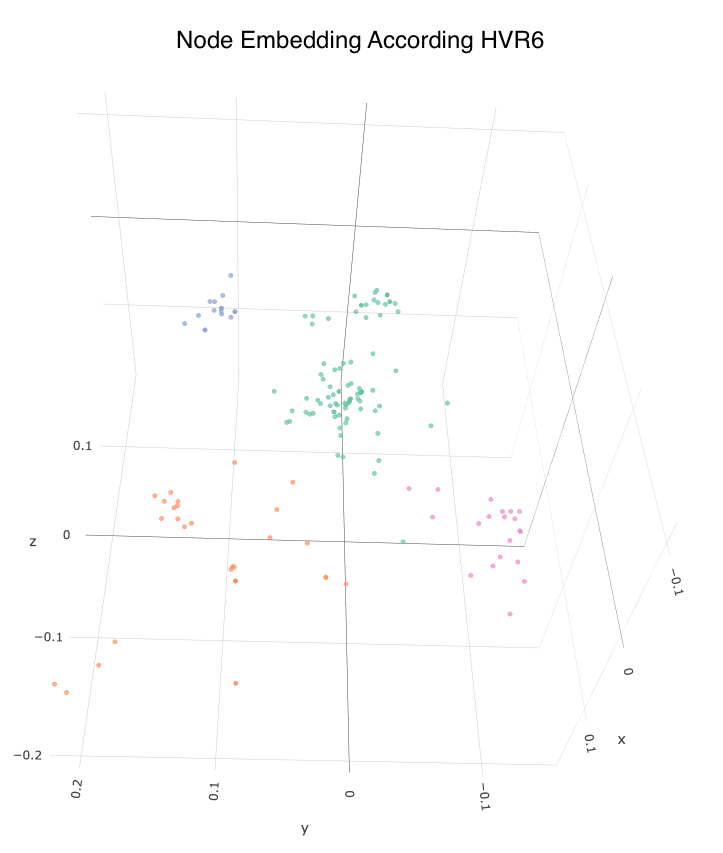}
		\caption{Spectral clustering applied to HVR 6}
	\end{subfigure}
	\caption{Nodes embedding using TWIST and spectral decompositions with $K=4$.}
	\label{fig:genes_embedding}
\end{figure}

We apply the TWIST to this $212\times 212\times 9$ tensor with the core tensor $15\times 15\times 3.$ The embedding of each layer is plotted in Figure \ref{fig:genes_layer_embedding}. We make the following comments. 
\begin{enumerate}
	\item The 9 HVRs fall into 4 groups (Figure \ref{fig:genes_layer_embedding} (a)):
	$
	\{1, 2, 3, 4, 5, 6\}, \{7\}, \{8\}, \{9\}. 
	$
	
	\smallskip
	
	By comparison, \cite{larremore2013network} found that the 6 HVRs fall into 4 groups (without layers 2-4): 
	$\{1, 5, 6\}, \{7\}, \{8\}, \{9\}.$  The two findings are consistent. 
	
	\smallskip
	
	\item  TWIST places sparse networks of layers 2-4 to the same group as layers 1 and 5. 
	
	\smallskip

	By comparison, the sparse layers 2-4 had to be discarded in \cite{larremore2013network}. The new result implies that the sequences remains mostly unchanged in the beginning (HVRs 1-6), and start to diversity from HVR 7 onward. 
	
	
	\smallskip
	
	\item Hierarchical structure of the 9 HVRs. 
	
	\smallskip
	
	If we zoom in the mini group $\{1, 2, 3, 4, 5, 6\}$ (Figure \ref{fig:genes_layer_embedding}(b)), we notice that the first 5 layers are more tied together, so we have a finer partition: $\{1, 2, 3, 4, 5\}, \{6\}$. This operation can be repeated. Therefore, TWIST can be easily used to form a hierarchical structure of the 9 HVRs (Figure \ref{fig:genes_layer_embedding} (c)). 
	
	\smallskip
	
	\item Computational ease of the TWIST
	
	\smallskip
	
	TWIST can easily cluster layers and nodes using $K$-means. This is much easier than the procedure in \cite{larremore2013network}, which first finds the community structure for each layer, and then computes their similarities. 
	
	\smallskip
	
	\item Better community structure is obtained by combining information from similar layers. 
	
	\smallskip
	
	TWIST is applied to the first 6 similar layers $\{1, 2, 3, 4, 5, 6\}$ to identify their common local structure, while spectral clustering is applied to HVR 6 to find its community structure, as was done in \cite{larremore2013network}, see Figure \ref{fig:genes_embedding} (a)-(b). 
	Clearly, the 4 local communities are much more separated in Figure \ref{fig:genes_embedding}(a) than in (b). 
	
\end{enumerate}  


\subsection{Worldwide food trading networks}

We consider the dataset on the worldwide food trading networks, which is collected by \cite{de2015structural}, and is available at \url{http://www.fao.org}. The data contains an economic network in which layers represent different products, nodes are countries and edges at each layer represent trading relationships of a specific food product among countries. 

We focus on the trading data in 2010 only. We convert the original directed networks to undirected ones by ignoring the directions. We delete the links with weight less than 8 (the first quartile) and abandon the layers whose largest component consists less than 150 nodes. These are done to to filter out the less important information. Finally we extract the intersections of the largest components of the remaining layers. 

After data preprocessing, we obtained a 30-layers network with 99 nodes at each layer. Each layer represents trading relationships between 99 countries/regions worldwide with respect to 30 different food products. Together they form a mixture multi-layer tensor of dimension $99\times99\times30$.

\begin{figure}
	\centering
	\begin{subfigure}[b]{.7\linewidth}
		\includegraphics[width=\linewidth]{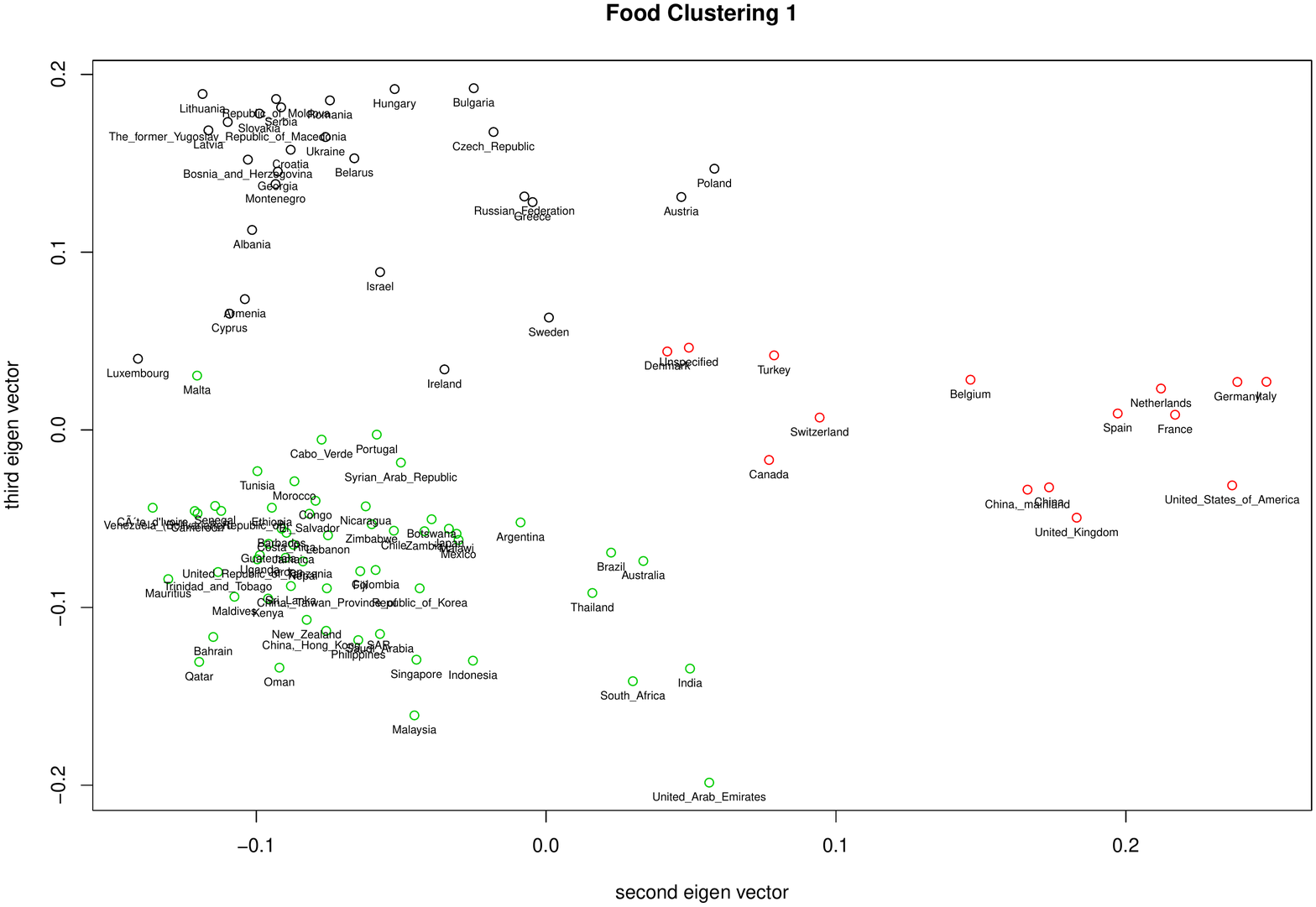}
		\caption{Embedding of countries for networks in cluster 1}
	\end{subfigure}
	\begin{subfigure}[b]{.7\linewidth}
		\includegraphics[width=\linewidth]{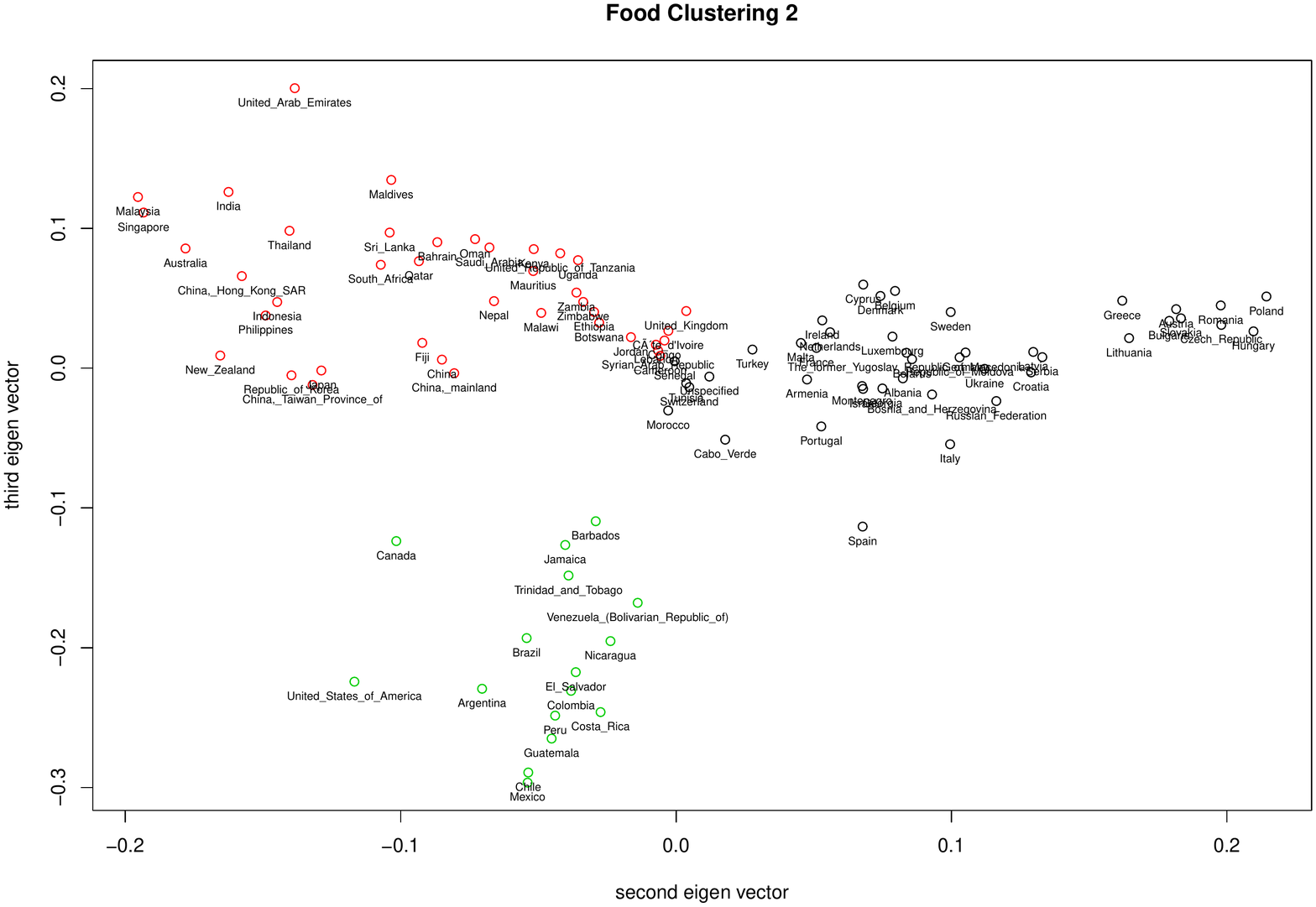}
		\caption{Embedding of countries for networks  in cluster 2}
	\end{subfigure}
	\caption{Embedding of countries on two different types of food trading networks.}
	\label{fig:country_embedding}
\end{figure}

\begin{figure}
	\centering
	\begin{subfigure}[b]{.46\linewidth}
		\includegraphics[width=\linewidth]{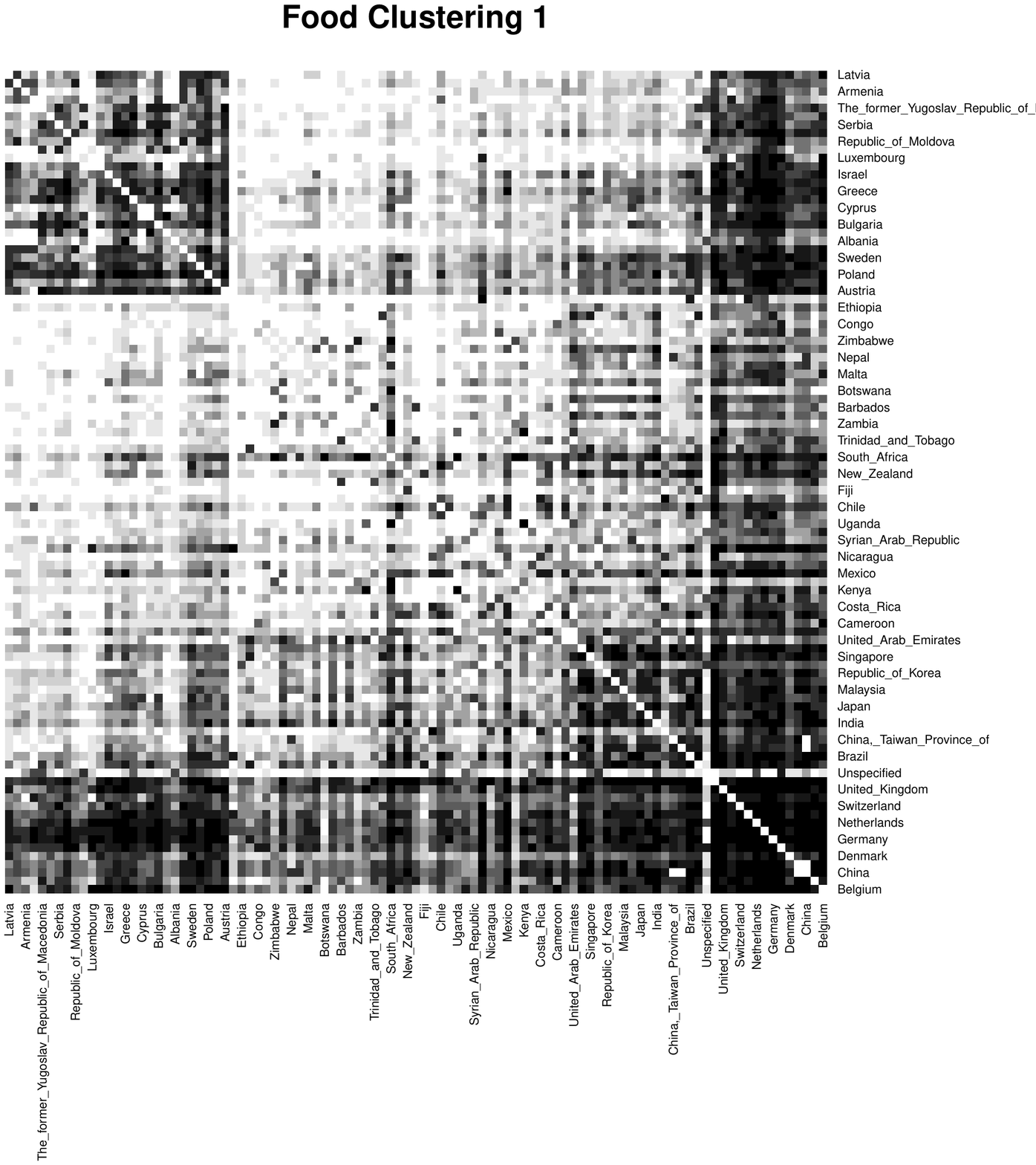}
		\caption{Heat map of networks in cluster 1.}
	\end{subfigure}
	\begin{subfigure}[b]{.46\linewidth}
		\includegraphics[width=\linewidth]{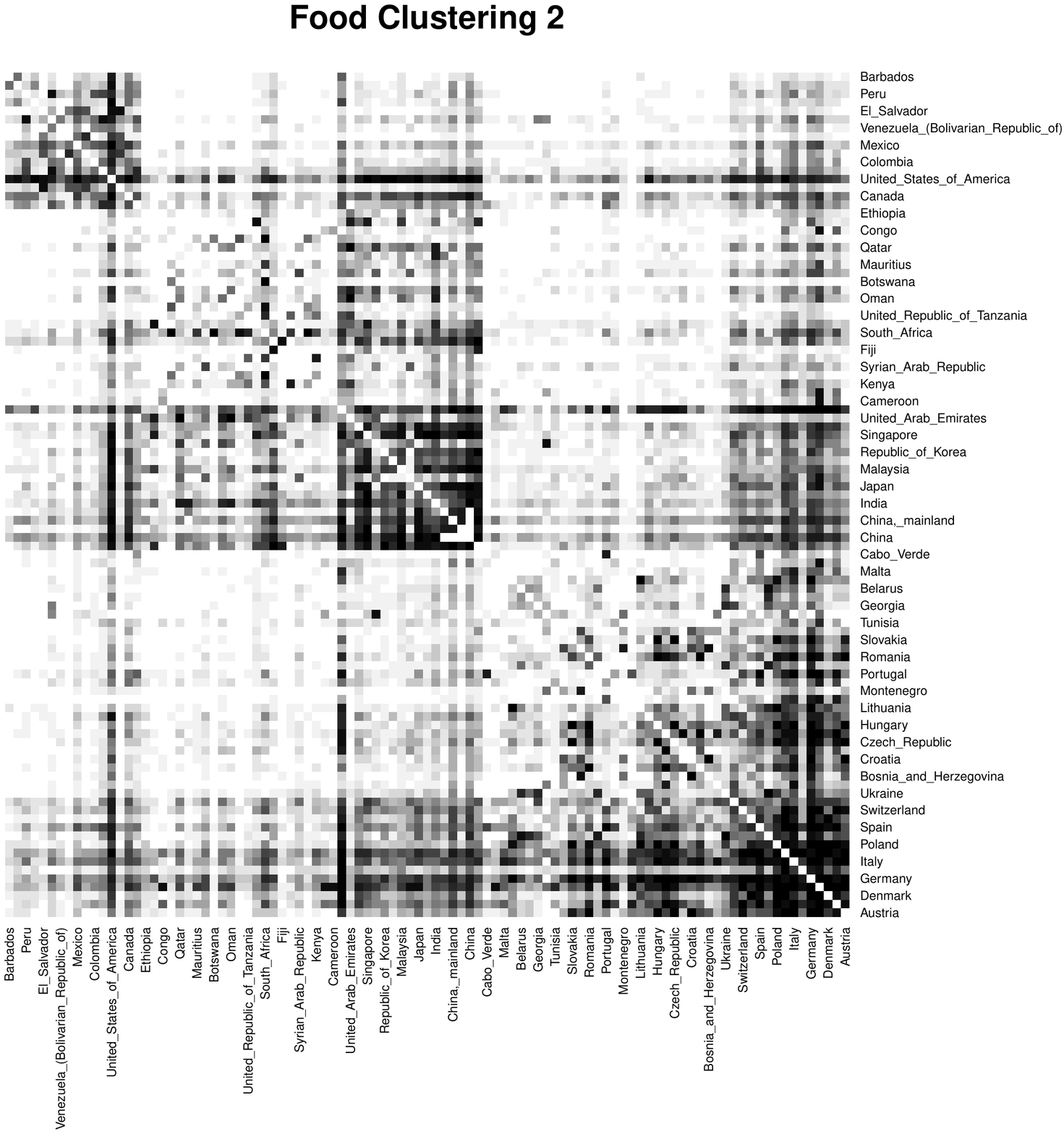}
		\caption{Heat map of networks in cluster 2.}
    \end{subfigure}
	\caption{Heat maps of two types of networks.}
	\label{fig:food_networks}
\end{figure}

We first apply Algorithm 1 in the TWIST procedure to the mixture multi-layer tensor, which results in a tensor decomposition with a core tensor of dimension $20\times 20\times 2$. The resulting two clusters of layers are listed in Table \ref{food_cluster}. We then apply Algorithm 2 in the TWIST procedure to each cluster separately (here we have two clusters) to find the community structures for each cluster, in order to obtain the clustering result of countries. This time, we take the core tensor of dimension $4\times 4\times 1.$ The embedding of 99 countries with clustering results from K-means are shown in Fig. \ref{fig:country_embedding}. 
For the two types of networks, we plot the sum of adjacency matrices with nodes arranged according the community labels in Figure \ref{fig:food_networks} to have a glance of different community structures of two network types.

\begin{table} 
\begin{center}
	\begin{tabular}{ | l  p{10cm} |}
		\hline
		Food cluster 1: &
		Beverages non alcoholic, Food prep nes,       Chocolate products nes, Crude materials, Fruit prepared nes, Beverages distilled alcoholic, Coffee green, Pastry, Sugar confectionery, Wine, Tobacco unmanufactured    \\ \hline
		Food cluster 2: & 	
	    Cheese whole cow milk, Cigarettes, Flour wheat, Beer of barley, Cereals breakfast, Milk skimmed dried, Juice fruit nes, Maize, Macaroni, Oil palm, Milk whole dried, Oil essential nes, Rice milled, Sugar refined, Tea, Spices nes, Vegetables preserved nes, Waters ice etc, Vegetables fresh nes  \\
		\hline
	\end{tabular}
\caption{List of two clusters of food.}
\label{food_cluster}
\end{center}
\end{table}

We make the following remarks from Table \ref{food_cluster}, Figures \ref{fig:country_embedding} and  \ref{fig:food_networks}. 
\begin{enumerate}
\item 
Trading patterns of food are different for unprocessed and processed foods.

\smallskip
 
Specifically, cluster 1 consists mainly of raw or unprocessed food (e.g., crude materials, coffee green, unmanufactured tobacco), while cluster 2 is mainly made of processed food (e.g., such as cigarettes, flour wheat, essential oil, milled rice, refined sugar). 

\smallskip

\item  For unprocessed food, global trading is the more dominant trading pattern than regional one. Some countries have closer trading ties with countries across the globe. 

\smallskip

From cluster 1, a small number of countries, such as China, Canada, United Kingdom, United States, France, Germany, are very active in trading with others as well as amongst themselves. This small group of countries is called a hub community. 
This implies reflects the fact that these large countries import unprocessed food from, and/or export unprocessed food to a great number of other countries worldwide. 

\smallskip

\item For processed foods, regional trading is very dominant. In fact, the world trading map is striking similar to the world geography map in Figure \ref{fig:country_embedding} (b).  

\smallskip 

I cluster 2, countries are mainly clustered by the geographical location, i.e.,  countries in the same continent have closer trading ties. Examples of these clusters include countries in America (United State, Canada, Mexico, Brazil, Chile), in Asia and Africa (China, Japan, Singapore, Thailand, Indonesia, Philippines, India), and in Europe (Germany, Italy, Poland, Spain, Denmark, Switzerland). Regional trading of processed food can have many advantages, e.g., keeping the food cost low due to lower transportation cost, and keeping food refresh due to faster delivery. 


\smallskip

There are some interesting "outliers" as well. For instance, United Kingdom has closer trading ties with African and Middle Eastern countries than its European neighbor, which might be interesting to delve into further. 


\end{enumerate}

\section{Conclusion and discussion}\label{sec:discussion}

In this paper, we have proposed a novel mixture multi-layer stochastic block model (MMSBM) to capture the intrinsic local as well as global community structures. A tensor-based algorithm, TWIST, was proposed to conduct community detection on multi-layer networks and shown to have near optimal error bounds under weak conditions in the MMSBM framework. In particular, the method allows for very sparse networks in many layers. The proposed method outperforms other state of the art methods both in nodes community detection and layers clustering by extensive simulation studies. We also applied the algorithm to two real dataset and found some interesting results. 

A number of future directions are worth exploring. As a natural extension, one can generalize the tensor-based representation to account for adjacency matrices capturing the degree heterogeneity of nodes.  The layers of networks could have the spacial and temporal structures of networks in many real applications, one could incorporate these into the model. On a more theoretical level, it is of interest to explore  theoretical properties in more sparse scenario. It is also important to develop scalable algorithms which can handle millions nodes with thousands layers in this big data era.

\bibliographystyle{plainnat}
\bibliography{manuscript}

\section{Proofs}

\subsection{Proof of Lemma~\ref{lem:EA_decomp}}

Write $\mathbf{X}=\mathbf{B}\times_1\bar{Z}\times_2\bar{Z}\times_3W$, then
\begin{equation*}
\begin{aligned}
\mathbf{X}(:,:,l)&=\mathbf{X}\times_3 e_l^{(L)}=\mathbf{B}\times_1\bar{Z}\times_2\bar{Z}\times_3W\times_3(e_l^{(L)})^T\\
&=\mathbf{B}\times_1\bar{Z}\times_2\bar{Z}\times_3(e_l^{(L)})^TW\\
&=\mathbf{B}\times_1\bar{Z}\times_2\bar{Z}\times_3(e_l^{(L)})^T({e}_{\ell_1}^{(m)},{e}_{\ell_2}^{(m)},...,{e}_{\ell_L}^{(m)})^T\\
&=\mathbf{B}\times_1\bar{Z}\times_2\bar{Z}\times_3({e}_{\ell_l}^{(m)})^T\\
&={B}(:,:,\ell_l)\times_1\bar{Z}\times_2\bar{Z}\\
&=\bar{Z}
\left(\begin{array}{rrrrrrr}
0_{K_1} &  &  &  &  &  &\\
 & \ddots &  &  &  &  &\\
\undermat{\ell_l-1}{~~~&~~~&0_{K_{\ell_l-1}}} &  &  & &\\
& & & B_{\ell_l} & & &\\
& & & & 0_{K_{\ell_l+1}} & &\\
& & & & & \ddots &\\
& & & & \undermat{m-\ell_l}{~~~&~~~& 0_{K_m}}\\
& & & & & \\
\end{array}
\right)
\bar{Z}^T\\
&=Z_{\ell_l}B_{\ell_l}Z_{\ell_l}^T=\mathbb{E}(A_l|\ell_l).
\end{aligned}
\end{equation*}
Hence we conclude that 
$\mathbb{E}(\mathbf{A}|\LL)=\mathbf{X}=\mathbf{B}\times_1\bar{Z}\times_2\bar{Z}\times_3W.$

\subsection{Proof of Lemma~\ref{lem:barU_sep}}

Since $k_1\ne k_2$,
\begin{equation*}
\begin{aligned}
1&\le\Vert \bar{Z}(i_1,:)-\bar{Z}(i_2,:)\Vert_{\ell_2}=\Vert\left(\bar{U}(i_1,:)-\bar{U}(i_2,:) \right)\bar{D}\bar{R}^T \Vert_{\ell_2}\\
&\le \Vert\left(\bar{U}(i_1,:)-\bar{U}(i_2,:) \right)\bar{D}\Vert_{\ell_2}\\
&\le \sigma_1(\bar{D})\Vert\bar{U}(i_1,:)-\bar{U}(i_2,:) \Vert_{\ell_2}.
\end{aligned}
\end{equation*}
It immediately implies the claim. 

\subsection{Proof of Lemma~\ref{lem:ss}}

By the definition of $\bar{\bB}$, it is clear that $\sigma_{\min}(\bar{\bC})\geq \sigma_{\min}(\bar{\bB})\cdot \sigma_r^2(\bar{D})\sqrt{L_{\min}}$. Recall that $\bar{D}$ consists of the singular values of $\bar{Z}$. It is obvious by definition that $\|\bar{Z}\|_{\rm F}=nm$. Therefore, $\sigma_r^2(\bar{D})\geq \kappa_0^{-2}r^{-1}\cdot nm$ which concludes the proof. 

\

\subsection{Proof of Lemma~\ref{lem:incoh}}

Recall that $\calM_3\big(\EE(\bA|\LL)\big)=\bar{W}\calM_3(\bar{\bC})(\bar{U}\otimes \bar{U})^{\top}$. Therefore,
$$
\|e_l^{\top}\bar{W}\|\sigma_{\min}(\bar{\bC})\leq \|e_l^{\top}\bar{W}\calM_3(\bar{\bC})(\bar{U}\otimes\bar{U})^{\top}\|=\|\EE A_l|\ell_l\|_{\rm F}\leq np_{\max}.
$$
By Lemma~\ref{lem:ss}, we conclude that $\max_l\|e_l^{\top}\bar{W}\|\leq \kappa_0^2r/(m\sqrt{L_{\min}})$. Similarly, recall that $\bar{Z}=\bar{U}\bar{D}\bar{R}^{\top}$ is the singular value decomposition of $\bar{Z}$. Then, 
$$
\|e_j^{\top}\bar{U}\|=\|e_j^{\top}\bar{Z}\bar{R}\bar{D}^{-1}\|\le \frac{1}{\sigma_r({\bar{D}})}\|e_j^{\top}\bar{Z}\|=\frac{\sqrt{m}}{\sigma_r(\bar{D})}\le\kappa_0\sqrt{\frac{r}{n}},
$$
where the last inequality is due to $\sigma_r(\bar{D})\geq \kappa_0^{-1}\sqrt{mn/r}$ by the proof of Lemma~\ref{lem:ss}. 

\subsection{Proof of Theorem~\ref{thm:power_iteration}}

For $t\geq 1$, we begin with the bound for ${\rm d}(\widehat U^{(t)},\bar{U})$ where $\widehat U^{(t)}$ is the output of a regularized power iteration with input $\widehat U^{(t-1)}$ and $\widehat W^{(t-1)}$. Without loss of generality, assume that ${\rm d}(\widehat U^{(t-1)},\bar{U})\vee {\rm d}(\widehat W^{(t-1)},\bar{W})\leq 1/4$. For all integers $t\geq 0$, denote
$$
{\rm Err}_t=\max\Big\{{\rm d}(\widehat U^{(t-1)},\bar{U}), {\rm d}(\widehat W^{(t-1)},\bar{W})\Big\}.
$$

 By Algorithm~\ref{algo:Power Iterations}, $\widehat U^{(t)}$ is the top-$r$ left singular vectors of $\mathcal{M}_1(\mathbf{A})\big(\widetilde U^{(t-1)}\otimes \widetilde W^{(t-1)} \big)$ where we abuse the notation and denote $\otimes$ the Kronecker product. By Algorithm~\ref{algo:Power Iterations} and property of regularization (see \cite{keshavan2010matrix} and \cite{ke2019community}),  
\begin{equation}\label{eq:tildeU}
{\rm d}(\widetilde U^{(t-1)},\bar{U})\leq 2\sqrt{2}\cdot {\rm d}(\widehat U^{(t-1)}, \bar{U})\quad {\rm and}\quad \max_j\|e_j^{\top}\widetilde U^{(t-1)}\|\leq \sqrt{2}\delta_1
\end{equation}
and 
\begin{equation}\label{eq:tildeW}
{\rm d}(\widetilde W^{(t-1)},\bar{W})\leq 2\sqrt{2}\cdot {\rm d}(\widehat W^{(t-1)}, \bar{W})\quad {\rm and}\quad \max_j\|e_j^{\top}\widetilde W^{(t-1)}\|\leq \sqrt{2}\delta_2.
\end{equation}
Write
$$
\mathcal{M}_1(\mathbf{A})\big(\widetilde U^{(t-1)}\otimes \widetilde W^{(t-1)} \big)=\mathcal{M}_1(\mathbb{E}\mathbf{A})\big(\widetilde U^{(t-1)}\otimes \widetilde W^{(t-1)} \big)+\mathcal{M}_1(\mathbf{A}-\mathbb{E}\mathbf{A})\big(\widetilde U^{(t-1)}\otimes \widetilde W^{(t-1)} \big).
$$
Observe that the left singular space of $\mathcal{M}_1(\mathbb{E}\mathbf{A})\big(\widetilde U^{(t-1)}\otimes \widetilde W^{(t-1)} \big)$ is the column space of $\bar{U}$. In addition,
\begin{align*}
\sigma_r\Big(\mathcal{M}_1(\mathbb{E}\mathbf{A})\big(\widetilde U^{(t-1)}\otimes \widetilde W^{(t-1)} \big)\Big)=\sigma_r\Big(\calM_1(\bar{\bC})\big((\bar{U}^{\top}\widetilde{U}^{(t-1)})\otimes (\bar{W}^{\top}\widetilde{W}^{(t-1)})\big)\Big)\\
\geq \sigma_{r}(\calM_1(\bar{\bC}))\cdot \sigma_{\min}(\bar{U}^{\top}\widetilde{U}^{(t-1)})\sigma_{\min}(\bar{W}^{\top}\widetilde{W}^{(t-1)})\geq \sigma_r(\calM_1(\bar{\bC}))/4,
\end{align*}
where we used the fact
$$
\|I-(\bar U^{\top}\widetilde U^{(t-1)})(\widetilde U^{(t-1)\top}U)\|\leq \frac{\sqrt{2}}{2},
$$
implying that $\sigma_{\min}(\bar U^{\top}\widetilde U^{(t-1)})\geq 1/2$. 

We next bound the operator norm of $\mathcal{M}_1(\mathbf{A}-\mathbb{E}\mathbf{A})\big(\widetilde U^{(t-1)}\otimes \widetilde W^{(t-1)} \big)$. For notational simplicity, denote $\bDelta=\bA-\EE\bA$. Write
\begin{equation*}
\begin{aligned}
\big\|\calM_1\big(&\bDelta\times_2(\widetilde U^{(t-1)})^{\top}\times_3(\widetilde W^{(t-1)})^{\top}\big) \big\|\\
\le&\big\|\mathcal{M}_1\big(\bDelta\times_2(\widetilde U^{(t-1)})^{\top}\times_3(\widetilde W^{(t-1)}-\bar{W}\widetilde{O}^{(t-1)}_W)^{\top}\big)\big\|+\big\|\mathcal{M}_1\big(\bDelta\times_2(\bar{U}\widetilde{O}_U^{(t-1)})^{\top}\times_3(\bar{W}\widetilde{O}_W^{(t-1)})^{\top}\big)\big\|\\
&+\big\|\mathcal{M}_1\big(\bDelta\times_2(\widetilde U^{(t-1)}-\bar{U}\widetilde{O}_U^{(t-1)})^{\top}\times_3(\bar{W}\widetilde{O}^{(t-1)}_W)^{\top}\big)\big\|
\end{aligned}
\end{equation*}
where $\widetilde{O}^{(t-1)}_U=\argmin_{O\in \mathbb{O}_r}\|\widetilde{U}^{(t-1)}-\bar{U}O\|$ and $\widetilde{O}^{(t-1)}_W=\argmin_{O\in \mathbb{O}_m}\|\widetilde{W}^{(t-1)}-\bar{W}O\|$.\\
By Lemma~\ref{lem:tensor_matrix_connection} (from \cite[Lemma~6]{xia2017statistically}), we obtain
\begin{equation*}
\begin{aligned}
\big\|\mathcal{M}_1\big(\bDelta\times_2&(\widetilde U^{(t-1)})^{\top}\times_3(\widetilde W^{(t-1)}-\bar{W}\widetilde{O}^{(t-1)}_W)^{\top}\big)\big\|\\
\leq&\sqrt{\frac{2rm}{\max\{r,2m\}}}\big\|\bDelta\times_2(\widetilde{U}^{(t-1)})^{\top}\times_3\big(\widetilde{W}^{(t-1)}-\bar{W}\widetilde{O}_W^{(t-1)}\big)^{\top}\big\|\\
&=\sqrt{r\wedge 2m}\cdot \big\|\bDelta\times_2(\widetilde{U}^{(t-1)})^{\top}\times_3\big(\widetilde{W}^{(t-1)}-\bar{W}\widetilde{O}_W^{(t-1)}\big)^{\top}\big\|.
\end{aligned}
\end{equation*}
The last term can be sharply bounded via the tensor incoherent norm. Indeed, for any $v_1\in\mathbb{R}^n,v_2\in\mathbb{R}^r,v_3\in\mathbb{R}^m$ with $\Vert v_j\Vert_{\ell_2}\le1,j=1,2,3$,
\begin{equation*}
\begin{aligned}
&\big<\bDelta\times_2(\widetilde{U}_2^{(t-1)})^{\top}\times_3\big(\widetilde{W}^{(t-1)}-\bar{W}\widetilde{O}_W^{(t-1)}\big)^{\top},v_1\otimes v_2\otimes v_3\big>\\
&=\big<\bDelta,\ v_1\otimes (\widetilde{U}^{(t-1)}v_2)\otimes \big((\widetilde{W}^{(t-1)}-\bar{W}\widetilde{O}_W^{(t-1)})v_3\big)\big>\\
&=\| \widetilde{W}^{(t-1)}-\bar{W}\widetilde{O}_W^{(t-1)}\|\cdot \Big<\bDelta,v_1\otimes (\widetilde{U}^{(t-1)}v_2)\otimes \Big(\big((\widetilde{W}^{(t-1)}-\bar{W}\widetilde{O}_W^{(t-1)})v_3\big)/\|\widetilde{W}^{(t-1)}-\bar{W}\widetilde{O}_W^{(t-1)}\|\Big)\Big>\\
&\le\big\| \widetilde{W}^{(t-1)}-\bar{W}\widetilde{O}_W^{(t-1)}\big\|\cdot \|\bDelta\|_{2,\sqrt{2}\delta_1}
\end{aligned}
\end{equation*}
where we use (\ref{eq:tildeU}) and the fact that 
$$
|e_j^{\top} \widetilde{U}^{(t-1)}v_2|\leq \|e_j^{\top} \widetilde{U}^{(t-1)}\|\leq \sqrt{2}\delta_1.
$$ 
Thus, 
\begin{equation}\label{eq:thm1bound1}
\big\|\mathcal{M}_1\big(\bDelta\times_2(\widetilde U^{(t-1)})^{\top}\times_3(\widetilde W^{(t-1)}-\bar{W}\widetilde{O}^{(t-1)}_W)^{\top}\big)\big\|\le\sqrt{r\wedge 2m}\cdot \| \widetilde{W}^{(t-1)}-\bar{W}\widetilde{O}_W^{(t-1)}\|\cdot \|\bDelta\|_{2,\sqrt{2}\delta_1}.
\end{equation}
In the same fashion, 
\begin{equation}\label{eq:thm1bound2}
\big\|\mathcal{M}_1\big(\bDelta\times_2(\widetilde U^{(t-1)}-\bar{U}\widetilde{O}_U^{(t-1)})^{\top}\times_3(\bar{W}\widetilde{O}^{(t-1)}_W)^{\top}\big)\big\| \leq\sqrt{r\wedge 2m}\cdot \|\widetilde U^{(t-1)}-\bar{U}\widetilde{O}_U^{(t-1)}\|\cdot \|\bDelta\|_{3,\sqrt{2}\delta_2}.
\end{equation}
Putting together (\ref{eq:thm1bound1}) and (\ref{eq:thm1bound2}), we obtain
\begin{align*}
\big\|\calM_1\big(\bDelta\times_2(\widetilde U^{(t-1)})^{\top}&\times_3(\widetilde W^{(t-1)})^{\top}\big) \big\|\leq \big\|\calM_1(\bDelta\times_2 \bar{U}^{\top}\times_3 \bar{W}^{\top}) \big\|\\
+&\sqrt{r\wedge 2m}\cdot \Big(\| \widetilde{W}^{(t-1)}-\bar{W}\widetilde{O}_W^{(t-1)}\|\cdot \|\bDelta\|_{2,\sqrt{2}\delta_1}+\|\widetilde U^{(t-1)}-\bar{U}\widetilde{O}_U^{(t-1)}\|\cdot \|\bDelta\|_{3,\sqrt{2}\delta_2}\Big).
\end{align*}
It remains to bound $\|\calM_1(\bDelta\times_2\bar{U}^{\top}\times_3\bar{W}^{\top})\|$ where $\bar{U}$ and $\bar{W}$ are deterministic singular vectors. Towards that end, a matrix Bernstein inequality could yield a sharp bound. Indeed, write
\begin{align*}
\big\|\calM_1(\bDelta\times_2\bar{U}^{\top}\times_3\bar{W}^{\top})\big\|\leq \Big\|\sum_{i_3=1}^{L}\sum_{(i_1,i_2)\in\mathfrak{I}_n^2}\big(A(i_1,i_2,i_3)-\EE A(i_1,i_2,i_3)\big)e_{i_1}\big((\bar{U}^{\top}e_{i_2})\otimes(\bar{W}^{\top} e_{i_3})\big)^{\top}\Big\|\\
+\Big\|\sum_{i_3=1}^{L}\sum_{(i_1,i_2)\in\mathfrak{I}_n^2}\big(A(i_2,i_1,i_3)-\EE A(i_2,i_1,i_3)\big)e_{i_2}\big((\bar{U}^{\top}e_{i_1})\otimes(\bar{W}^{\top} e_{i_3})\big)^{\top}\Big\|
\end{align*}
where $\mathfrak{I}_n^2=\{(i_1,i_2),1\leq i_1<i_2\leq n\}$. Clearly, it is equivalent to bound the spectral norm of the sum of independent random matrices. The following bounds are obvious.
$$
\big\|\big(A(i_1,i_2,i_3)-\EE A(i_1,i_2,i_3)\big)e_{i_1}\big((\bar{U}^{\top}e_{i_2})\otimes(\bar{W}^{\top} e_{i_3})\big)^{\top}\big\|\leq \delta_1\delta_2
$$
and 
\begin{align*}
\Big\|\sum_{i_3=1}^L\sum_{(i_1,i_2)\in\mathfrak{I}_n^2}{\rm Var}\big(A(i_1,i_2,i_3)-\EE A(i_1,i_2,i_3)\big)\cdot \big((\bar{U}^{\top}e_{i_2})\otimes(\bar{W}^{\top} e_{i_3})\big)\big((\bar{U}^{\top}e_{i_2})\otimes(\bar{W}^{\top} e_{i_3})\big)^{\top} \Big\|\\
\leq np_{\max}.
\end{align*}
Therefore, by matrix Bernstein inequality (\cite{tropp2012user} and \cite{koltchinskii2015optimal}), with probability at least $1-n^{-2}$,
\begin{align*}
\big\|\calM_1(\bDelta\times_2\bar{U}^{\top}\times_3\bar{W}^{\top})\big\|\leq C_1\sqrt{np_{\max}\log n}+C_2\delta_1\delta_2\log n
\end{align*}
for some absolute constants $C_1,C_2>0$.
By Davis-Kahan Theorem, 
\begin{equation*}
\begin{aligned}
{\rm d}(\widehat U^{(t)},\bar{U})\le8\sqrt{2}\cdot \frac{\sqrt{r\wedge 2m}\cdot \big({\rm d}(\widehat W^{(t-1)},\bar W)\|\bDelta\|_{2,\sqrt{2}\delta_1}+{\rm d}\big(\widehat U^{(t-1)},\bar{U}\big)\|\bDelta\|_{3,\sqrt{2}\delta_2}\big)}{\sigma_{r}\big(\calM_1(\bar{\bC})\big)}\\
+\frac{C_1\sqrt{np_{\max}\log n}+C_2\delta_1\delta_2\log n}{\sigma_r(\calM_1(\bar{\bC}))}.
\end{aligned}
\end{equation*}
By Theorem~\ref{thm:tensor_CI}, if $Lnp_{\max}\geq \log n$,  there exist absolute constants $C_3,C_4>0$ such that 
$$
\|\bDelta\|_{2,\sqrt{2}\delta_1}\le C_3\sqrt{np_{\text{max}}}\log(n)\log(\delta_1^2n)+C_4\delta_1\sqrt{n^2p_{\text{max}}}\log^2(n)\log(\delta_1^2n)
$$
and
$$
\|\bDelta\|_{3,\sqrt{2}\delta_2}\leq C_3\sqrt{np_{\text{max}}}\log(n)\log(\delta_2^2L)+C_4\delta_2\sqrt{nLp_{\text{max}}}\log^2(n)\log(\delta_2^2L)
$$
with probability at least $1-n^{-2}$.

As a result, we get 
\begin{align*}
{\rm d}(&\widehat U^{(t)},\bar{U})\\
&\le {\rm Err}_{t-1}\cdot \frac{\sqrt{r\wedge 2m}\Big(C_3\sqrt{np_{\max}}\log(n)+C_4\sqrt{np_{\max}}\big((\delta_1\sqrt{n})\vee (\delta_2\sqrt{L})\big)\log^2n\Big)\log\big(\delta_1^2n\vee \delta_2^2 L\big)}{\sigma_r\big(\calM_1(\bar{\bC})\big)}\\
&\hspace{3cm}+\frac{C_1\sqrt{np_{\max}\log n}+C_2\delta_1\delta_2\log n}{\sigma_r\big(\calM_1(\bar{\bC})\big)}.
\end{align*}
Similarly, under the same event (only $\|\bDelta\|_{2,\delta}$, $\|\bDelta\|_{3,\delta}$ and $\|\calM_1(\bDelta)(\bar{U}\otimes \bar{W})\|$ involve probabilities), the bound holds for ${\rm d}(\widehat{W}^{(t)},\bar{W})$. To this end, we conclude with 
\begin{align*}
{\rm Err}_t
&\le {\rm Err}_{t-1}\cdot \frac{\sqrt{r\wedge 2m}\Big(C_3\sqrt{np_{\max}}\log(n)+C_4\sqrt{np_{\max}}\big((\delta_1\sqrt{n})\vee (\delta_2\sqrt{L})\big)\log^2n\Big)\log\big(\delta_1^2n\vee \delta_2^2 L\big)}{\sigma_{\min}(\bar{\bC})}\\
&\hspace{3cm}+\frac{C_1\sqrt{np_{\max}\log n}+C_2\delta_1\delta_2\log n}{\sigma_{\min}(\bar{\bC})}
\end{align*}
where we denote $\sigma_{\min}(\bar{\bC})=\min\big\{\sigma_r\big(\calM_1(\bar{\bC})\big),\sigma_m\big(\calM_3(\bar{\bC})\big)\big\}$. 
To guarantee a contraction property, we assume that 
\begin{equation}\label{eqf:PI_cond}
\sigma_{\min}(\bar{\bC})\geq 2\sqrt{r\wedge 2m}\Big(C_3\sqrt{np_{\max}}\log(n)+C_4\sqrt{np_{\max}}\big((\delta_1\sqrt{n})\vee (\delta_2\sqrt{L})\big)\log^2n\Big)\log\big((\delta_1\vee \delta_2)^2n\big)
\end{equation}
for some large enough absolute constants $C_3,C_4>0$. If condition (\ref{eqf:PI_cond}), then for all $t=1,2,\cdots, t_{\max}$, 
$$
{\rm Err}_t\leq \frac{1}{2}\cdot {\rm Err}_{t-1}+\frac{C_1\sqrt{np_{\max}\log n}+C_2\delta_1\delta_2\log n}{\sigma_{\min}(\bar{\bC})}
$$
which holds with probability at least $1-2n^{-2}$ and $C_1,C_2>0$ are also some absolute constants. The above contraction inequality implies that after 
$$
t_{\max}=O\Big(\log\big(\sigma_{\min}(\bar{\bC})/ (\sqrt{np_{\max}}+\delta_1\delta_2)\big)\vee 1\Big)
$$ 
iterations, with probability at least $1-2n^{-2}$,
$$
{\rm Err}_{t_{\max}}\leq \frac{C_1\sqrt{np_{\max}\log n}+C_2\delta_1\delta_2\log n}{\sigma_{\min}(\bar{\bC})}
$$
for some absolute constants $C_1,C_2>0$ which concludes the proof. 

\

\subsection{Proof of Thoerem~\ref{thm:tensor_CI}}

Without loss of generality, we only prove $\|\bA-\EE\bA\|_{1,\delta}$. The spirit of proving $\|\bA-\EE\bA\|_{3,\delta}$ is similar. The main ideal of proving sharp concentration inequality for tensor incoherent norms is a combination of the techqniues in \cite{yuan2016tensor, yuan2017incoherent} and \cite{ke2019community} (see also \cite{xia2017statistically}).

Let $\bE$ be an $n\times n\times L$ random tensor with each entry being  a Rademacher random variable such that for $\forall i_1,i_2\in [n],i_3\in [L]$
$$
\mathbb{P}(E(i_1,i_2,i_3)=+1)=\mathbb{P}({E}(i_1,i_2,i_3)=-1)=\frac{1}{2}
$$
and $\bE$ is partially symmetric such that
$$
E(i_1,i_2,i_3)=E(i_2,i_1,i_3).
$$
By a standard symmetrization argument (e.g., \cite{gine1984some}), we get for any $t>0$,
$$
\mathbb{P}\left\{\Vert \mathbf{A}-\mathbb{E}\mathbf{A} \Vert_{1,\delta}\ge 3t\right\}\le\max_{u_{1} \otimes u_{2} \otimes u_3 \in \mathcal{U}_1(\delta)}\mathbb{P}\{\langle\mathbf{A}-\mathbb{E}\mathbf{A},u_{1} \otimes u_{2} \otimes u_3\rangle\ge t\}+4\mathbb{P}\{\Vert\bE\odot \mathbf{A}\Vert_{1,\delta} \ge t\}
$$
where $\odot$ denotes a Hadamard product of matrices. 

We begin with the probabilistic upper bound of $|\langle\mathbf{A}-\mathbb{E}\mathbf{A},u_{1} \otimes u_{2} \otimes u_3\rangle|$. Fix  any $u_{1} \otimes u_{2} \otimes u_3 \in \mathcal{U}_1(\delta)$, we write $\bDelta=\bA-\EE\bA$ and 
$$
\langle\mathbf{A}-\mathbb{E}\mathbf{A},u_{1} \otimes u_{2} \otimes u_3\rangle=\sum_{i_3=1}^L\sum_{(i_1,i_2)\in\mathfrak{I}_n^2}\Delta(i_1,i_2,i_3)\sum_{(i_1^{\prime},i_2^{\prime})\in\{(i_1,i_2),(i_2,i_1)\}}u_{1i_1^\prime}u_{2i_2^\prime}u_{3i_3}
$$
where we denote $\mathfrak{I}_n^2=\{(i_1,i_2):1\leq i_1\leq i_2\leq n\}$. The following bounds are clear. 
$$
\Big||\Delta(i_1,i_2,i_3)\sum_{(i_1^{\prime},i_2^{\prime})\in\{(i_1,i_2),(i_2,i_1)\}}u_{1i_1^\prime}u_{2i_2^\prime}u_{3i_3}\Big|\le 2\delta
$$
and
\begin{equation*}
\begin{aligned}
{\rm Var}& \Big(\sum_{i_3=1}^L\sum_{i_1,i_2\in\mathfrak{I}_n^2}\Delta(i_1,i_2,i_3)\sum_{(i_1^{\prime},i_2^{\prime})\in\{(i_1,i_2),(i_2,i_1)\}}u_{1i_1^\prime}u_{2i_2^\prime}u_{3i_3}\Big)\\
&=\sum_{i_3=1}^L\sum_{i_1,i_2\in\mathfrak{I}_n^2}\mathbb{E}\left( \Delta(i_1,i_2,i_3)\right)^2\Big(\sum_{(i_1^{\prime},i_2^{\prime})\in\{(i_1,i_2),(i_2,i_1)\}}u_{1i_1^\prime}u_{2i_2^\prime}u_{3i_3}\Big)^2\\
&\le \sum_{i_3=1}^L\sum_{i_1,i_2\in\mathfrak{I}_n^2}p_{\text{max}}\cdot2\sum_{(i_1^{\prime},i_2^{\prime})\in\{(i_1,i_2),(i_2,i_1)\}}u_{1i_1^\prime}^2u_{2i_2^\prime}^2u_{3i_3}^2\le2p_{\text{max}}.
\end{aligned}
\end{equation*}
By Bernstein inequality, we obtain
$$
\max_{u_{1} \otimes u_{2} \otimes u_3 \in \mathcal{U}_1(\delta)}\mathbb{P}\{\langle\mathbf{A}-\mathbb{E}\mathbf{A},u_{1} \otimes u_{2} \otimes u_3\rangle\ge t\}\le \exp\left(\frac{-t^2}{8p_{\text{max}}} \right)+\exp\left(\frac{-3t}{8\delta} \right).
$$
Now, we are to bound $\mathbb{P}\{\Vert\bE\odot \mathbf{A}\Vert_{1,\delta} \ge t\}$. For notational simplicity, we write $n_1:=n, n_2:=n, n_3:=L$. Define a discretized version of $\mathcal{U}_1(\delta)$ as
$$
\mathcal{U}_1^*(\delta):=\{u_{1} \otimes u_{2} \otimes u_3 \in \mathcal{U}_1(\delta):\Vert u_j\Vert_{\ell_2}\le c_j,u_j\in\{\pm2^{k/2}c_j/\sqrt{2n_j},k=0,1,...,m_j \}^{n_j}, \forall j=1,2,3\}
$$
with $m_j=\lceil\log_2 (\delta^2n_j)-1\rceil$ for $j=1$ and $m_j=\lceil\log_2 (n_j)-1\rceil$ for $j=2,3$. By choosing $1/2\le c_1\le1$ and $1/\sqrt{2}\le c_j\le1$ for $j=2,3$, we can guarantee, for $u_1\otimes u_2\otimes u_3\in\calU_1^*(\delta)$, that
$$
u_1\in\{\pm2^{-k/2}, k=2+\lceil\log_2 (\delta^{-2})\rceil,...,2+\lceil\log_2 (\delta^{-2})\rceil+\lceil\log_2 (\delta^2n_1)-1\rceil \}^{n_1}
$$
and  such that for $j=2,3$,
$$
u_j\in\{\pm2^{-k/2},k=2,...,\lceil\log_2 (n_j)-1\rceil+2 \}^{n_j}.
$$
It is well known that (see, e.g., \cite{yuan2017incoherent})
\begin{equation*}
\begin{aligned}
\Vert\bE\odot \mathbf{A}\Vert_{1,\delta}&=\max_{u_{1} \otimes u_{2} \otimes u_3 \in \mathcal{U}_1(\delta)}\langle\bE\odot \mathbf{A},u_{1} \otimes u_{2} \otimes u_3\rangle\le\frac{2^3}{\prod_{j=1}^3 c_j}\max_{u_{1} \otimes u_{2} \otimes u_3 \in \mathcal{U}^*_1(\delta)}\langle\bE\odot \mathbf{A},u_{1} \otimes u_{2} \otimes u_3\rangle\\
&\le 2^5\max_{u_{1} \otimes u_{2} \otimes u_3 \in \mathcal{U}^*_1(\delta)}\langle\bE\odot \mathbf{A},u_{1} \otimes u_{2} \otimes u_3\rangle
\end{aligned}
\end{equation*}
implying that it suffices to bound $\max_{u_{1} \otimes u_{2} \otimes u_3 \in \mathcal{U}^*_1(\delta)}\langle\bE\odot \mathbf{A},u_{1} \otimes u_{2} \otimes u_3\rangle$. 

A simple fact of cardinality bound is $\log\left|\mathcal{U}^*_1(\delta) \right|\le 4(2n\vee L)\leq 8n$ where we assume $L\leq n$. For a sharper union bound, define, for $\forall \bU\in\mathcal{U}^*_1(\delta)$, that
$$
A_k(\bU)=\{(i_2,i_3):\left|u_{2i_2}u_{3i_3}\right|=2^{-k/2}  \}
$$
for $\forall k=4,\cdots, \lceil\log_2 (n_2)-1\rceil+\lceil\log_2 (n_3)-1\rceil+4 $ and
$$
B_{k,s}(\bU)=\{i_1:(i_2,i_3)\in A_k(U), \Omega(i_1,i_2,i_3)\ne0, \left|u_{1i_1}\right|=2^{-s/2}  \}
$$
for $\forall s=2+\lceil\log_2 (\delta^{-2})\rceil,...,2+\lceil\log_2 (\delta^{-2})\rceil+\lceil\log_2 (\delta^2n_1)-1\rceil$ and where $\bOmega$ denotes the support of $\bA$, i.e., $\bOmega(i_1,i_2,i_3)={\bf 1}(\bA(i_1,i_2,i_3)>0)$. 

Moreover, for a positive integer $k^\star$ (whose value is determined later), define
$$
S_{2,3,k^\star}(\bU)=\{(i_2,i_3): |u_{2i_2}u_{3i_3}|\le2^{-k^\star/2-1/2}\}.
$$
For notational simplicity, let us omit the dependence of $A_k(\bU)$, $B_{k,s}(\bU)$, $S_{2,3,k^\star}(\bU)$ on $\bU$ when no confusion occurs. 

For $\forall \bU\in\mathcal{U}^*_1(\delta)$ and $\forall k^\star\in \mathbb{N}_+$, the following decomposition holds
\begin{equation*}
\begin{aligned}
\langle\bE\odot \mathbf{A},u_{1} \otimes u_{2} \otimes u_3\rangle&=\langle\bE\odot \mathbf{A}, u_1\otimes \mathcal{P}_{S_{2,3,k^\star}}(U_{2,3})\rangle\\
&+\sum_{4\le k\le k^\star}\sum_{s=2+\lceil\log_2 (\delta^{-2})\rceil}^{2+\lceil\log_2 (\delta^{-2})\rceil+\lceil\log_2 (\delta^2n_1)-1\rceil}\langle\bE\odot \mathbf{A}, \mathcal{P}_{B_{k,s}}(u_1) \otimes \mathcal{P}_{A_k}(U_{2,3}) \rangle\\
\end{aligned}
\end{equation*}
where we write $U_{2,3}:=u_{2} \otimes u_{3}$. The notation $\mathcal{P}_{C}(u)$ is the projection operator which projects $u$ onto the support of $C$. 

Let us write $\mathbf{Y}:=\bE\odot \mathbf{A}$ for brevity in the following discussion.

\paragraph*{Bounding for $\langle\mathbf{Y}, \mathcal{P}_{B_{k,s}}(u_1) \otimes \mathcal{P}_{A_k}(U_{2,3})\rangle$}
To take the advantage of the sparsity of $\bY$, we define the aspect ratio (see, e.g., \cite{yuan2016tensor} and \cite{xia2017statistically}) for a tensor $\mathbf{X}\in\mathbb{R}^{n_1\times...\times n_m}$ such that for $1\le j_1<...<j_l\le m$ and $1\le l\le m$,
$$
\nu_{j_1,...,j_l}(\mathbf{X})=\max_{i_{j_1}\in[n_{j_1}],...,i_{j_l}\in[n_{j_l}]}\left|\{(i_1,i_2,...,i_m): X(i_1,i_2,...,i_m)\ne 0, i_k\in[n_k], k\in[m]\backslash\{j_1,...,j_l\} \} \right|.
$$
By Chernoff bound (see, e.g., \cite{yuan2016tensor}), we get
$$
\mathbb{P}\{\nu_{2,3}(\mathbf{Y})\ge13(n_1p_{\text{max}}+\log n) \}\le n^{-2}
$$
and
$$
\mathbb{P}\{\nu_{3}(\mathbf{Y})\ge13(n_1n_2p_{\text{max}}+\log n) \}\le n^{-2}
$$
Write $\nu_{2,3}^\star = 26\max\{n_1p_{\text{max}},\log n\}$ and $\nu_{3}^\star = 26\max\{n_1n_2p_{\text{max}},\log n\}$. Denote the event $\calE_{2,3}=\{\nu_{2,3}(\mathbf{Y})\le \nu_{2,3}^\star\}$ and $\calE_{3}=\{\nu_{3}(\mathbf{Y})\le \nu_{3}^\star\}$. We discuss two cases.\\[2ex]
(1) If $n_1p_{\text{max}}\ge\log n$ (recall that $n_1=n$), then $\nu_{2,3}^\star = 26n_1p_{\text{max}}$. Under the event $\calE_{2,3}$, define
$$
\mathfrak{B}_1(k,l):=\{\bV=\mathcal{P}_{B}(u_1)\otimes\mathcal{P}_{A_k}(U_{2,3}):|A_k|\le2^{k-l},|B|\le\nu_{2,3}^\star2^{k-l}, u_1\otimes U_{2,3}\in\mathcal{U}^*_1(\delta)\}
$$
for $\forall k\in \mathbb{N}$ and $\forall 0\le l\le k$. 

Then, under $\calE_{2,3}$, for any $\bU\in\mathcal{U}^*_1(\delta)$, we have $\mathcal{P}_{B_{k,s}}(u_1)\otimes\mathcal{P}_{A_k}(U_{2,3})\in \mathfrak{B}_1(k,l)$ for some $l$. Therefore, we write
\begin{equation*}
\begin{aligned}
&\sum_{ 4\le k\le k^\star}\sum_{s=2+\lceil\log_2 (\delta^{-2})\rceil}^{2+\lceil\log_2 (\delta^{-2})\rceil+\lceil\log_2 (\delta^2n_1)-1\rceil}\langle\mathbf{Y}, \mathcal{P}_{B_{k,s}}(u_1)\otimes\mathcal{P}_{A_k}(U_{2,3})\rangle\\
&\le k^\star\lceil\log_2 (\delta^2n_1)\rceil\max_{\substack{4\le k\le k^\star \\2+\lceil\log_2 (\delta^{-2})\rceil\le s\le2+\lceil\log_2 (\delta^{-2})\rceil+\lceil\log_2 (\delta^2n)-1\rceil}}\langle\mathbf{Y}, \mathcal{P}_{B_{k,s}}(u_1)\otimes\mathcal{P}_{A_k}(U_{2,3})\rangle\\
&\le k^\star\lceil\log_2 (\delta^2n_1)\rceil\max_{\substack{4\le k\le k^\star }}\max_{0\le l\le k}\max_{\bV\in\mathfrak{B}_1(k,l)}\langle\mathbf{Y}, \bV\rangle
\end{aligned}
\end{equation*}
As shown by Yuan and Zhang (2016) \cite{yuan2017incoherent},
$$
\log|\mathfrak{B}_1(k,l)|\le3\cdot2^{(k-l)/2}\sqrt{4\nu_{2,3}^\star(\bigvee_jn_j)\log(\bigvee_jn_j)}=3\cdot2^{(k-l)/2}\sqrt{4\nu_{2,3}^\star n\log n}.
$$
Now, we bound $\sup_{\bV\in\mathfrak{B}_1(k,l)}\langle\mathbf{Y},\bV\rangle$. For $\forall \bV\in\mathfrak{B}_1(k,l)$, 
$$
\langle\mathbf{Y},\bV\rangle=\sum_{i_3=1}^L\sum_{(i_1,i_2)\in\mathfrak{I}_n^2}Y(i_1,i_2,i_3)\sum_{(i_1^{\prime},i_2^{\prime})\in\{(i_1,i_2),(i_2,i_1)\}}V(i_1^{\prime},i_2^{\prime},i_3).
$$
Then,
$$
\Big|Y(i_1,i_2,i_3)\sum_{(i_1^{\prime},i_2^{\prime})\in\{(i_1,i_2),(i_2,i_1)\}}V(i_1^{\prime},i_2^{\prime},i_3)\Big|\le2\Vert V\Vert_{\text{max}}\le2^{-k/2+1}\delta
$$
and
\begin{equation*}
\begin{aligned}
&{\rm Var}\Big(\sum_{i_3=1}^L\sum_{(i_1,i_2)\in\mathfrak{I}_n^2}Y(i_1,i_2,i_3)\sum_{(i_1^{\prime},i_2^{\prime})\in\{(i_1,i_2),(i_2,i_1)\}}V(i_1^{\prime},i_2^{\prime},i_3)\Big)\\
&=\sum_{i_3=1}^L\sum_{(i_1,i_2)\in\mathfrak{I}_n^2}\mathbb{E}\left(A(i_1,i_2,i_3)^2{E}(i_1,i_2,i_3)^2\right)\Big(\sum_{(i_1^{\prime},i_2^{\prime})\in\{(i_1,i_2),(i_2,i_1)\}}V(i_1^{\prime},i_2^{\prime},i_3) \Big)^2\\
&\le 2\sum_{i_3=1}^L\sum_{(i_1,i_2)\in\mathfrak{I}_n^2}p_{\text{max}}\sum_{(i_1^{\prime},i_2^{\prime})\in\{(i_1,i_2),(i_2,i_1)\}}V(i_1^{\prime},i_2^{\prime},i_3)^2\le2p_{\text{max}}\Vert \bV\Vert_F^2\le2^{-l+1}p_{\text{max}}
\end{aligned}
\end{equation*}
where we use the fact $\Vert \bV\Vert_{\rm F}^2=\left\Vert\mathcal{P}_{B_{k,s}}(u_1) \right\Vert_{\rm F}^2\left\Vert \mathcal{P}_{A_k}(U_{2,3})\right\Vert_{\rm F}^2 \le2^{k-l}\cdot(2^{-k/2})^2\le2^{-l}$.

By Bernstein's inequality, for each fixed $\bV\in\mathfrak{B}_1(k,l)$, 
\begin{equation*}
\begin{aligned}
\mathbb{P}\left\{\langle\mathbf{Y},\bV\rangle\ge\frac{2^{-5}t}{2k^\star} \right\}&\le \exp\left(\frac{-t^2}{4\cdot 2^{12}{k^{\star}}^2\cdot2^{-l+1}p_{\text{max}}}\right)+\exp\left(\frac{-3t}{4\cdot2^6k^\star\cdot2^{-k/2+1}\delta}  \right)\\
&\le\exp\left(\frac{-t^2}{2^{15}\cdot2^{-l}{k^{\star}}^2p_{\text{max}}}\right)+\exp\left(\frac{-3t}{2^9\cdot2^{-k/2}k^\star\delta}  \right).
\end{aligned}
\end{equation*}
Therefore,
\begin{equation*}
\begin{aligned}
&\mathbb{P}\left\{\max_{\bV\in\mathfrak{B}_1(k,l)}\langle\mathbf{Y},\bV\rangle\ge\frac{2^{-5}t}{2k^\star\log(\delta^2n)} \right\}\\
&\le\exp\left(8n-\frac{t^2}{2^{15}\cdot2^{-l}k^{\star2}p_{\text{max}}\log^2(\delta^2n)}\right)+\exp\left(3\cdot2^{(k-l)/2}\sqrt{4\nu_{2,3}^\star n\log n}-\frac{3t}{2^9\cdot2^{-k/2}\delta k^{\star}\log(\delta^2n)}  \right).
\end{aligned}
\end{equation*}
It turns out that for
$$
t\ge\max\left\{ 2^9k^\star\sqrt{p_{\text{max}}n}\log(\delta^2n), 2^{11}k^\star\delta\sqrt{\nu_{2,3}^\star n\log n}\log(\delta^2n)\right\},
$$
we get
\begin{equation*}
\begin{aligned}
&\mathbb{P}\left\{\max_{\bV\in\mathfrak{B}_1(k,l)}\langle\mathbf{Y},\bV\rangle\ge\frac{2^{-5}t}{2k^\star\log(\delta^2n)} \right\}\le\exp\left(-\frac{t^2}{2^{16}{k^{\star}}^2p_{\text{max}}\log^2(\delta^2n)}\right)+\exp\left(-\frac{3t}{2^9k^\star\delta\log(\delta^2n)}  \right).
\end{aligned}
\end{equation*}
Recall that $\nu_{2,3}^\star = 26n_1p_{\text{max}}$. In this case,  the condition becomes 
\begin{equation*}
t\ge\max\left\{ 2^9k^\star\sqrt{p_{\text{max}}n}\log(\delta^2n),\ \  2^{11}\sqrt{26}k^\star\delta\sqrt{ n^2p_{\text{max}}\log n}\log(\delta^2n)\right\}.
\end{equation*}

(2) If $n_1p_{\text{max}}<\log n$, then $\nu_{2,3}^\star = 26\log n$. In this case, the network can be extremely sparse. 
Then, we need a sharper analysis for $\max_{\bU\in\mathcal{U}^*_1(\delta)}\langle\bY, \calP_{B_{k,s}}(u_1)\otimes \calP_{A_k}(U_{2,3}) \rangle$ for any positive integers $k\leq k^{\star}$ and $s\leq 2+\log(\delta^{-2})+\log(\delta^2 n)$. Now, define
\begin{equation*}
\begin{aligned}
\mathfrak{B}_1(\bOmega,k,s,l):=\big\{V=\mathcal{P}_{B_{k,s}}(u_1)\otimes \mathcal{P}_{A_k}(U_{2,3}):|A_k|\le2^{k-l},|B_{k,s}|\le\nu_{2,3}(\bOmega)2^{k-l}\wedge n\\
, u_1\otimes U_{2,3}\in\mathcal{U}^*_1(\delta)\big\}
\end{aligned}
\end{equation*}
for which we need a sharper analysis of the cardinality of $\mathfrak{B}_1(\bOmega,k,s,l)$ with respect to the sparsity of $\bOmega$.

We start with 
\begin{align*}
\calU_{2,3}(k,l):=\big\{\calP_{A_k}(U_{2,3}): U_{2,3}=u_2&\otimes u_3, |A_k|\leq 2^{k-l}\\
,&u_j\in\{\pm 2^{-k/2}, k=2,\cdots,\lceil \log(n_j)-1\rceil+2\}^{n_j}, j=2,3\big\}.
\end{align*}
To investigate the  respective sparsity on the fibers of $\calP_{A_k}(U_{2,3})$, recall the inner product 
$$
\langle \bY, \calP_{B_{k,s}}(u_1)\otimes \calP_{A_k}(U_{2,3}) \rangle=\langle \bY\times_1 \calP_{B_{k,s}}^{\top}(u_1), \calP_{A_k}(U_{2,3})\rangle
$$ 
where $\bY\times_1 \calP_{B_{k,s}}^{\top}(u_1)$ is a $n\times L$ matrix. It is easy to check that the sparsity of fibers (row and column vector) of $\bY\times_1 \calP_{B_{k,s}}^{\top}(u_1)$ is bounded by $\nu_3(\bOmega)$. As a result, it suffices to restrict to $\calP_{A_k}(U_{2,3})$ whose fiber sparsity is bounded by $\nu_3(\bA)$. See, for instance, \cite[Lemma~11]{yuan2016tensor}.

To this end, define
$$
\nu(\mathcal{P}_{A_k}(U_{2,3})):=\max_{i=1,2}\nu_i(\mathcal{P}_{A_k}(U_{2,3}))\le \nu_3(\bOmega)
$$
and
$$
\mathfrak{B}_1^{(2,3)}(\bOmega,k,l):=\{V=\mathcal{P}_{A_k}(U_{2,3}): V\in\calU_{2,3}(k,l), \nu(V)\le \nu_3(\bOmega)\}.
$$
To eliminate the dependence of $\mathfrak{B}_1^{(2,3)}(\bOmega,k,l)$ on $\bOmega$, we condition on event $\calE_3$ so that $\nu_3(\bOmega)\leq \nu_3^{\star}=26n^2p_{\rm max}$ (where we assumed $nLp_{\max}\geq \log n$). Then, define 
$$
\mathfrak{B}_1^{(2,3)\star}(k,l):=\bigcup_{\bOmega: \nu_{3}(\bOmega)\leq \nu_3^{\star}}\mathfrak{B}_1^{2,3}(\bOmega,k,l). 
$$
As shown in \cite[Lemma~11]{yuan2016tensor}, conditioned on $\bOmega$ (or $\bA$), for all $\calP_{B_{k,s}}^{\top}(u_1)$, the following holds
\begin{equation}
\label{eq:maxB1}
\max_{V\in \calU_{2,3}(k,l)}\langle  \bY\times_1\calP_{B_{k,s}}^{\top}(u_1), V\rangle\leq \max_{V\in \mathfrak{B}_1^{(2,3)\star}(k,l)}\langle \bY\times_1\calP_{B_{k,s}}^{\top}(u_1), V \rangle. 
\end{equation}

Now we consider the cardinality of $\mathfrak{B}_1^{(2,3)\star}(k,l)$. Note that the aspect ratio (on each fiber) of $\mathfrak{B}_1^{(2,3)\star}(k,l)$ is bounded by $\nu_3^\star$ and there are at most $2^{k-l}$ non-zero entries with each equal to $2^{-k/2}$. 
 By Lemma 12 in \cite{yuan2016tensor} (or \cite[Lemma~1]{xia2017statistically}), 
\begin{equation*}
\begin{aligned}
\log\left|\mathfrak{B}_1^{(2,3)\star}(k,l)\right|\leq (21/4)(k+2)\sqrt{\nu_3^{\star}2^{k-l}}\cdot L(\sqrt{\nu_3^{\star}2^{k-l}}, (k+2)n)
\end{aligned}
\end{equation*}
where the function $L(x,y)=\max\{1,\log(ey/x)\}$. Therefore, 
$$
\log\left|\mathfrak{B}_1^{(2,3)\star}(k,l)\right|\leq 10(k+2)\sqrt{\nu_3^{\star}2^{k-l}}\cdot \log n.
$$
Next, we need to study the respective cardinality for the set of $\calP_{B_{k,s}}(u_1)$ in the right hand side of (\ref{eq:maxB1}). A sharper analysis of sparsity of $\calP_{B_{k,s}}(u_1)$ is needed. For each fixed $V\in \mathfrak{B}_1^{(2,3)\star}(k,l)$, $|{\rm supp}(V)|\leq 2^{k-l}$ and 
$$
\langle \bY\times_1\calP_{B_{k,s}}^{\top}(u_1), V \rangle=\sum_{\omega\in{\rm supp}(V)}\big<Y(:, \omega)V(\omega), \calP_{B_{k,s}}(u_1)\big>
$$
implying that sparse $\calP_{B_{k,s}}(u_1)$ suffices to realize the maximum above when choosing $u_1$. For each fixed $V\in  \mathfrak{B}_1^{(2,3)\star}(k,l)$ with $|{\rm supp}(V)|\leq 2^{k-l}$, by Bernstein inequality of the sum of Bernoulli random variables (for each $\omega\in{\rm supp}(V)$, $\Omega(:,\omega)$ has $n$ independent Bernoulli random variable), 
$$
\PP\Big(\Big|\big\{\Omega(:,\omega): \omega\in {\rm supp}(V)\big\}\Big|\geq 4\cdot 2^{k-l}np_{\max}+4t\Big)\leq e^{-t}
$$
for any $t\geq 1$. Now, taking the union bound for all $V\in\mathfrak{B}_1^{(2,3)\star}(k,l)$, we obtain
\begin{align*}
\PP\Big(\max_{V\in\mathfrak{B}_1^{(2,3)\star}(k,l)}\Big|\big\{\Omega(:,\omega): \omega\in {\rm supp}(V)\big\}\Big|\geq 4\cdot 2^{k-l}np_{\max}+4t\Big)\leq {\rm Card}\big(\mathfrak{B}_1^{(2,3)\star}(k,l)\big)\cdot e^{-t}\\
\leq e^{-t+10(k+2)\sqrt{\nu_3^{\star}2^{k-l}}\log n}.
\end{align*}
As a result, we conclude that with probability at least $1-n^{-3}$, 
\begin{equation*}
\max_{V\in\mathfrak{B}_1^{(2,3)\star}(k,l)}\Big|\big\{\Omega(:,\omega): \omega\in {\rm supp}(V)\big\}\Big|\leq 4\cdot 2^{k-l}np_{\max}+10(k+2)\sqrt{\nu_3^{\star}2^{k-l}}\log n+3\log n.
\end{equation*}
There are three terms on the above right hand side. If $4\cdot 2^{k-l}np_{\max}$ dominates, then it suffices to consider $\calP_{B_{k,s}}(u_1)$ with $|B_{k,s}|\leq 8\cdot 2^{k-l}np_{\max}$. This is exactly the case (1) where in the definition of $\mathfrak{B}_{1}(k,l)$ the cardinality is bounded as $|B|=O(2^{k-l}np_{\max})$. Therefore, there is no need to consider the case when $4\cdot 2^{k-l}np_{\max}$ dominates. To this end, we assume that $10(k+2)\sqrt{\nu_3^{\star}2^{k-l}}\log n+3\log n$ dominates and then with probability at least $1-n^{-3}$,
\begin{equation}\label{eq:B1sparsity}
\max_{V\in\mathfrak{B}_1^{(2,3)\star}(k,l)}\Big|\big\{\Omega(:,\omega): \omega\in {\rm supp}(V)\big\}\Big|\leq 20(k+2)\sqrt{\nu_3^{\star}2^{k-l}}\log n+6\log n.
\end{equation}
Denote the event $\calE_{2,3}'$ and define 
\begin{align*}
\mathfrak{B}_1^{(1)\star}(s)=&\Big\{D_s(u_1): \|D_s(u_1)\|_{\ell_0}\leq 20(k+2)\sqrt{\nu_3^{\star}2^{k-l}}\log n\\
,&u_1\in\{\pm 2^{-k/2}, k=2+\lceil\log(\delta^{-2})\rceil,\cdots,2+\lceil\log(\delta^{-2})\rceil+\lceil\log(\delta^2 n_1)-1 \rceil\}^{n_1}\Big\}
\end{align*}
where the operator $D_s(\cdot)$ zeros the entries whose absolute values are not $2^{-s/2}$. 
As a result (\cite[Lemma~11]{yuan2016tensor}), we conclude that, conditioned on $\calE_{2,3}\cap \calE_3\cap \calE'_{2,3}$,
$$
\max_{\bU\in\calU_{1}^{\star}(\delta)}\big<\bY, \calP_{B_{k,s}}(u_1)\otimes \calP_{A_k}(U_{2,3})\big>\leq \max_{v_1\in \mathfrak{B}_1^{(1)\star}(s), V_{2,3}\in \mathfrak{B}_1^{(2,3)\star}(k,l)} \langle \bY, v_1\otimes V_{2,3}\rangle.
$$
Meanwhile, the cardinality of the product sets
\begin{equation*}
\begin{aligned}
\log\left|\mathfrak{B}_1^{(1)\star}(s)\times \mathfrak{B}_1^{(2,3)\star}(k,l)\right|&\le\log\left|\mathfrak{B}_1^{(1)\star}(s)\right|+\log\left| \mathfrak{B}_1^{(2,3)\star}(k,l)\right|\\
&\le C_1\sqrt{n^2p_{\max}2^{k-l}}\log^2n
\end{aligned}
\end{equation*}
where $C_1>0$ is an absolute constants and we used the fact that the set $\mathfrak{B}_1^{(1)\star}(s)$ contains sparse vectors whose sparsity is bounded by (\ref{eq:B1sparsity}) . 

For each $\bV=v_1\otimes V_{2,3}$ with $v_1\in \mathfrak{B}_1^{(1)\star}(s), V_{2,3}\in \mathfrak{B}_1^{(2,3)\star}(k,l)$, we apply Bernstein inequality to bound $\langle \bY, v_1\otimes V_{2,3}\rangle$.
Clearly (similar to the first case),
$$
\left|Y(i_1,i_2,i_3)\sum_{(i_1^{\prime},i_2^{\prime})\in\{(i_1,i_2),(i_2,i_1)\}}V(i_1^{\prime},i_2^{\prime},i_3)\right|\le2\Vert \bV\Vert_{\text{max}}\le2^{-(k+s)/2+1}
$$
and
\begin{equation*}
\begin{aligned}
&{\rm Var}\left(\sum_{i_3=1}^L\sum_{(i_1,i_2)\in\mathfrak{I}_n^2}Y(i_1,i_2,i_3)\sum_{(i_1^{\prime},i_2^{\prime})\in\{(i_1,i_2),(i_2,i_1)\}}V(i_1^{\prime},i_2^{\prime},i_3)\right)\le2^{-l+1}p_{\text{max}}
\end{aligned}
\end{equation*}
By Bernstein's inequality and the union bound (also $n^2p_{\max}\geq \log n$), when
\begin{equation*}
\begin{aligned}
t\ge\max\left\{ 2^9\sqrt{C_1np_{\text{max}}}, C_2\delta\sqrt{n^2p_{\max}}\log^2n\right\}
\end{aligned}
\end{equation*}
where $C_1,C_2>0$ are absolute constants, we have 
\begin{equation*}
\begin{aligned}
\mathbb{P}\left\{\max_{U\in\calU_1^{\star}(\delta)}\langle\mathbf{Y},\calP_{B_{k,s}}(u_1)\otimes \calP_{A_k}(U_{2,3})\rangle\ge2^{-6}t \right\}\le\exp\left(-\frac{t^2}{2^{16}{p_{\text{max}}}}\right)+\exp\left(-\frac{3t}{2^7\delta}  \right)\\
\end{aligned}
\end{equation*}
where we use the fact that $2^{-s/2}\le\delta/2$ and $2^{-k/2}\le2^{-1}$. 

Combining (1) and (2) together and applying the union bound on $l,k,s$, we can conclude that for 
\begin{equation}\label{eq:t_1}
t\ge\max\left\{ C_1\sqrt{np_{\text{max}}}\log n, C_2\delta\sqrt{n^2p_{\text{max}}}\log^2n\right\}
\end{equation}
\begin{equation*}
\begin{aligned}
\mathbb{P}&\left\{\max_{U\in \mathcal{U}_1^\star(\delta)}\sum_{ 4\le k\le k^\star}\sum_{s=2+\lceil\log_2 (\delta^{-2})\rceil}^{2+\lceil\log_2 (\delta^{-2})\rceil+\lceil\log_2 (\delta^2n_1)-1\rceil}\langle\mathbf{Y}, \mathcal{P}_{B_{k,s}}(u_1)\otimes \mathcal{P}_{A_k}(U_{2,3})\rangle\ge 2^{-5}t \right\}\\
&\hspace{2cm}\le9(\log n)^2\lceil \log_2(\delta^2n)\rceil\left[\exp\left(-\frac{t^2}{C_3^2 p_{\text{max}}}\right)+\exp\left(-\frac{3t}{C_4\delta}  \right)\right]\\
\end{aligned}
\end{equation*}
where $C_1,C_2,C_3,C_4>0$ are some absolute constants. 

\paragraph*{Bounding for $\langle\mathbf{Y}, u_1\otimes \mathcal{P}_{S_{2,3,k^\star}}(U_{2,3})\rangle$} For $\forall \bV:= u_1\otimes \mathcal{P}_{S_{2,3,k^\star}}(U_{2,3})$, write
$$
\langle\mathbf{Y},\bV\rangle=\sum_{i_3=1}^L\sum_{(i_1,i_2)\in\mathfrak{I}_n^2}Y(i_1,i_2,i_3)\sum_{(i_1^{\prime},i_2^{\prime})\in\{(i_1,i_2),(i_2,i_1)\}}V(i_1^{\prime},i_2^{\prime},i_3).
$$
The following bounds hold
$$\left|Y(i_1,i_2,i_3)\sum_{(i_1^{\prime},i_2^{\prime})\in\{(i_1,i_2),(i_2,i_1)\}}V(i_1^{\prime},i_2^{\prime},i_3)\right|\le2\Vert \bV\Vert_{\text{max}}\le2^{-k^\star/2+1/2}\delta$$
\begin{equation*}
\begin{aligned}
&Var\left(\sum_{i_3=1}^L\sum_{(i_1,i_2)\in\mathfrak{I}_n^2}Y(i_1,i_2,i_3)\sum_{(i_1^{\prime},i_2^{\prime})\in\{(i_1,i_2),(i_2,i_1)\}}V(i_1^{\prime},i_2^{\prime},i_3)\right)\le2p_{\text{max}}\Vert \bV\Vert_{\rm F}^2\le2p_{\text{max}}
\end{aligned}
\end{equation*}
By Bernstein's inequality and applying the union bound, we get
\begin{align*}
\mathbb{P}\Big\{\max_{u_{1} \otimes u_{2} \otimes u_3 \in \mathcal{U}_1(\delta)}&\langle\mathbf{Y},u_1\otimes \mathcal{P}_{S_{2,3,k^\star}}(U_{2,3})\rangle\ge\frac{t}{2^6} \Big\}\\
&\le\exp\left(12n-\frac{t^2}{2^{15}p_{\text{max}}} \right)+\exp\left(12n-\frac{3t}{2^{8}2^{-k^\star/2+1/2}\delta} \right)
\end{align*}
We choose $k^\star=\lceil2\log_2n\rceil$ so that $2^{-k^\star/2}\le1/n$, then if 
\begin{equation}\label{eq:t_2}
t\ge\max\left\{ C_1\sqrt{p_{\text{max}}n}, C_2\delta\right\},
\end{equation}
we get
\begin{equation*}
\begin{aligned}
\mathbb{P}\left\{\max_{U \in \mathcal{U}_1^\star(\delta)}\langle\mathbf{Y},u_1\otimes \mathcal{P}_{S_{2,3,k^\star}}(U_{2,3})\rangle\ge\frac{t}{2^6} \right\}\le\exp\left(-\frac{t^2}{C_3p_{\text{max}}} \right)+\exp\left(-\frac{3nt}{C_4\delta} \right)
\end{aligned}
\end{equation*}
for some absolute constants $C_1,C_2,C_3,C_4>0$. 

\paragraph*{Combining bounds together} 
Combining \eqref{eq:t_1} and \eqref{eq:t_2} gives the condition on $t$:
\begin{equation}\label{eq:t_final}
t\ge\max\left\{ C_1\sqrt{np_{\text{max}}}(\log n)\log(\delta^2n), C_2\delta\sqrt{n^2p_{\text{max}}}(\log n)^{2}\log(\delta^2n)\right\}
\end{equation}
where $C_1, C_2>0$ are absolute constants, and (on event $\calE_{2,3}\cap \calE_3\cap \calE_{2,3}'$) we have 
\begin{equation*}
\begin{aligned}
\mathbb{P}\{\Vert\bE\odot \mathbf{A}\Vert_{1,\delta} \ge t\}&\le \mathbb{P}\left\{\max_{\bU\in \mathcal{U}^*_1(\delta)}\langle\bE\odot \mathbf{A},u_{1} \otimes u_{2} \otimes u_3\rangle\ge 2^{-5}t\right\}\\
&\le \mathbb{P}\left\{\max_{\bU\in \mathcal{U}^*_1(\delta)}\langle\bE\odot \mathbf{A}, u_1\otimes\mathcal{P}_{S_{2,3,k^\star}}(U_{2,3})\rangle\ge\frac{2^{-5}t}{2}\right\}\\
&+\mathbb{P}\left\{\max_{\bU\in \mathcal{U}^*_1(\delta)}\sum_{4\le k\le k^\star}\sum_{s=2+\lceil\log_2 (\delta^{-2})\rceil}^{2+\lceil\log_2 (\delta^{-2})\rceil+\lceil\log_2 (\delta^2n)-1\rceil}\langle\bE\odot \mathbf{A}, \mathcal{P}_{B_{k,s}}(u_1)\otimes\mathcal{P}_{A_k}(U_{2,3}) \rangle\ge\frac{2^{-5}t}{2}\right\}\\
&\le10(\log n)^2\lceil\log_2\delta^2n\rceil\left[\exp\left(-\frac{t^2}{C_3p_{\text{max}}}\right)+\exp\left(-\frac{3t}{C_4\delta}\right)\right]
\end{aligned}
\end{equation*}
Finally, we can decompose the probability with respect to event $\calE_{2,3}\cap \calE_3\cap \calE'_{2,3}$ and $(\calE_{2,3}\cap \calE_3\cap \calE_{2,3}')^{\rm c}$ and get  
\begin{equation*}
\begin{aligned}
\mathbb{P}\left\{\Vert \mathbf{A}-\mathbb{E}\mathbf{A} \Vert_{1,\delta}\ge 3t\right\}
\le\frac{2}{n^2}+ 10(\log n)^2\lceil\log_2\delta^2n\rceil\left[\exp\left(-\frac{t^2}{C_3p_{\text{max}}}\right)+\exp\left(-\frac{3t}{C_4\delta}\right)\right]
\end{aligned}
\end{equation*}
when \eqref{eq:t_final} holds.


\

\subsection{Proof of Thoerem~\ref{thm:global_consistency}}

Under the conditions of Corollary~\ref{cor:PI}, we get with probability at least $1-n^{-2}$ that,
$$
\| \widehat{U}-\bar{U}\widehat{O}\|\le R^{\dagger}
$$
with $\widehat{O}=\argmin_{O\in \mathbb{O}_r}\| \widehat{U}-\bar{U}O\|$ and $R^{\dagger} =  C_3\kappa_0^2(r/\sqrt{m})\cdot \sqrt{\log n}/\sqrt{Lnp_{\max}}$ for some absolute constant $C_3>0$.

Since the rank of $\widehat{U}-\bar{U}\widehat{O}$ is at most $2r$, we have
$$
\| \widehat{U}-\bar{U}\widehat{O}\|_{\rm F}\le\sqrt{2r}\|\widehat{U}-\bar{U}\widehat{O}\|\le\sqrt{2r}R^{\dagger}.
$$
Write $\widehat{U}=[\hat{u}_1,\cdots,\hat{u}_n]^{\top}$ and $\bar{U}=[\bar{u}_1,\cdots,\bar{u}_n]^{\top}$, where $\hat{u}_j^{\top}$ and $\bar{u}_j^{\top}$ denote the $j$-th row of $\widehat{U}$ and $\bar{U}$ respectively. Hence 
$$
\sum_{i=1}^n \| \hat{u}_i-\widehat{O}^{\top}\bar{u}_i \|^2=\|\widehat{U}-\bar{U}\widehat{O}\|_{\rm F}^2\le 2rR^{\dagger 2}.
$$
We claim that $\bar{U}$ has $\bar{K}$ distinct rows. To see this, define a $\bar{K}\times \mathring{K}$ matrix
$
Z^*=[z_1^{*},\cdots,z_{\bar{K}}^{*}]^{\top}
$
where $z_k^{*{\top}}=\bar{Z}(j_k,:)$ for some $j_k\in\mathcal{\bar{V}}_k$, $k\in[\bar{K}]$. Then by the definition of $Z^*$, we have
$$
\bar{u}_j=\bar{U}(j,:)=\bar{Z}(j,:)\bar{R}\bar{D}^{-1}=z_k^{*\top}\bar{R}\bar{D}^{-1}=:\bar{v}_k^*,\quad j\in\calV_k.
$$
It implies that the rows of $\bar{U}$ in the same global community take the same value. Therefore,  $\{\bar{u}_i \}_{i=1}^n$ can only take $\bar{K}$ distinct values from $\{\bar{v}_k^*\}_{k=1}^{\bar{K}}$.

To investigate the performance of k-means, we first consider putting the $K$ clustering centers at $\{\widehat{O}^{\top}\bar{v}_k^*\}_{k=1}^{\bar{K}}$. The objective value (within-cluster sum of squares) of K-means algorithm, denoted by ${\rm WCSS}^*$, is
\begin{equation}\label{eq:WCSS}
\begin{aligned}
{\rm WCSS}^*=\sum_{k=1}^{\bar{K}}\sum_{j\in\mathcal{\bar{V}}_k}\|\hat{u}_j-\widehat{O}^{\top}\bar{v}_k^*\|^2=\sum_{i=1}^n\|\hat{u}_i-\widehat{O}^{\top}\bar{u}_i\|^2\le 2rR^{\dagger2}.
\end{aligned}
\end{equation}
Define the following index set
$$
J=\{i\in[n]:\|\hat{u}_i-\widehat{O}^{\top}\bar{u}_i\|\le \alpha\sqrt{r}/3 \}
$$
where $\alpha=\kappa_0^{-1}/\sqrt{nm}$. Clearly, $\left|J^c\right|\left(\alpha\sqrt{r}/3\right)^2\le\sum_{i\in J^c}\Vert\hat{u}_i-\widehat{O}^{\top}\bar{u}_i\Vert^2\le 2rR^{\dagger2}$ and hence
\begin{equation}\label{eq:J_complement}
\begin{aligned}
\left|J^c\right|\le\frac{18R^{\dagger2}}{\alpha^2}.
\end{aligned}
\end{equation}
We now denote the objective value of K-means algorithm on $\widehat{U}$ by $\widehat{{\rm WCSS}}$. We make the following claim:
\begin{claim}\label{claim:1}
For each $k\in[\bar{K}]$, there exists a unique clustering center within a distance of $\alpha\sqrt{r}$ to $\widehat{O}^{\top}\bar{v}_k^*$.
\end{claim}
To prove Claim~\ref{claim:1}, we first prove the existence of such a clustering center. Otherwise, assume for some $k\in[\bar{K}]$, K-means algorithm assigns no center within a distance of $\alpha\sqrt{r}$ to $\widehat{O}^{\top}\bar{v}_k^*$. For any $j\in \mathcal{\bar{V}}_k\bigcap J$, denote the closest center to $\hat{u}_j$ by $\hat{c}_j$, 
$$
\|\hat{u}_j-\hat{c}_j\|\ge\| \hat{c}_j-\widehat{O}^{\top}\bar{u}_j\|-\|\hat{u}_j-\widehat{O}^{\top}\bar{u}_j \|\ge\alpha\sqrt{r}-\frac{\alpha\sqrt{r}}{3}=\frac{2\alpha\sqrt{r}}{3}.
$$
By the conditions of Theorem~\ref{thm:global_consistency}, $|\mathcal{\bar{V}}_k|\gtrsim n/\bar{K}$. By \eqref{eq:J_complement}, we get $|\mathcal{\bar{V}}_k\backslash J|\le 18R^{\dagger 2}/\alpha^2$. Then, 
$$
\left|\mathcal{\bar{V}}_k\cap J\right|=\left|\mathcal{\bar{V}}_k\right|-\left|\mathcal{\bar{V}}_k\backslash J\right|\gtrsim \frac{n}{\bar{K}}-\frac{18R^{\dagger 2}}{\alpha^2}=O\left(\frac{n}{\bar{K}}\right)
$$
where we use condition (\ref{eq:global_cluster_cond1}). Hence,
\begin{equation*}
\begin{aligned}
\widehat{\rm WCSS}&\ge\left|\mathcal{\bar{V}}_k\cap J\right|\cdot \|\hat{u}_j-\hat{c}_j\|^2\gtrsim \frac{n}{\bar{K}}\cdot \left(\frac{2\alpha\sqrt{r}}{3} \right)^2
\gtrsim \frac{nr\alpha^2}{\bar{K}}=\frac{Cnr}{\bar{K}}\cdot\frac{1}{\kappa_0^2mn}.
\end{aligned}
\end{equation*}
However, \eqref{eq:WCSS} implies that ${\rm WCSS}^*\le2rR^{\dagger 2}=o(r/(\kappa_0^2m\bar{K}))$ under condition (\ref{eq:global_cluster_cond1}), which is a contradiction. This proves the existence of such clustering centers. 

Next for any $k\ne l$, by Lemma~\ref{lem:barU_sep},
$$
\Vert \widehat O^{\top}\bar{v}_k^*-\widehat{O}^{\top}\bar{v}_l^*\|=\| \bar{v}_k^*-\bar{v}_l^*\|\ge\frac{1}{\sigma_1(\bar{D})}
$$
and recall that 
$$
\sigma_1(\bar{D})\le\kappa_0\sigma_r(\bar{D})\le\frac{\kappa_0\Vert\bar{Z}\Vert_{\rm F}}{\sqrt{r}}\le\kappa_0\sqrt{\frac{nm}{r}}
$$
implying that $\| \widehat{O}^{\top}\bar{v}_k^*-\widehat{O}^{\top}\bar{v}_l^*\|\ge 3\alpha\sqrt{r}$. Then one clustering center cannot be within a distance of $\alpha\sqrt{r}$ to two different clusters $\widehat{O}^{\top}\bar{v}_k^*$ and $\widehat{O}^{\top}\bar{v}_l^*$ at the same time. Therefore, for each $k\in[\bar{K}]$, the clustering center within a distance of $\alpha\sqrt{r}$ to $\widehat{O}^{\top}\bar{v}_k^*$ is unique. This completes the proof for Claim~\ref{claim:1}.\\
Now denote the clustering centers (in Claim~\ref{claim:1}) that minimize the K-means objective by $\{\hat{v}_k\}_{k=1}^{[\bar{K}]}$. Then for $\forall j\in\mathcal{\bar{V}}_k\bigcap J$,
\begin{equation*}
\begin{aligned}
\Vert \hat{u}_j-\hat{v}_k\Vert \le\Vert\hat{u}_j-\widehat{O}^{\top}\bar{v}_k^*\Vert+\Vert \widehat{O}^{\top}\bar{v}_k^*-\hat{v}_k \Vert\le\frac{\alpha\sqrt{r}}{3}+\alpha\sqrt{r}\le\frac{4\alpha\sqrt{r}}{3}.
\end{aligned}
\end{equation*}
For any $l\ne k$, $\| \widehat{O}^{\top}\bar{v}_k^*-\hat{v}_l\|\ge\| \bar{v}_k^{\star}-\bar{v}_l^{\star}\| -\Vert\hat{v}_l- \widehat{O}^{\top}\bar{v}_l^* \|\ge3\alpha\sqrt{r}-\alpha\sqrt{r}=2\alpha\sqrt{r}$, then 
\begin{equation*}
\begin{aligned}
\Vert \hat{u}_j-\hat{v}_l\| \ge\| \widehat{O}^{\top}\bar{v}_k^*-\hat{v}_l \Vert-\Vert \widehat{O}^{\top}\bar{v}_k^*-\hat{u}_j\Vert \ge2\alpha\sqrt{r}-\frac{\alpha\sqrt{r}}{3}\ge\frac{5\alpha\sqrt{r}}{3}.
\end{aligned}
\end{equation*}
Then $\hat{u}_j$ can only be assigned to the center $\hat{v}_k$, which indicates that nodes in $J$ are all correctly clustered and those wrongly clustered can only happened in $J^{\rm c}$. Therefore we conclude by \eqref{eq:J_complement} that 
$$
\left|J^{\rm c}\right|\le\frac{18R^{\dagger2}}{\alpha^2}=C_1\kappa_0^2(nm)\cdot R^{\dagger^2}=C_1\kappa_0^6\frac{r^2n\log n}{Lnp_{\max}}
$$
for some absolute constant $C_1>0$. Therefore,
$$
n^{-1}\cdot \calL(\widehat{\bar \VV}, \bar{\VV})\leq n^{-1}|J^{\rm c}|\leq \frac{C_1\kappa_0^6r^2\log n}{Lnp_{\max}}.
$$

\

\subsection{Proof of Thoerem~\ref{thm:exact_recovery}}

The proof of the first claim is identical to the proof of Theorem~\ref{thm:global_consistency} by observing that
\begin{equation}\label{eq:Wsep}
\|(e_{i_1}-e_{i_2})^{\top}\bar{W}\|\geq c_1\sqrt{\frac{m}{L}}\quad{\rm if}\quad \ell_{i_1}\neq \ell_{i_2}
\end{equation}
for some absolute constants $c_1\in(0,1)$. We only prove the second claim. 

Without loss of generality, denote $\widehat W$ the left singular vectors of 
\begin{align*}
\calM_3(\bA)(\widetilde U\otimes \widetilde U)=\calM_3(\EE\bA)(\widetilde{U}\otimes\widetilde{U})+\calM_3(\bDelta)(\widetilde{U}\otimes\widetilde{U})
\end{align*}
where $\bDelta=\bA-\EE\bA$ and $\max_j\|e_j^{\top}\widetilde{U}\|\leq \sqrt{2}\delta_1$ and (by Corollary~\ref{cor:PI}) with probability at least $1-n^{-2}$,
$$
{\rm d}(\widetilde{U},\bar{U})\leq C_3\kappa_0^2\cdot\frac{(r/\sqrt{m})\sqrt{\log n}}{\sqrt{Lnp_{\max}}}
$$
where $C_3>0$ is some absolute constant. Write 
$$
\calM_3(\EE\bA)(\widetilde{U}\otimes\widetilde{U})=\bar{W}\calM_3(\bar{\bC})\big((\bar{U}^{\top}\widetilde{U})\otimes (\bar{U}^{\top}\widetilde{U})\big).
$$
By the fact $\sigma_{\min}(\bar{U}^{\top}\widetilde{U})\geq 1/\sqrt{2}$ (under the lower bound condition of $\sqrt{Lnp_{\max}}$), we have 
$$
\sigma_m\Big(\calM_3(\bar{\bC})\big((\bar{U}^{\top}\widetilde{U})\otimes (\bar{U}^{\top}\widetilde{U})\big)\Big)\geq \sigma_m\big(\calM_3(\bar{\bC})\big)/2\geq \frac{n\sqrt{Lm}p_{\max}}{2r\kappa_0^2}
$$
where the last inequality is due to Lemma~\ref{lem:ss}. Denote the thin singular value decomposition of $\calM_3(\EE\bA)(\widetilde{U}\otimes\widetilde{U})$ by 
$$
\calM_3(\EE\bA)(\widetilde{U}\otimes\widetilde{U})=\calM_3(\bar{\bC})\big((\bar{U}^{\top}\widetilde{U})\otimes (\bar{U}^{\top}\widetilde{U})\big)=\bar{W}\widetilde{D}\widetilde{K}^{\top}
$$
where $\widetilde{D}$ is an $m\times m$ diagonal matrix and $\widetilde{K}\in\OO_{(2r)\times m}$. Meanwhile, denote the thin SVD of $\calM_3(\bA)(\widetilde{U}\otimes\widetilde{U})$ by
$$
\calM_3(\bA)(\widetilde{U}\otimes\widetilde{U})=\widehat{W}\widehat{D}\widehat{K}^{\top}
$$
where $\widehat{D}$ is an $m\times m$ diagonal matrix and $\widehat{K}\in\OO_{(2r)\times m}$. Therefore, 
$$
\widehat{W}\widehat{D}\widehat{K}^{\top}=\bar{W}\widetilde{D}\widetilde{K}^{\top}+\calM_3(\bDelta)(\widetilde{U}\otimes\widetilde{U}).
$$
Recall that $\max_j\|e_j^{\top}\widetilde{U}\|\leq \sqrt{2}\delta_1\leq \kappa_0\sqrt{2r/n}$. By Theorem~\ref{thm:tensor_CI} and Lemma~\ref{lem:tensor_matrix_connection},  with probability at least $1-3n^{-2}$, $\big\|\calM_3(\bDelta)(\widetilde{U}\otimes\widetilde{U})\big\|\leq \sqrt{r\wedge 2m}\|\bDelta\|_{1,\sqrt{2}\delta_1}\leq C_2\kappa_0\sqrt{mrnp_{\max}}\log^2(n)\log(r\kappa_0)$ for some absolute constant $C_2>0$. On the same event, by Davis-Kahan theorem, there exists an orthonormal matrix $\widetilde{O}_1,\widetilde{O}_2\in\OO_{r}$ so that 
$$
\max\big\{\|\widehat{K}-\widetilde{K}\widetilde{O}_2\|,\ \|\widehat{W}-\bar{W}\widetilde{O}_1\|\}\leq C_2\kappa_0^3r^{3/2}\log^2(n)\log(r\kappa_0)/\sqrt{Lnp_{\max}}.
$$
and as a result
$$
\|\widehat{D}-\widetilde{O}_1^{\top}\widetilde{D}\widetilde{O}_2\|\leq C_3\kappa_0\sqrt{mrnp_{\max}}\log^2(n)\log(r\kappa_0).
$$

\noindent Therefore, 
\begin{align*}
\widehat{W}=&\bar{W}\widetilde{D}\widetilde{K}^{\top}\widehat{K}\widehat{D}^{-1}+\calM_3(\bDelta)(\widetilde{U}\otimes\widetilde{U})\widehat{K}\widehat{D}^{-1}\\
=&\bar{W}\widetilde{O}_1(\widetilde{O}_1^{\top}\widetilde{D}\widetilde{O}_2)(\widetilde{K}\widetilde{O}_2)^{\top}\widehat{K}\widehat{D}^{-1}+\calM_3(\bDelta)(\widetilde{U}\otimes\widetilde{U})\widehat{K}\widehat{D}^{-1}\
\end{align*}
implying that 
\begin{align*}
\widehat{W}-\bar{W}\widetilde{O}_1=\bar{W}\widetilde{O}_1\big((\widetilde{O}_1^{\top}\widetilde{D}\widetilde{O}_2)(\widetilde{K}\widetilde{O}_2)^{\top}\widehat{K}\widehat{D}^{-1}-I_m\big)+\calM_3(\bDelta)(\widetilde{U}\otimes\widetilde{U})\widehat{K}\widehat{D}^{-1}.
\end{align*}
Therefore, for any $l\in[L]$, 
\begin{align*}
\big\|e_{l}^{\top}(&\widehat{W}-\bar{W}\widetilde{O}_1) \big\|\\
\leq& \|e_l^{\top}\bar{W}\|\cdot \big\| (\widetilde{O}_1^{\top}\widetilde{D}\widetilde{O}_2)(\widetilde{K}\widetilde{O}_2)^{\top}\widehat{K}\widehat{D}^{-1}-I_m\big\|+\|e_l^{\top}\calM_3(\bDelta)(\widetilde{U}\otimes\widetilde{U})\widehat{K}\widehat{D}^{-1}\|\\
\leq& \|e_l^{\top}\bar{W}\|\cdot \frac{C_2\kappa_0^3r^{3/2}\log^2(n)\log(r\kappa_0)}{\sqrt{Lnp_{\max}}}+\|e_l^{\top}\calM_3(\bDelta)(\widetilde{U}\otimes\widetilde{U})\|\cdot \|\widehat{D}^{-1}\|.
\end{align*}
To bound the last term, write 
\begin{align*}
\|e_l^{\top}\calM_3(\bDelta)(\widetilde{U}\otimes\widetilde{U})\|\leq \|e_l^{\top}\calM_3(\bDelta)(\bar{U}\otimes\bar{U})\|+\|e_l^{\top}\calM_3(\bDelta)\big(\bar{U}\otimes(\widetilde{U}-\bar{U}\widehat{O})\big)\|\\
+\|e_l^{\top}\calM_3(\bDelta)\big((\widetilde{U}-\bar{U}\widehat{O})\otimes \widetilde{U}\big)\|
\end{align*}
where $\widehat{O}=\argmin_{O\in\OO_r}\|\widehat{U}-\bar{U}O\|$. Since $\max_l\|e_l^{\top}\bar{U}\|\leq \delta_1$ and $\max_l\|e_l^{\top}\widehat{U}\|\leq \sqrt{2}\delta_1$, 
\begin{align*}
\|e_l^{\top}\calM_3(\bDelta)(\widetilde{U}\otimes\widetilde{U})\|\leq  \|e_l^{\top}\calM_3(\bDelta)(\bar{U}\otimes\bar{U})\|+2\sqrt{2m}\|\bDelta\|_{1,\sqrt{2}\delta_1}\cdot {\rm d}(\widehat{U},\bar{U})\\
\leq \|e_l^{\top}\calM_3(\bDelta)(\bar{U}\otimes\bar{U})\|+C_3\kappa_0^3r^{3/2}\log^2(n)\log(r\kappa_0)\sqrt{\log(n)/L}
\end{align*}
where the last inequality is due to Corollary~\ref{cor:PI} and Theorem~\ref{thm:tensor_CI}. By Bernstein inequality, it is easy to get that 
$$
\max_l \|e_l^{\top}\calM_3(\bDelta)(\bar{U}\otimes\bar{U})\|\leq C_3r\sqrt{p_{\max}\log n}+C_4\delta_1^2\log n\leq C_3'r\sqrt{p_{\max}\log n}
$$
which holds with probability at least $1-n^{-2}$ and where the last inequality holds for $\delta_1=O(\kappa_0\sqrt{r/n})$ and $Lnp_{\max}\geq \log n$. 

As a result, under the condition (\ref{eq:sparsity_cond}), for all $l\in[L]$,
\begin{align*}
\big\|e_{l}^{\top}(\widehat{W}-\bar{W}\widetilde{O}_1) \big\|\leq  \|e_l^{\top}\bar{W}\|\cdot \frac{C_2\kappa_0^3r^{3/2}\log^2(n)\log(r\kappa_0)}{\sqrt{Lnp_{\max}}}+\frac{C_3r^2\kappa_0^2\sqrt{\log n}}{n\sqrt{Lmp_{\max}}}\\
+\frac{C_4\kappa_0^5r^{5/2}\log^{5/2}(n)\log(r\kappa_0)}{Ln\sqrt{m}p_{\max}}\\
\leq c_1/6\cdot\sqrt{\frac{m}{L}}+\frac{C_3r^2\kappa_0^2\sqrt{\log n}}{n\sqrt{Lmp_{\max}}}+\frac{C_4\kappa_0^5r^{5/2}\log^{5/2}(n)\log(r\kappa_0)}{Ln\sqrt{m}p_{\max}}
\end{align*}
where the constant $c_1$ is the same constant in (\ref{eq:Wsep}). Therefore, if condition (\ref{eq:sparsity_cond}) holds and for some large enough constant $C_1>0$
$$
\sqrt{L}np_{\max}\geq C_1m^{-1}\kappa_0^5r^{5/2}\log^{5/2}(n)\log(r\kappa_0),
$$
then for all $l\in[L]$, 
$$
\|e_l^{\top}(\widehat{W}-\bar{W}\widetilde{O}_1)\|< c_1/5\cdot \sqrt{m/L}
$$
with probability at least $1-3n^{-2}$. On this event, if $\ell_{i_1}=\ell_{i_2}$ for $i_1\neq i_2\in[L]$, then $\|(e_{i_1}-e_{i_2})^{\top}\widehat{W}\|< 2c_1/5\cdot\sqrt{m/L}$.  On the other hand, if $\ell_{i_1}\neq \ell_{i_2}$, then $\|(e_{i_1}-e_{i_2})^{\top}\widehat{W}\|> 3c_1/5\cdot \sqrt{m/L}$.  It suggests that if $\varepsilon\in[0.4c_1\sqrt{m/L}, 0.6c_2\sqrt{m/L}]$, then Algorithm~\ref{algo:network} with parameter $\varepsilon$  and $m$ can exactly recover the network classes. 

\subsection{Proof of Lemma~\ref{lem:initialization}}

We begin with ${\rm d}(\widehat{U}^{(0)},\bar{U})$. By definition, $\widehat{U}^{(0)}$ are the top-$r$ left singular vectors of 
$$
\bA\times_3 {\bf 1}_L^{\top}=\EE\bA\times_3 {\bf 1}_L^{\top}+\bDelta\times_3{\bf 1}_L^{\top}
$$
where $\bDelta=\bA-\EE\bA$. Recall the decomposition (\ref{eq:A_tucker}), $\bA=\bar{\bC}\times_1\bar{U}\times_2\bar{U}\times_3\bar{W}$ and then
\begin{align*}
\bA\times_3 {\bf 1}_L^{\top}=\bar{U}\big(\bar{\bC}\times_3({\bf 1}_L^{\top}\bar{W})\big)\bar{U}^{\top}+\bDelta\times_3{\bf 1}_L^{\top}.
\end{align*}
By definition of $\bar{W}$, it is clear that ${\bf 1}_L^{\top}\bar{W}=\sqrt{\bd_L}$ where $\bd_L=(L_1,\cdots,L_m)$. Therefore, 
$$
\sigma_r\big(\bar{\bC}\times_3({\bf 1}_L^{\top}\bar{W})\big)=\sigma_r(\bar{\bC}\times_3\sqrt{\bd_L}).
$$
By definition, it is obvious that $\|\bDelta\times_3{\bf 1}_L^{\top}\|\leq \sqrt{L}\cdot \|\bDelta\|_{3,1/\sqrt{L}}$. By Theorem~\ref{thm:tensor_CI}, with probability at least $1-3n^{-2}$,
$$
\|\bDelta\times_3{\bf 1}_L^{\top}\|\leq \sqrt{L}\cdot \|\bDelta\|_{3,1/\sqrt{L}}\leq C_2\sqrt{Lnp_{\max}}\log^2n.
$$
Therefore, by Davis-Kahan theorem, 
$$
{\rm d}(\widehat{U}^{(0)},\bar{U})\leq \min\Big\{C_3\frac{\sqrt{np_{\max}}\log^2n}{\sigma_r\big(\bar{\bC}\times_3(\bd_L/L)^{1/2}\big)},\ 2\Big\}
$$
which holds with probability at least $1-3n^{-2}$.

Next, we investigate ${\rm d}(\widehat{W}^{(0)},\bar{W})$. As shown in Theorem~\ref{thm:power_iteration}, as long as 
\begin{equation}\label{eq:init_cond1}
\sigma_r\big(\bar{\bC}\times_3(\bd_L/L)^{1/2}\big)\geq 4C_3\sqrt{np_{\,max}}\log^2n,
\end{equation}
then the regularization can guarantee ${\rm d}(\widetilde{U}^{(0)},U)\leq \sqrt{2}{\rm d}(\widehat{U}^{(0)},\bar{U})$ and $\max_j\|e_j^{\top}\widetilde{U}^{(0)}\|\leq \sqrt{2}\delta_1$. Recall that $\widehat{W}^{(0)}$ are the top-$m$ left singular vectors of 
$$
\calM_3(\bA)(\widetilde{U}^{(0)}\otimes \widetilde{U}^{(0)})=\calM_3(\EE\bA)(\widetilde{U}^{(0)}\otimes \widetilde{U}^{(0)})+\calM_3(\bDelta)(\widetilde{U}^{(0)}\otimes \widetilde{U}^{(0)}).
$$
As shown in the proof of Theorem~\ref{thm:power_iteration}, under Condition (\ref{eq:init_cond1}),
\begin{align*}
\sigma_m\big(\calM_3(\EE\bA)(\widetilde{U}^{(0)}\otimes \widetilde{U}^{(0)})\big)\geq \sigma_m\big(\calM_3(\bar{\bC})\big)/4\geq \sigma_{\min}(\bar{\bC})/4.
\end{align*}
To bound the operator norm of $\calM_3(\bDelta)(\widetilde{U}^{(0)}\otimes \widetilde{U}^{(0)})$, we use the following lemma (\cite[Lemma~6]{xia2017statistically}):
\begin{lemma}\label{lem:tensor_matrix_connection}
For a tensor $\mathbf{A}\in\mathbb{R}^{n_1\times n_2\times n_3}$ with multilinear ranks $(r_1,r_2,r_3)$, the following fact holds for $j=1,2,3$:
$$\Vert\mathcal{M}_j(\mathbf{A})\Vert\le\Vert\mathbf{A}\Vert\sqrt{\frac{(r_1r_2r_3)/r_j}{\max_{j^\prime\ne j}r_j}}$$
\end{lemma}
\noindent By Lemma~\ref{lem:tensor_matrix_connection}, 
$$
\big\|\calM_3(\bDelta)(\widetilde{U}^{(0)}\otimes \widetilde{U}^{(0)})\big\|=\big\|\calM_3(\bDelta\times_1\widetilde{U}^{(0)\top}\times_2\widetilde{U}^{(0)\top})\big\|\leq \sqrt{m}\cdot\|\bDelta\times_1\widetilde{U}^{(0)\top}\times_2\widetilde{U}^{(0)\top}\|.
$$
By the incoherence property of $\widetilde{U}^{(0)}$,
$$
\big\|\calM_3(\bDelta)(\widetilde{U}^{(0)}\otimes \widetilde{U}^{(0)})\big\|\leq \sqrt{m}\|\bDelta\|_{1,\sqrt{2}\delta_1}\leq C_1\delta_1\sqrt{mn^2p_{\max}}\log^2(n)\log(\delta_1^2n)
$$
where the last inequality, due to Theorem~\ref{thm:tensor_CI}, holds with probability at least $1-3n^{-2}$. By Davis-Kahan theorem (\cite{davis1970rotation}), we obtain 
$$
{\rm d}\big(\widehat{W}^{(0)},\bar{W}\big)\leq \min\Big\{C_4\frac{\delta_1\sqrt{mn^2p_{\max}}\log^2(n)\log(\delta_1^2n)}{\sigma_{\min}(\bar{\bC})},\ 2\Big\}
$$
which completes the proof. 

\


\end{document}